%% file: 11ver_pulsation_in_lambda_Boo_stars.tex
\documentclass[a4paper,fleqn,usenatbib]{mnras}
\usepackage{newtxtext,newtxmath}
\usepackage[T1]{fontenc}
\usepackage{ae,aecompl}
\usepackage{graphicx}	% Including figure files
\usepackage{amsmath}	% Advanced maths commands
\usepackage{amssymb}	% Extra maths symbols
\usepackage{booktabs}	% Pretty tables for all!
\usepackage{url}

\usepackage{tablefootnote,footnotehyper} % footnotes in tables
\usepackage{array}
\newcolumntype{H}{>{\setbox0=\hbox\bgroup}c<{\egroup}@{}} % for hiding columns in tables

\title[Pulsation in $\lambda$\,Boo stars]{The Pulsation Properties of $\lambda$ Bootis Stars I. The Southern TESS Sample}

\author[Simon J. Murphy et al.]{
S. J. Murphy$^{1,2}$\thanks{E-mail: simon.murphy@sydney.edu.au (SJM)}, E. Paunzen$^{3}$, T. R. Bedding$^{1,2}$, P. Walczak$^{4}$, and D. Huber$^{5}$
\\
$^{1}$ Sydney Institute for Astronomy (SIfA), School of Physics, University of Sydney, NSW 2006, Australia\\
$^{2}$ Stellar Astrophysics Centre, Department of Physics and Astronomy, Aarhus University, 8000 Aarhus C, Denmark\\
$^{3}$ Department of Theoretical Physics and Astrophysics, Masaryk University, Kotl\'a\v{r}sk\'a 2, 611 37 Brno, Czech Republic\\
$^{4}$ Instytut Astronomiczny, Uniwersytet Wroc\l{}awski, Kopernika 11, 51-622, Wroc\l{}aw, Poland\\
$^{5}$ Institute for Astronomy, University of Hawai`i, 2680 Woodlawn Drive, Honolulu, HI 96822, USA\\
} 

\date{Accepted XXX. Received YYY; in original form ZZZ}
\pubyear{2019}
\begin{document}

\label{firstpage}
\pagerange{\pageref{firstpage}--\pageref{lastpage}}
\maketitle

\begin{abstract}
We analyse TESS light curves for 70 southern $\lambda$\,Boo stars to identify binaries and to determine which of them pulsate as $\delta$\,Scuti stars. We find two heartbeat stars and two eclipsing binaries among the sample. We calculate that 81 percent of $\lambda$\,Boo stars pulsate as $\delta$\,Sct variables, which is about twice that of normal stars over the same parameter space.
We determine the temperatures and luminosities of the $\lambda$\,Boo stars from photometry and \textit{Gaia} DR2 parallaxes.
A subset of 40 $\lambda$\,Boo stars have 2-min TESS data, reliable temperatures and luminosities, and $\delta$\,Sct pulsation. We use Petersen diagrams (period ratios), \'echelle diagrams and the period--luminosity relation to identify the fundamental mode in 20 of those 40 stars and conclude that a further 8 stars are not pulsating in this mode. For the remaining 12, the fundamental mode cannot be unambiguously identified. Further mode identification is possible for 12 of the fundamental mode pulsators that have regular sequences of pulsation overtones in their \'echelle diagrams.
We use stellar evolution models to determine statistically that the $\lambda$\,Boo stars are only superficially metal weak. Simple pulsation models also better fit the observations at a metallicity of $Z=0.01$ than at $Z=0.001$. The TESS observations reveal the great potential of asteroseismology on $\lambda$\,Boo stars, for determining precise stellar ages and shedding light on the origin(s) of the $\lambda$\,Boo phenomenon.
\end{abstract}


%%%%%%%%%%%%%%%%%%%%%%%%%%%%%%%%%%%%%%%%%%%%%
%%%%%%%%%%%%%%%%% BODY OF PAPER %%%%%%%%%%%%%%%%%%

\section{Introduction}
\label{sec:intro}

The present era of space-based photometry has heralded a wave of advances in our understanding of stars. Thanks to asteroseismology, processes that are unconstrainable by other means, such as angular momentum transport \citep{aertsetal2019}, the extent of overshooting around the convective core \citep{johnstonetal2019a} and the existence of strong magnetic fields in stellar cores \citep{stelloetal2016}, are all now much better understood.  Many of these results were made possible by the Kepler Mission \citep{boruckietal2010,gillilandetal2010a}, which was designed to stare at a single, fixed patch of sky. The Transiting Exoplanet Survey Satellite (TESS; \citealt{rickeretal2015}) differs, in that it surveys the sky in segments of 27-d duration, that will cover 85\% of the sky during the two-year nominal mission.  This allows different problems to be addressed by targeting particular classes of stars. TESS is already enabling research on a broad range of stellar astrophysics, including compact pulsators (\citealt{belletal2019}, Charpinet et al. 2019 submitted), chemically peculiar variable stars \citep{cunhaetal2019,sikoraetal2019}, massive stars \citep{pedersenetal2019} and $\delta$\,Sct stars (\citealt{antocietal2019}; Bedding et al. 2020, in press). In this paper, we use TESS to investigate $\lambda$\,Boo stars.
 
The $\lambda$\,Boo stars have spectral types from late-B to early-F and their absorption spectra carry the fingerprints of selective accretion \citep{venn_lambert1990}. They are heavily depleted in refractory (iron-peak) elements by up to 2\,dex \citep{stuerenburg1993}, but have solar abundances of the volatile elements C, N, O, and S \citep{paunzenetal1999a,kampetal2001}. \citet{venn_lambert1990} hypothesized that the peculiarities originate when circumstellar dust grains are separated from gas, and only the latter is accreted by the star. However, the origin of the accreted material is a long-standing puzzle with several contenders that include residual circumstellar or protoplanetary discs \citep{kamaetal2015}, over-dense pockets of the interstellar medium \citep{kamp&paunzen2002,martinez-galarzaetal2009}, and ablated hot Jupiters \citep{jura2015}. Since $\lambda$\,Boo stars appear to span a wide range of stellar ages, there might be multiple processes responsible for the phenomenon \citep{murphy&paunzen2017}. Many young $\lambda$\,Boo stars have been observed to have protoplanetary discs (\citealt{malfaitetal1998b, fisheretal2000, meeusetal2001, mawetetal2017, ligietal2018, cugnoetal2019, grattonetal2019, maciasetal2019, bruzzoneetal2020, perezetal2020}), though the fraction of $\lambda$\,Boo stars with debris discs is not statistically different from that of normal stars \citep{grayetal2017}. Disc accretion, potentially moderated by embedded planets, gives photospheric abundances in good agreement with observations \citep{jermyn&kama2018}.

An important piece of the $\lambda$\,Boo puzzle is the question of whether $\lambda$\,Boo stars are only superficially metal weak (that is, they are peculiar stars), rather than being globally metal weak like the Population II stars. Here, asteroseismology is a valuable tool, probing the stellar interior rather than just the surface. The $\lambda$\,Boo stars are found in and around the instability strips of $\delta$\,Sct and $\gamma$\,Dor variables, which pulsate in pressure and gravity modes, respectively \citep{aertsetal2010}, and some also show Rossby modes \citep{saioetal2018a}. For the $\gamma$\,Dor stars, \textit{Kepler} has been transformational, with 611 class members identified and their pulsation properties characterised \citep{vanreethetal2015b,glietal2020a}. However, the shorter TESS datasets do not have the frequency resolution required for the same level of study. Hence, in this work, we use the TESS data to identify and characterise the intersection of the $\lambda$\,Boo stars and $\delta$\,Sct stars.

The $\delta$\,Sct instability strip extends from mid-A to early-F spectral types \citep{dupretetal2004,murphyetal2019}, but the pulsational driving and damping mechanisms are not fully understood. Is not possible to determine {\it a priori} which of the theoretically possible oscillation modes will be excited, for a given stellar temperature, luminosity and metallicity. In addition, there are many $\delta$\,Sct stars located beyond the theoretical blue edge of the instability strip \citep{bowman&kurtz2018}. A particular curiosity is that even in the middle of the $\delta$\,Sct instability strip, only 70\% of A stars pulsate as $\delta$\,Sct stars, and this number declines rapidly towards the instability strip edges \citep{murphyetal2019}. For the $\lambda$\,Boo stars, the pulsator fraction is expected and observed to be higher, because the He\,{\sc ii} partial ionisation zone that drives the $\delta$\,Sct pulsations is enhanced in helium from the accreted gas \citep{paunzenetal2002a,murphy2014}.

The fundamental mode in normal $\delta$\,Sct stars follows a period--luminosity (P--L) relation \citep{poleskietal2010,mcnamara2011,ziaalietal2019,jayasingheetal2020}, but it is not strict. A colour term is sometimes included to account for the relatively wide temperature range of $\delta$\,Sct stars which, at least in theory, excites different oscillation modes \citep{dupretetal2004}. \citet{paunzenetal2002a} showed that $\lambda$\,Boo stars follow the same period--luminosity--colour (PLC) relation as normal $\delta$\,Sct stars, and a separate metallicity term was not needed. By computing pulsation $Q$ values, they also showed that $\lambda$\,Boo stars generally pulsate in high-overtone modes. Bedding et al. (2020, in press) have found that $\lambda$\,Boo stars are over-represented among their sample of $\delta$\,Sct stars pulsating with regular overtone sequences at high frequencies.

TESS data for southern $\lambda$\,Boo stars constitute the most homogeneous, precise, and continuous time-series photometry available for the class. We describe these data in Sec.\,\ref{sec:data}, along with the stellar parameters that allow us to place these stars in the HR diagram. These data offer the first opportunity to systematically assess the pulsation properties of $\lambda$\,Boo stars. We analyse the pulsator fraction and compare it to that of normal A stars in Sec.\,\ref{sec:puls_frac}. We analyse the Fourier transforms of $\lambda$\,Boo light curves and attempt mode identification in Sec.\,\ref{sec:PLC}. In Sec.\,\ref{sec:models}, we readdress the question of whether the $\lambda$\,Boo stars are globally metal poor. A summary is given in Sec.\,\ref{sec:results} and notes on the pulsation properties of individual stars are given in Appendix\,\ref{sec:notes}.

%%%%%%%%%%%%%%%%%%%%%%%%%%%%%
%%%%%%%%%%%%%%%%%%%%%%%%%%%%%

\section{Data collection and processing}
\label{sec:data}

\subsection{Target selection}

The 70 $\lambda$\,Boo stars studied here originate from four sources: 27 confirmed $\lambda$\,Boo stars from the \citet{murphyetal2015b} catalogue, omitting the probable or uncertain members, and omitting HD\,198161 which is not resolved from its neighbour HD\,198160 in TESS pixels; 20 $\lambda$\,Boo stars from the Tycho sample in \citet{grayetal2017}; nine from the `Misc' sample in the same paper; and 14 from a survey for southern $\lambda$\,Boo stars from Siding Spring Observatory (Murphy et al., submitted). A few targets belong to more than one source.

\subsection{TESS light curves}
\label{ssec:light_curves}

TESS observes each sector for 27\,d and the sectors have overlap near the ecliptic poles. Stars can therefore fall in more than one sector and have light curves of longer durations (in some multiple of 27\,d). Of our 70 targets, 21 have observations in more than one sector. The majority (56) of our TESS targets were observed at 2-min cadence and the rest only have full-frame image (FFI) data at 30-minute cadence. We used all available TESS sectors for each $\lambda$\,Boo star unless they had 2-min data in some sectors but only FFI data in others, in which case we analysed only the 2-min data. 

\citet{jenkinsetal2016} described the TESS science processing operations centre (SPOC) pipeline that produces the 2-min light curves. We downloaded these 2-min data using the {\sc python} package {\sc lightkurve} \citep{lightkurvecollaboration2018}. We used `PDCSAP' flux, which generally produces smooth light curves without any large drifts in flux from systematic effects on the detector. We performed no additional processing on these light curves.

For the 14 $\lambda$\,Boo stars that only have 30-minute FFI data, SPOC light curves are not currently available for download. We used the {\sc python} package {\sc eleanor} \citep{feinsteinetal2019} to extract time series from the FFIs, and kept data with {\it quality flag =} 0. We made light curves using the `corrected' flux, which we found to produce smoother light curves than either the `raw' or `psf' flux options. There were often drifts in flux following the data downlink events, typically lasting for $\sim$1\,d. We excluded these data points from the analysis if we deemed them severe because their inclusion generates strong peaks at low frequencies in the Fourier transforms of the light curves, which can mask lower-amplitude pulsations.

\begin{figure*}
\begin{center}
\includegraphics[width=0.96\textwidth]{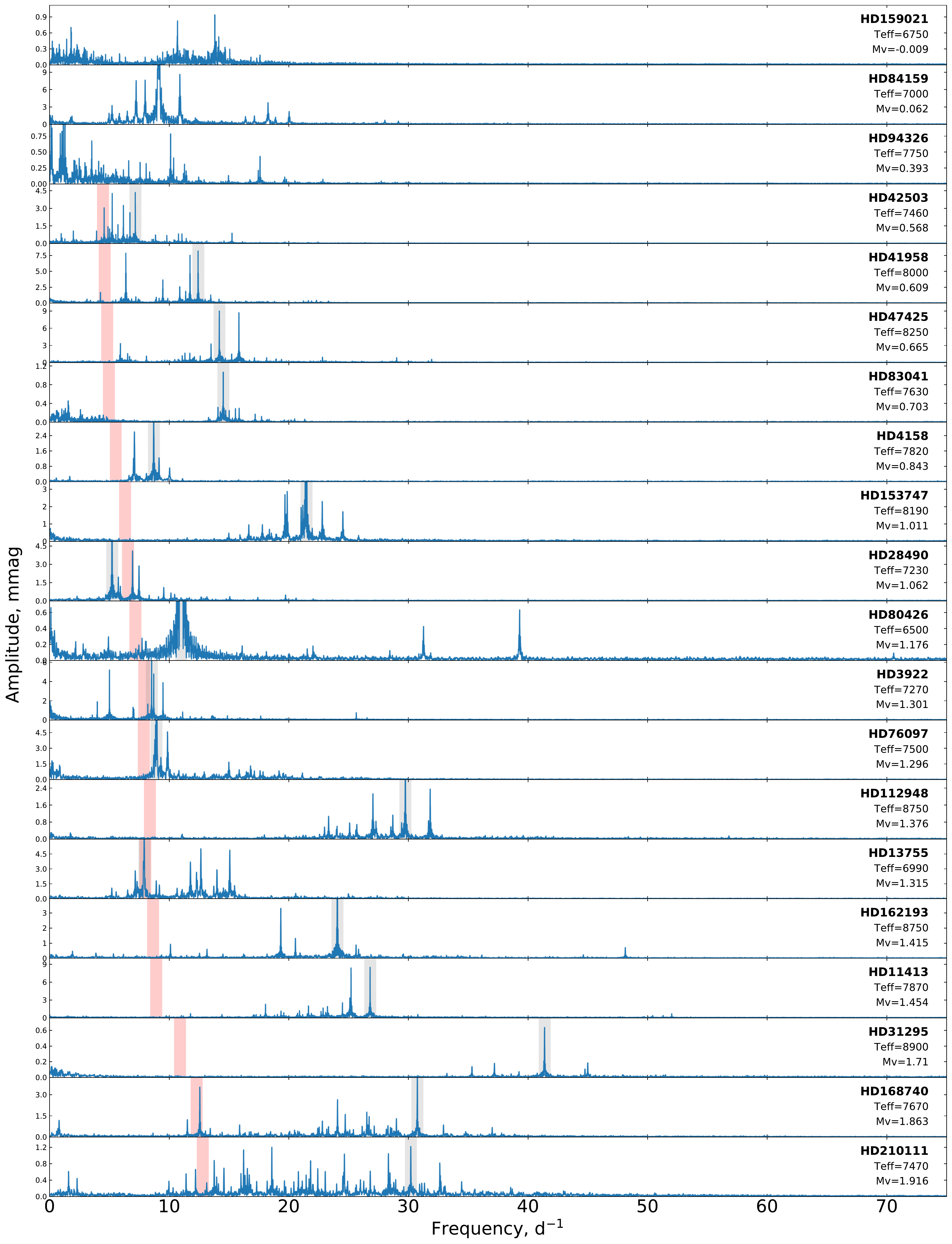}\\
\caption{Fourier amplitude spectra of TESS 2-min data for 20 $\lambda$\,Boo stars that are also $\delta$\,Sct stars, ordered by luminosity (brightest at the top). Only stars with well-determined atmospheric parameters are shown (see Sec.\,\ref{ssec:params}). The strongest peak (Table\:\ref{tab:puls}) is indicated with a grey rectangle; the red rectangles give the predicted location of the fundamental mode based on the period--luminosity relation for $\delta$\,Sct stars (Sec.\,\ref{sec:PLC}). Each panel has its own vertical axis range, which aids visibility of smaller peaks, occasionally at the expense of taller peaks. Asterisks denote stars studied in Bedding et al. (2020, in press) and double asterisks denote stars with \'echelle diagrams shown in that paper. The Fourier transforms of all stars are plotted individually in Appendix.\,\ref{sec:FTs}. Continued in Fig.\,\ref{fig:FT_montage_2}.}
\label{fig:FT_montage}
\end{center}
\end{figure*}

Using the {\sc astropy.stats} package, we computed the Lomb-Scargle periodogram of each TESS light curve, from which we extracted the frequencies and amplitudes of the strongest three peaks above 4\,d$^{-1}$ (Table\:\ref{tab:puls}). We considered peaks with a signal-to-noise ratio (SNR) above 4 to be significant and did not tabulate peaks below this threshold. We computed the SNR by dividing the amplitude of the peak by the average amplitude in a 2-d$^{-1}$ window around that peak. 
Of the 56 $\lambda$\,Boo stars with 2-min TESS data, 47 pulsate as $\delta$\,Sct stars and only 40 of those have reliable temperatures and luminosities (Sec.\,\ref{ssec:params}). We show the Fourier transforms of those 40 stars in Fig.\,\ref{fig:FT_montage} and \ref{fig:FT_montage_2}, ordered by luminosity. Fourier transforms for all 70 stars are provided in Appendix\,\ref{sec:FTs} with the extracted peaks in Table\:\ref{tab:puls} highlighted.

\begin{figure*}
\begin{center}
\includegraphics[width=0.96\textwidth]{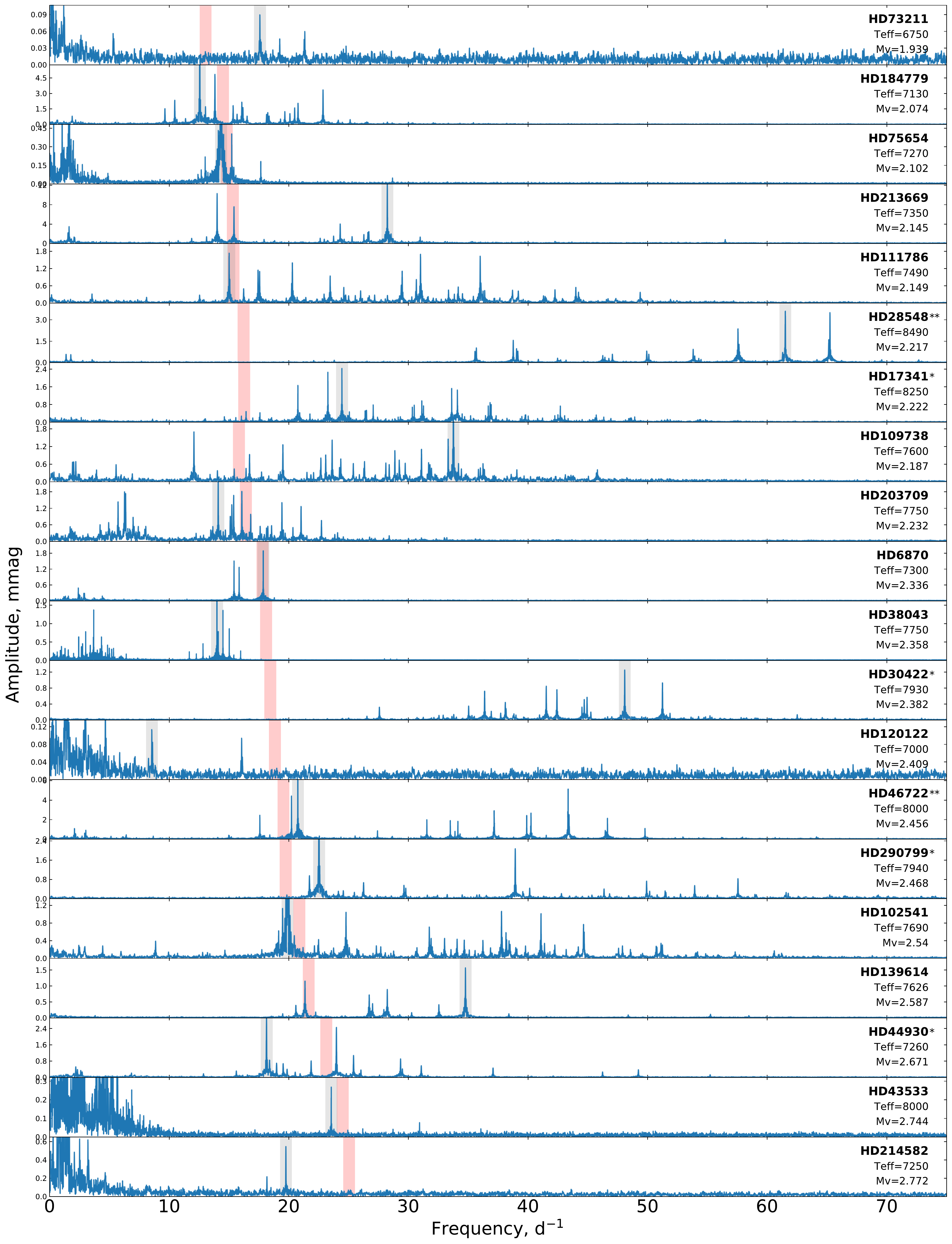}\\
\caption{Same as Fig.\,\ref{fig:FT_montage}, for the less luminous half of the sample. The Fourier transforms of all stars are plotted individually in Appendix.\,\ref{sec:FTs}.}
\label{fig:FT_montage_2}
\end{center}
\end{figure*}

Peaks at frequencies below $4$\,d$^{-1}$ are not pressure modes and are not included in Table\:\ref{tab:puls}. However, if such peaks were present, we reported them in the notes on individual stars (Appendix\,\ref{sec:notes}). Although substantial progress has recently been made on the interpretation of g and r\:modes in A/F stars \citep{kurtzetal2015,vanreethetal2016,saioetal2018a,saioetal2018b,glietal2019b,ouazzanietal2019}, this requires light curves with a time-base of at least a few years for enough frequency resolution to study the pulsation modes. We considered $\lambda$\,Boo stars that have data from Year 1 of TESS observations (Sectors 1--13). Figures\,\ref{fig:FT_montage} and \ref{fig:FT_montage_2} show power excesses at these frequencies in many stars, but their detailed analysis is not possible given the relatively short durations of the light curves.

\begin{table*}
\caption{Pulsation properties, including frequencies ($f_i$) and amplitudes ($a_i$), of the 70 $\lambda$\,Boo stars in this study. The final three columns indicate the 14 stars with only 30-min FFI data, the upper limit on any undetected pulsation modes for the 16 $\lambda$\,Boo stars that are not $\delta$\,Sct pulsators, and the TESS sectors in which each target was observed.}
\input{main_table.tex}
\label{tab:puls}
\end{table*}

\setcounter{table}{0}
\begin{table*}
\caption{{\bf continued.} Note that HD\,198160 and HD\,198161 are separated by less than one TESS pixel and are considered as a single object in this paper (see also Appendix\,\ref{sec:notes}).}
\input{main_table_part2.tex}
\end{table*}

\subsection{Stellar parameters}
\label{ssec:params}

The position of $\lambda$\,Boo stars on the HR diagram is important in many contexts. If $\lambda$\,Boo stars are young and accreting material left over from star or planet formation then they should lie near the ZAMS. In addition, and regardless of their age, their position with respect to the $\delta$\,Sct instability strip is important for understanding mode excitation and for an asteroseismic interpretation of the global stellar metallicity and age. Finally, accurate luminosities allow us to use the period--luminosity relation to identify the fundamental mode in some stars.

To determine stellar luminosities, we used Gaia DR2 parallaxes where available \citep{lindegrenetal2018}. Two program stars have no parallax data in either of the Gaia and Hipparcos catalogues. We still discuss the pulsation properties of these stars, but they are not included in analyses where a luminosity is required (Sections\,\ref{sec:puls_frac}--\ref{sec:models}). 

We determined $T_{\rm eff}$ from Geneva 7-colour photometry for the 44 stars for which this was available, since this approach results in the smallest uncertainties. To calibrate the Geneva photometry, we used the grids published by \citet{kunzlietal1997}, which are valid for B- to G-type stars. Note that for stars cooler than B9, no reddening-free indices exist within this photometric system. Therefore, to get reliable temperature values, reddening values must be provided as input. We also determined $T_{\rm eff}$ from spectral energy distributions (SEDs) using the VOSA (Virtual Observatory Sed Analyser) tool v6.0 \citep{bayoetal2008}, which was also the primary source of $T_{\rm eff}$ for targets without Geneva photometry. SEDs offer a precision of $\sim$250\,K. No outliers were detected among the 44 targets with $T_{\rm eff}$ determined by both methods: the two methods agree to within 1$\sigma$ for 67\% of the overlapping sample. Our final adopted uncertainties on the derived effective temperatures were fixed to be at least 2\%, which is typical of the uncertainty set by the accuracy of measured interferometric angular diameters \citep{casagrandeetal2014,whiteetal2018}.

The extinction in the $V$ band, $A_V$, was computed using the same photometry that we used to calculate the effective temperatures. For the 27 stars where this photometry was lacking, we used the python package {\sc mwdust} \citep{bovyetal2016} to extract extinctions from the combined (default) dust map using the target coordinates and distances.

Using the stellar temperatures, parallaxes, $A_V$, and $V$-band bolometric corrections, we calculated bolometric luminosities using the ``direct mode'' of the {\sc isoclassify} package \citep{huberetal2017}. We adopted 0.03\,mag uncertainties in reddening and bolometric corrections. To estimate stellar masses, we fitted the effective temperatures and luminosities derived in the previous step to MIST isochrones (MESA Isochrones and Stellar Tracks; \citealt{dotter2016,choietal2016}) using the ``grid mode'' of \texttt{isoclassify}, assuming a solar-neighbourhood metallicity prior. The procedure also yielded estimates of surface gravities, which were used for the interpolation of bolometric corrections in the previous step. To ensure consistency we performed one iteration between the ``direct mode'' and ``grid mode'' calculations. The resulting stellar parameters are given in Table\:\ref{tab:stellar_params}.

\begin{figure}
\begin{center}
\includegraphics[width=0.48\textwidth]{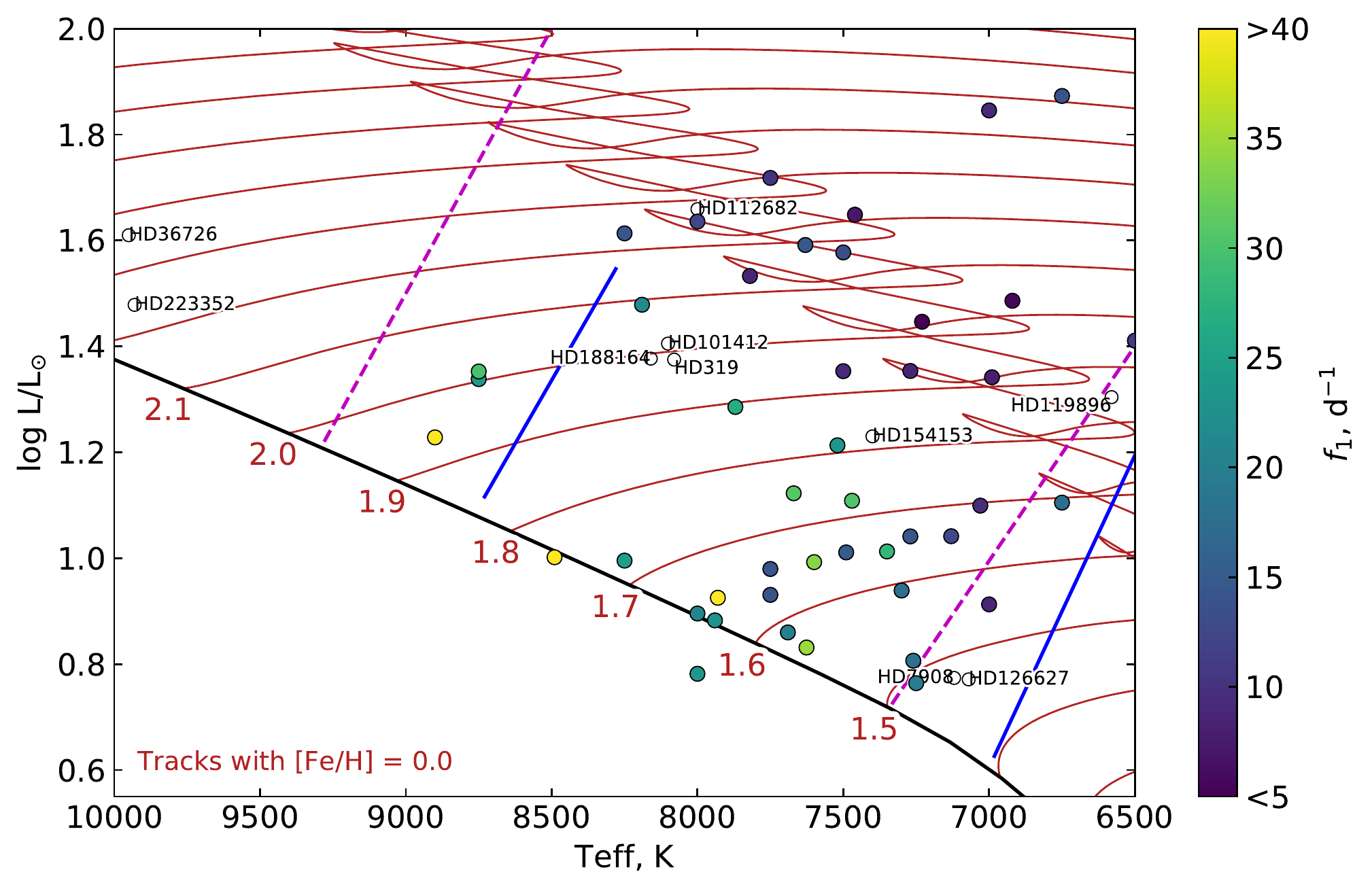}\\
\includegraphics[width=0.48\textwidth]{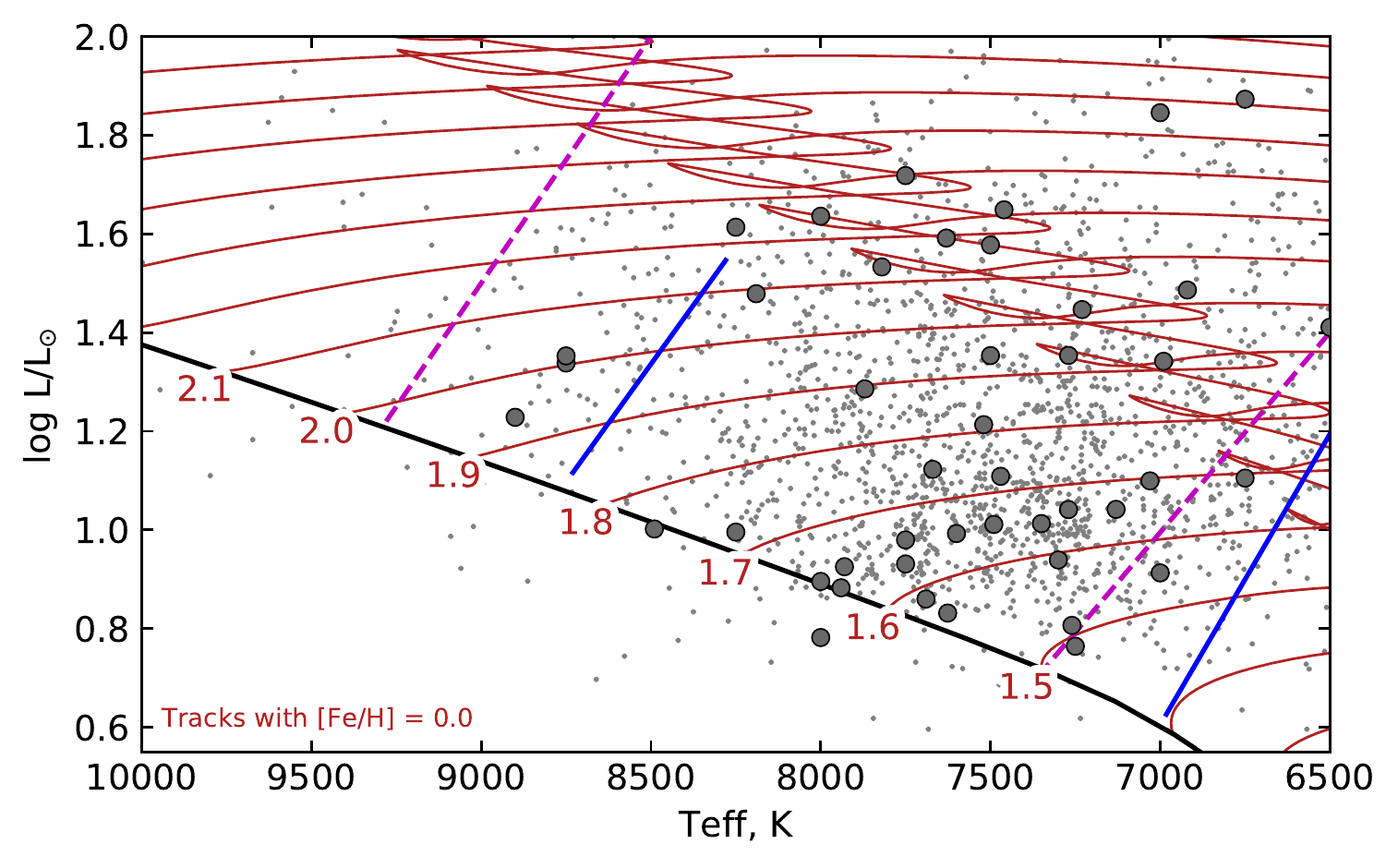}\\
\caption{Top: HR diagram of the southern $\lambda$\,Boo stars observed by TESS. Empty circles show non-pulsators while pulsators are shown as filled circles colour-coded by the strongest oscillation frequency. Stars with $A_V>1.0$\,mag, RUWE>2.0, or a nearby bright companion in the WDS catalogue are not shown. Bottom: The same pulsating $\lambda$\,Boo stars are shown with the \textit{Kepler} $\delta$\,Sct stars for comparison, corrected for differences in sensitivity (see Sec.\,\ref{sec:puls_frac}).}
\label{fig:HRD}
\end{center}
\end{figure}

We identified four binary systems in the sample from their TESS light curves: HD\,31508 and HD\,94390, which we discovered to be heartbeat stars (Appendix\,\ref{sec:notes}); and HD\,168947 and HD\,74423, which we found to be eclipsing binaries (the latter was also discussed by \citealt{handleretal2020}). We searched the Washington Double Star (WDS) catalog \citep{masonetal2001} for close binaries (<10$''$ separation), and found three objects with magnitude differences less than 2.5\,mag (HD\,15164, HD\,198160, and HD\,170680). Finally, we also used the renormalised unit weight error (RUWE) from Gaia DR2 to identify probable binaries. Objects with RUWE$>$2.0 are likely to be binaries \citep{evans2018,rizzutoetal2018} and three such binaries were identified, all of which were identified in our WDS search. Other quality controls, such as a filter on targets with more than 1.0\,mag of extinction, did not identify any additional stars whose parameters might be questionable. We plot the HR diagram of 58 stars in Fig.\,\ref{fig:HRD}, excluding from the original 70 stars the 7 stars found to be binaries, four additional stars above 10\,000\,K that are too hot to be $\delta$\,Sct pulsators (HD\,23392, HD\,100546, HD\,127659 and HD\,294253), and HD169142, which has anomalous stellar parameters. It is this sample of 58 stars that we use in Sections\,\ref{sec:puls_frac}--\ref{sec:models}; the boolean column ``good'' in Table\:\ref{tab:stellar_params} indicates which stars passed the filter.

\begin{table*}
\caption{Stellar parameters for the southern $\lambda$\,Boo stars used in this paper. Columns are: the HD number or other identifier, $V$-band magnitude, effective temperature, luminosity, absolute V magnitude, surface gravity, metallicity, mass, extinction in the $V$ band, Renormalised Unit Weight Error, a boolean flag indicating whether the parameters are `good' (see text), the magnitude difference from any companion in the WDS catalog, and the corresponding WDS separation.}
\input{southern_tess_LB_params_v5_part1.tex}
\label{tab:stellar_params}
\end{table*}

\setcounter{table}{1}
\begin{table*}
\caption{{\bf continued.} Stellar parameters for the southern $\lambda$\,Boo stars used in this paper.}
\input{southern_tess_LB_params_v5_part2.tex}
\end{table*}

%%%%%%%%%%%%%%%%%%%%%%%%%%%%%
%%%%%%%%%%%%%%%%%%%%%%%%%%%%%

\section{Pulsator fraction analysis}
\label{sec:puls_frac}

It has been known for some time that the fraction of $\lambda$\,Boo stars showing $\delta$\,Sct pulsations is high, but this has been difficult to quantify. Collecting precise, continuous and lengthy time-series photometry from the ground is difficult, requiring multi-site coordination and being limited by the effects of Earth's atmosphere. This resulted in heterogeneous surveys for variability, with short time-series of differing precision that did not include all group members. Despite these challenges, \citet{paunzenetal2002a} were able to conclude that the pulsator fraction of $\lambda$\,Boo stars is at least 70\%. It was difficult to assess whether this pulsator fraction is abnormally high due to the lack of a suitable reference against which to evaluate it because variability surveys of normal $\delta$\,Sct stars suffered from the same issues, but in greater numbers. Now, thanks to \textit{Kepler} and TESS, homogeneous samples exist that make meaningful comparisons possible.

To assess whether $\lambda$\,Boo stars have a higher pulsator fraction than normal A/F stars, we consider their position on the HR diagram using the stellar parameters determined in Sec.\,\ref{ssec:params}. Several $\lambda$\,Boo stars lie outside the observed and theoretical $\delta$\,Sct instability strips, so it is unsurprising that they would not pulsate. 
 \citet{murphyetal2019} recently assessed the pulsator fraction of normal A/F stars in the \textit{Kepler} field as a function of position on the HR diagram, providing a benchmark against which to assess the $\lambda$\,Boo stars.
 
The sensitivity of TESS is different from \textit{Kepler} for two principal reasons: (i) the effective passband (governed by the CCD response function) is redder for TESS, which leads to TESS pulsation amplitudes being 16.5\% lower than their \textit{Kepler} equivalents \citep{lund2019}; and (ii) TESS light curves are shorter in duration so the Fourier noise is higher. To correct for this sensitivity difference, we estimated the average Fourier noise of \textit{Kepler} and TESS light curves by firstly finding the median Fourier amplitude for each target (this is a good estimate of the noise level, \citealt{murphyetal2019}), and then taking the median of the distribution of these noise levels. We found that the median noise level is 13 times greater in the TESS data. Following our significance criterion of SNR=4 for detecting p\:modes in the TESS data, we concluded that \textit{Kepler} pulsators would typically be detectable with TESS if the \textit{Kepler} pulsation SNR is at least ($1.165 \times 13.3 \times 4 =$) 62.0. Since the median \textit{Kepler} noise is 0.00133\,mmag, this corresponds to an amplitude limit of 0.082\,mmag. That is, a peak with an observed amplitude above 0.082\,mmag in \textit{Kepler} data should be detectable in TESS data.\footnote{Note that this calculation does not correct for amplitude smearing caused by the integrations being a large fraction of the pulsation period in 30-min data \citep{chaplinetal2011,murphy2012a}; for our targets, which have a median $f_1$ value of 18\,d$^{-1}$ the correction is only 2.25\%. Further, for FFI data, the exposure times are almost identical to the \textit{Kepler} ones so no correction would apply.} Adopting this amplitude limit, we found that 1826 of the 1988 \textit{Kepler} $\delta$\,Sct stars in the \citet{murphyetal2019} sample would have been detectable at TESS precision, and we used this sample of 1826 \textit{Kepler} stars as our comparison sample in the following pulsator fraction calculations.
 
Using the same method as \citet{murphyetal2019}, we used the 1826 $\delta$\,Sct stars to calculate the fraction of pulsators as a function of position in the HR diagram. Using grid cells of approximately 200\,K in $T_{\rm eff}$ and 0.1\,dex in $\log L$, we computed the pulsator fraction at the position of each $\lambda$\,Boo star and treated this as the probability that the star would be observed to pulsate by TESS. Uncertainties were estimated by taking the standard deviation of the pulsator fraction in neighbouring cells, which are approximately 1$\sigma$ away in the observed quantities, $T_{\rm eff}$ and $\log L$. Using this model, we computed an expected pulsator fraction of 0.41 $\pm$ 0.05 for our stars. The observed fraction is much higher: 47 of the 58 $\lambda$\,Boo stars with reliable atmospheric parameters are observed to pulsate, giving a pulsator fraction of 0.81. Thus the fraction of $\lambda$\,Boo stars that show $\delta$\,Sct pulsations is about twice that of normal stars.

%%%%%%%%%%%%%%%%%%%%%%%%%%%%%
%%%%%%%%%%%%%%%%%%%%%%%%%%%%%

\section{Mode Identification for $\lambda$\,Boo stars}
\label{sec:PLC}

There has been significant recent progress in mode identification for $\delta$\,Sct stars with the discovery of regular patterns of peaks in high-frequency $\delta$\,Sct stars (Bedding et al. 2020, in press). In these stars, it is possible to see ridges in an \'echelle diagram when the correct frequency separation, $\Delta\nu$, is used, and this separation is approximately proportional to the square root of the mean stellar density \citep{aertsetal2010}. Bedding et al (2020, in press) have shown that ridges of $\ell=0$ and $\ell=1$ modes can be identified, and the radial order $n$ can be inferred. They were able to match the pulsation modes to models to determine stellar ages much more precisely than by positioning the star on an HR diagram using its {\it Gaia} parallax.

Of particular interest is the fact that $\lambda$\,Boo stars are often high-frequency pulsators \citep{paunzenetal2002a} and are also overrepresented in the Bedding et al. (2020, in press) sample (6 out of 60). This makes application of the \'echelle diagram to the $\lambda$\,Boo stars in this paper attractive. Although we do not attempt to model individual oscillation modes here, identifying candidates for future asteroseismic modelling and establishing a provisional mode identification is an important first step.

\begin{figure}
\begin{center}
\includegraphics[width=0.4\textwidth]{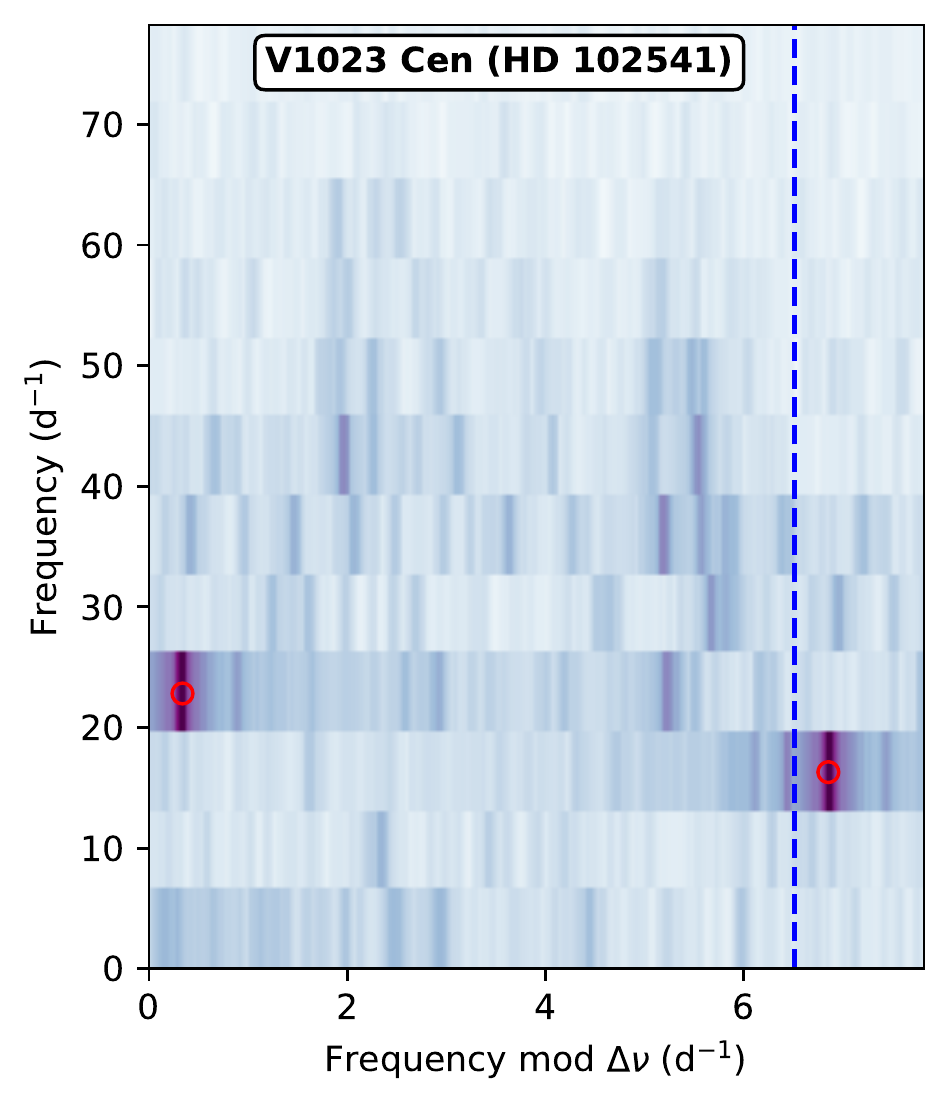}\\
\caption{\'Echelle diagram of HD\,102541, showing nearly vertical ridges at $x\approx2$ and $x\approx5.5$\,d$^{-1}$. The region right of the dashed blue is duplicated for clarity. The red circle at the left of order 4 highlights the fundamental mode.}
\label{fig:echelle_102541}
\end{center}
\end{figure}

While there remains progress to be made in interpreting the subtleties of the $\ell=0$ and $\ell=1$ ridges, the \'echelle diagrams can be used to identify the fundamental mode in some stars. For the 18\,$\delta$\,Sct stars in Bedding et al. (2020, in press) for which the fundamental mode was identifiable, the frequency of this mode was close to three times the large separation (their extended data figure 2). In other words, the fundamental mode lies at the right of the third order or the left of the fourth order in the \'echelle. Twelve of the 40 stars in Figures\,\ref{fig:FT_montage} and \ref{fig:FT_montage_2} show ridges in their \'echelle diagrams that enabled $\Delta\nu$ to be measured and the fundamental mode to be determined. An example is shown in Fig.\,\ref{fig:echelle_102541}, created with the {\sc echelle} package \citep{hey&ball2020}.

A second tool for identifying the fundamental mode in stars with well-determined parallaxes is via the period--luminosity relation, already discussed in the Introduction. In this work, we use a P--L relation to help identify the fundamental mode.

Finally, some $\delta$\,Sct stars pulsate in low-overtone radial modes that have a known period ratio. Particularly for high-amplitude pulsators, period ratios have been used to identify the fundamental mode and the first- and second-overtone modes using so-called Petersen diagrams \citep{petersen&christensen-dalsgaard1996}. Broadly speaking, the fundamental and first-overtone modes have a period ratio of $\sim$0.77 and the first- and second-overtone modes have a ratio close to 0.80. However, the ratios depend on: (i)~the stellar density, and by extension, on the period of the fundamental mode; (ii)~the stellar metallicity, with metal-poor stars having slightly higher ratios \citep{petersen&christensen-dalsgaard1996}; (iii)~the stellar rotation, which shifts the ratios by half a percent even for modest rotation of $v=50$\,km\,s$^{-1}$ \citep{suarezetal2006}; and (iv)~near-degeneracy effects such as mode coupling, which imparts extra scatter in the ratios \citep{suarezetal2007}.

When used alone, the above three methods sometimes leave considerable uncertainty in mode identification, but when two or more methods agree the identification is stronger. In addition, the fundamental mode usually has a high amplitude if it is excited, and it is the lowest frequency pressure mode that can be excited, though mixed modes and g\:modes can occur at lower frequencies. Hence, a dominant peak of low frequency ($\sim$4--25\,d$^{-1}$) identifiable with one or more of the three methods above is likely to be the fundamental mode.

Using these tools, in Table\:\ref{tab:fundamental} we provide the most likely identification for the fundamental mode in the 40 pulsating $\lambda$\,Boo stars shown in Figures\,\ref{fig:FT_montage} and \ref{fig:FT_montage_2}. These are the stars with good parallaxes, on which the P--L relation is likely to be useful, and with 2-min data, so we can be sure that the observed frequencies are not Nyquist aliases. Not all stars pulsate in the fundamental mode, and in some stars there is no clear candidate from the many excited modes. Nonetheless, the fundamental mode was identifiable in 20 of the 40 stars, and for an additional 8 stars we can conclude that the the fundamental mode is not excited. Our results show that the majority of $\lambda$\,Boo stars do pulsate in the fundamental mode.

\begin{table}
\centering
\caption{Most likely identification of the fundamental mode for the 40 pulsating $\lambda$\,Boo stars in Figures\,\ref{fig:FT_montage} and \ref{fig:FT_montage_2}. Stars are listed in the same order as the figures. The right column summarises the method(s) used in the identification; more detail is given in the notes on individual stars in Appendix\,\ref{sec:notes}. Some stars clearly do not pulsate in the fundamental mode ``<not excited>'' and for some stars it is not possible to identify the mode ``<unclear>''.}
\begin{tabular}{l r l}
\toprule
HD & Fundamental & Method(s)\\
 & d$^{-1}$ &\\
\midrule
HD\,159021 & 10.707 & Petersen\\
HD\,84159 & <unclear> & \\
HD\,94326 & <unclear> &\\
HD\,42503 & <unclear> & \\
HD\,41958 & <unclear> & \\
HD\,47425 & <unclear> & \\
HD\,83041 & <unclear> & \\
HD\,4158 & 7.106 & Petersen\\
HD\,153747 & <not excited> & P--L\\
HD\,28490 & 6.948 & \'echelle, P--L\\
HD\,80426 & <unclear> & \\
HD\,3922 & <unclear> & \\
HD\,76097 & <unclear> & \\
HD\,112948 & <not excited> & P--L\\
HD\,13755 & 7.925 & P--L, Petersen\\
HD\,162193 & 10.129 & P--L\\
HD\,11413 & 11.794 & \'echelle\\
HD\,31295 & <not excited> & P--L\\
HD\,168740 & 12.576 & \'echelle, P--L\\
HD\,210111 & 12.215 & \'echelle, P--L\\
HD\,73211 & <not excited> & P--L, Petersen\\
HD\,184779 & 12.568 & P--L, Petersen\\
HD\,75654 & 14.340 & P--L\\
HD\,213669 & 14.017 & P--L\\
HD\,111786 & 15.023 & P--L\\
HD\,28548 & <not excited> & \'echelle, P--L\\
HD\,17341 & <unclear> & \\
HD\,109738 & <unclear> & \\
HD\,203709 & <unclear> & \\
HD\,6870 & 17.873 & P--L\\
HD\,38043 & <not excited> & multiplets\\
HD\,30422 & <not excited> & \'echelle\\
HD\,120122 & <not excited> & P--L\\
HD\,46722 & 20.764 & \'echelle, P--L\\
HD\,290799 & 22.534 & \'echelle, P--L\\
HD\,102541 & 19.898 & \'echelle, P--L\\
HD\,139614 & 21.365 & \'echelle, P--L\\
HD\,44930 & 18.152 & \'echelle\\
HD\,43533 & 23.558 & P--L\\
HD\,214582 & 19.763 & P--L\\
\bottomrule
\end{tabular}
\label{tab:fundamental}
\end{table}

%%%%%%%%%%%%%%%%%%%%%%%%%%%%%
%%%%%%%%%%%%%%%%%%%%%%%%%%%%%

\section{Are $\lambda$\,Boo stars globally metal poor?}
\label{sec:models}

Spectroscopy and multi-colour photometry are sensitive to the metallicity of the photosphere, whereas evolutionary and pulsation models are sensitive to the global metallicity of the star. In combination, these techniques can determine whether $\lambda$\,Boo stars are superficially or globally metal weak. The most detailed analysis of the interior metallicity of a $\lambda$\,Boo star was done by \citet{casasetal2009}, who used asteroseismology to determine that 29\,Cyg is globally metal poor. However, the models were quite poorly constrained, with the 1$\sigma$ results spanning a mass range of 1.67 to 2.09\,M$_{\odot}$ in the solar-metallicity case, or 1.42 to 1.81\,M$_{\odot}$ in the globally metal poor case. The \citet{casasetal2009} result also remains at odds with the analysis by \citet{paunzenetal2015}, who computed stellar evolution and pulsation models of some $\lambda$\,Boo stars that had SuperWASP photometry, reaching the opposite conclusion. Since then, extremely precise distances have become available from {\it Gaia} DR2, allowing much better constrained models of $\lambda$\,Boo stars to be constructed.

We revisit the topic of global metallicities for $\lambda$\,Boo stars using a similar two-pronged methodology to \citet{paunzenetal2015}. Firstly, in Sec.\,\ref{ssec:evo_models} we compare the positions on the HR diagram of the 58 $\lambda$\,Boo stars having reliable atmospheric parameters (see Sec.\,\ref{ssec:params}) compared to stellar evolution models of two different metallicities. Secondly, in Sec.\,\ref{ssec:puls_models} we investigate pulsation models for a subsample of eight well-studied southern $\lambda$\,Boo stars (listed in Table\:\ref{tab:starlist}), chosen for having detailed abundance analyses in the literature. 

\begin{table*}
	\begin{center}
		\caption{Parameters of the eight stars chosen for pulsational analysis. Columns are the Henry Draper catalogue number, visual magnitude, projected surface rotation velocity (references in final column), Gaia parallax \citep{gaiacollaboration2016,gaiacollaboration2018a}, effective temperature and luminosity (Table\:\ref{tab:stellar_params}), stellar radius (calculated), and literature references to spectroscopic abundance analyses.}
		\begin{tabular}{rcrccccl}
			\toprule
HD & $V$  &  $v\sin{i}$  &  $\pi$ & $\log (T_{\rm eff}/{\rm K}$) & $\log{(L/{\rm L}_{\sun})}$ & $R$ & Literature abundances \\
			 & mag & km\,s$^{-1}$ &  mas & & & R$_{\odot}$  &\\
			\midrule			
319 & 5.923 & 60 &$ 11.81 \pm 0.14 $&$ 3.9074 \pm 0.0086 $&$ 1.375 \pm 0.020 $&$ 2.48 \pm 0.13 $ & \citet{stuerenburg1993,paunzenetal1999a}\\
11413 & 5.929 & 125 &$ 12.726 \pm 0.045 $&$ 3.8960 \pm 0.0086 $&$ 1.286 \pm 0.017 $&$ 2.36 \pm 0.12 $ & \citet{stuerenburg1993,paunzenetal1999a}\\
31295 & 4.648 & 115 &$ 29.21 \pm 0.53 $&$ 3.9494 \pm 0.0111 $&$ 1.228 \pm 0.033 $&$ 1.73 \pm 0.13 $ & \citet{venn_lambert1990,stuerenburg1993}\\
&&&&&&&\citet{paunzenetal1999a,paunzenetal1999b,kampetal2001}\\
36726 & 8.84 & 80 &$ \phantom{1}3.017 \pm 0.064 $&$ 3.9978 \pm 0.0179 $&$ 1.610 \pm 0.025 $&$ 2.15 \pm 0.20 $ & \citet{andrievskyetal2002}\\
75654 & 6.368 & 45 &$ 13.972 \pm 0.032 $&$ 3.8615 \pm 0.0086 $&$ 1.041 \pm 0.017 $&$ 2.09 \pm 0.10 $ & \citet{solanoetal2001,kampetal2001}\\
210111 & 6.386 & 30 &$ 12.714 \pm 0.056 $&$ 3.8733 \pm 0.0086 $&$ 1.109 \pm 0.017 $&$ 2.14 \pm 0.10 $ & \citet{stuerenburg1993,paunzenetal1999a}\\
&&&&&&&\citet{solanoetal2001}\\
290799 & 10.8 & 70 &$ \phantom{1}2.273 \pm 0.063 $&$ 3.8998 \pm 0.0086 $&$ 0.883 \pm 0.036 $&$ 1.46 \pm 0.11 $ & \citet{andrievskyetal2002}\\
294253 & 9.67 & 70 &$ \phantom{1}2.381 \pm 0.081 $&$ 4.0245 \pm 0.0086 $&$ 1.509 \pm 0.034 $&$ 1.69 \pm 0.12 $ & \citet{andrievskyetal2002}\\
    			\bottomrule
			\label{tab:starlist}
		\end{tabular}
	\end{center}
\end{table*}

\subsection{Stellar Evolution Models}
\label{ssec:evo_models}

We first consider the location of our target stars on the Hertzsprung-Russell (HR) diagram, with respect to evolutionary tracks of two different metallicities ($Z=0.010$ and 0.001) and two values of overshooting from the hydrogen-burning convective core ($f_{\rm{ov}}=0.00$ and 0.02;  \citealt{herwig2000}). We calculated the evolutionary tracks with {\sc mesa} \citep{paxtonetal2011, paxtonetal2013, paxtonetal2015, paxtonetal2018, paxtonetal2019} ver. 11554. We adopted an initial hydrogen abundance of $X_{\rm{ini}}=0.7$, and used OPAL opacity tables \citep{iglesias&rogers1996} supplemented with the data provided by \cite{fergusonetal2005} for low temperatures. The solar chemical element mixture was used, as determined by \cite{grevesse&noels1993}. We used the {\sc mesa} equation of state, which is a blend of the equation of state from OPAL \citep{rogers&nayfonov2002}, SCVH \citep{saumonetal1995}, PTEH \citep{polsetal1995}, HELM \citep{timmes&swetsy2000}, and PC \citep{pohtekhin&chabrier2010}. Nuclear reaction rates were from the JINA REACLIB database \citep{cyburtetal2010} plus additional tabulated weak reaction rates \citep{fuller1985, odaetal1994, langanke&martinezpinedo2000}. Screening was included via the prescriptions of \citet{salpeter1954, dewittetal1973, alastuey&jancovici1978} and \citet{itohetal1979}. Thermal neutrino loss rates were from \citet{itohetal1996}.

\begin{figure*}
\begin{center}
\includegraphics[width=1\columnwidth]{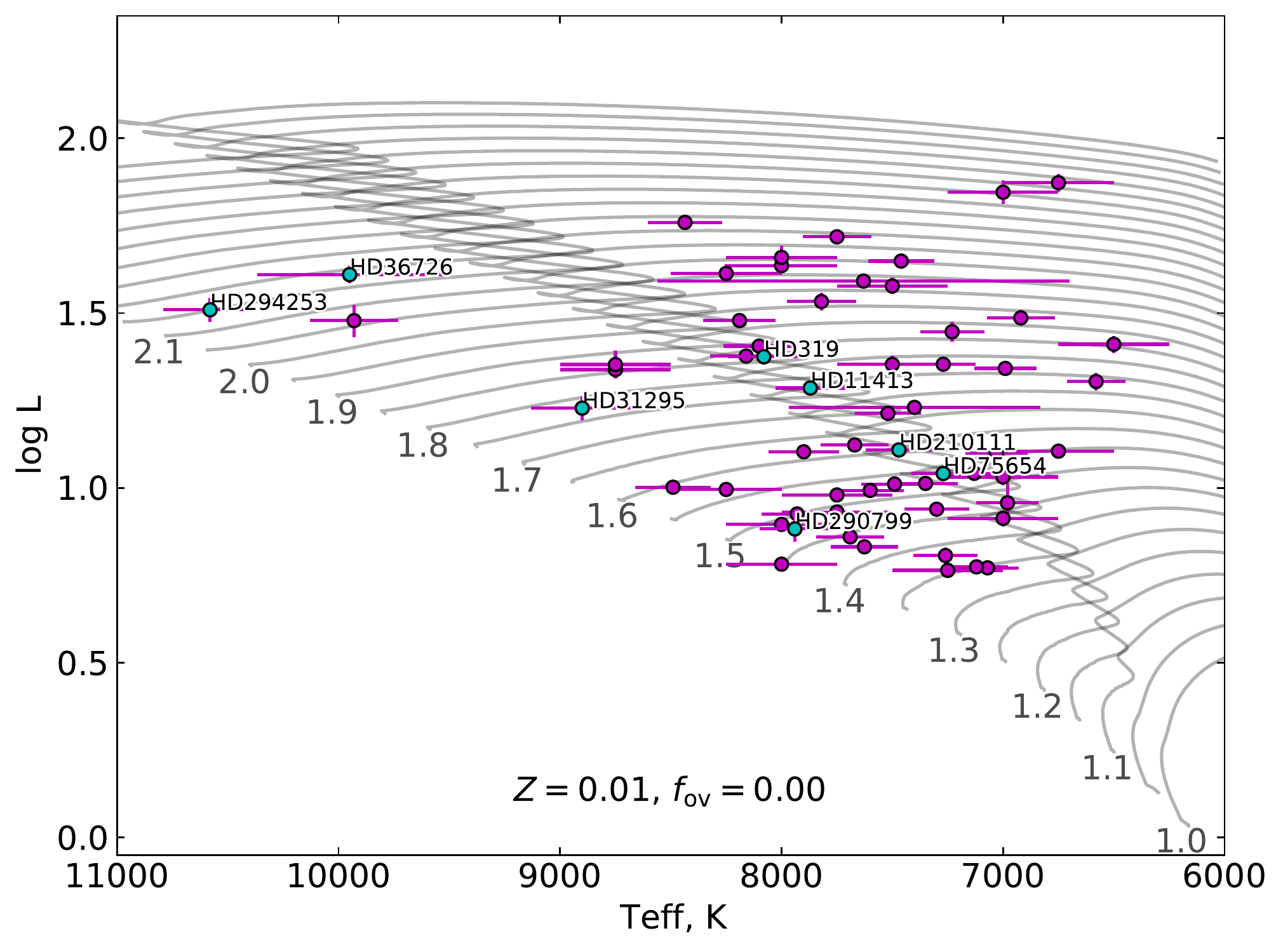}
\includegraphics[width=1\columnwidth]{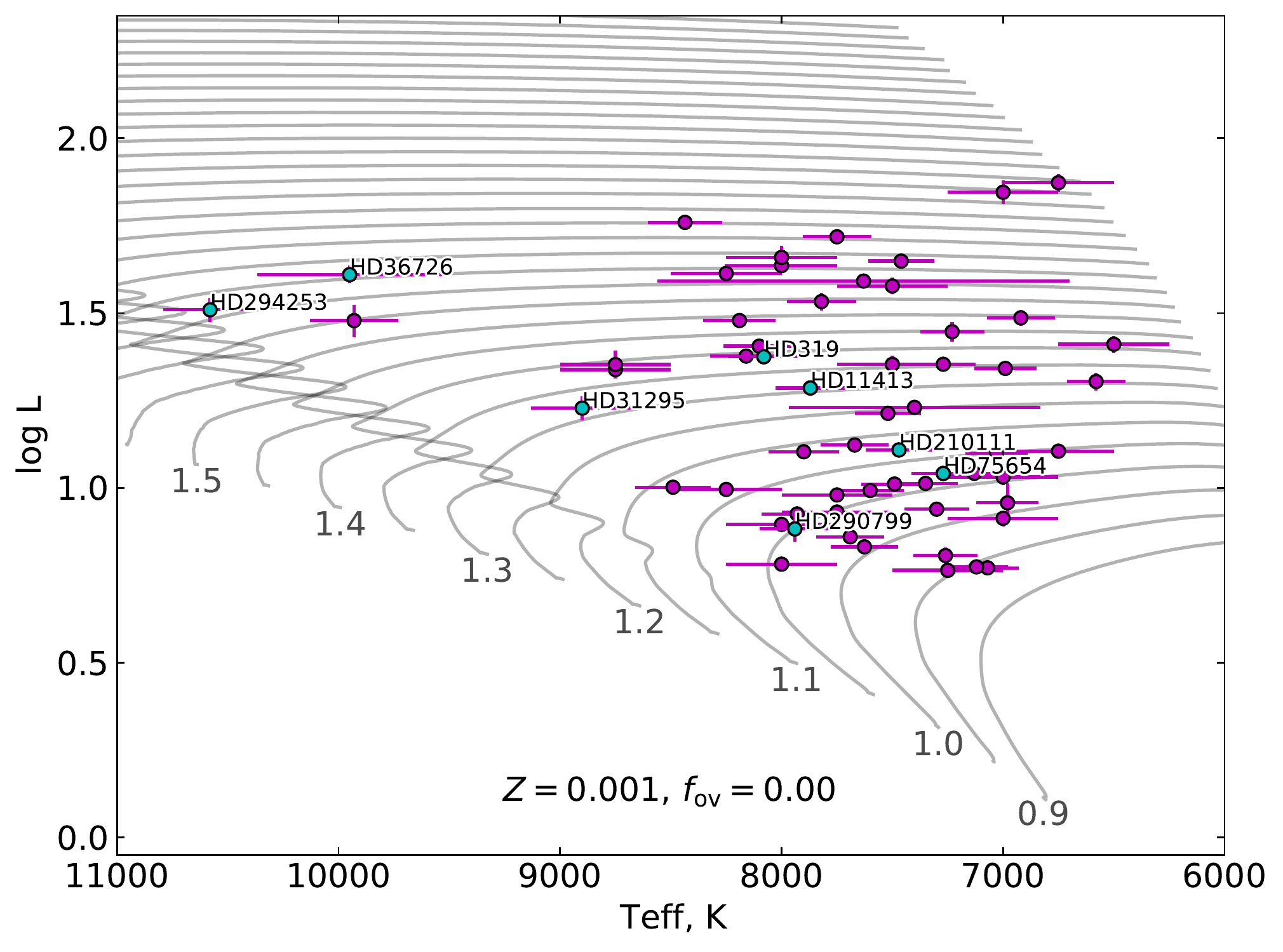}\\
\includegraphics[width=1\columnwidth]{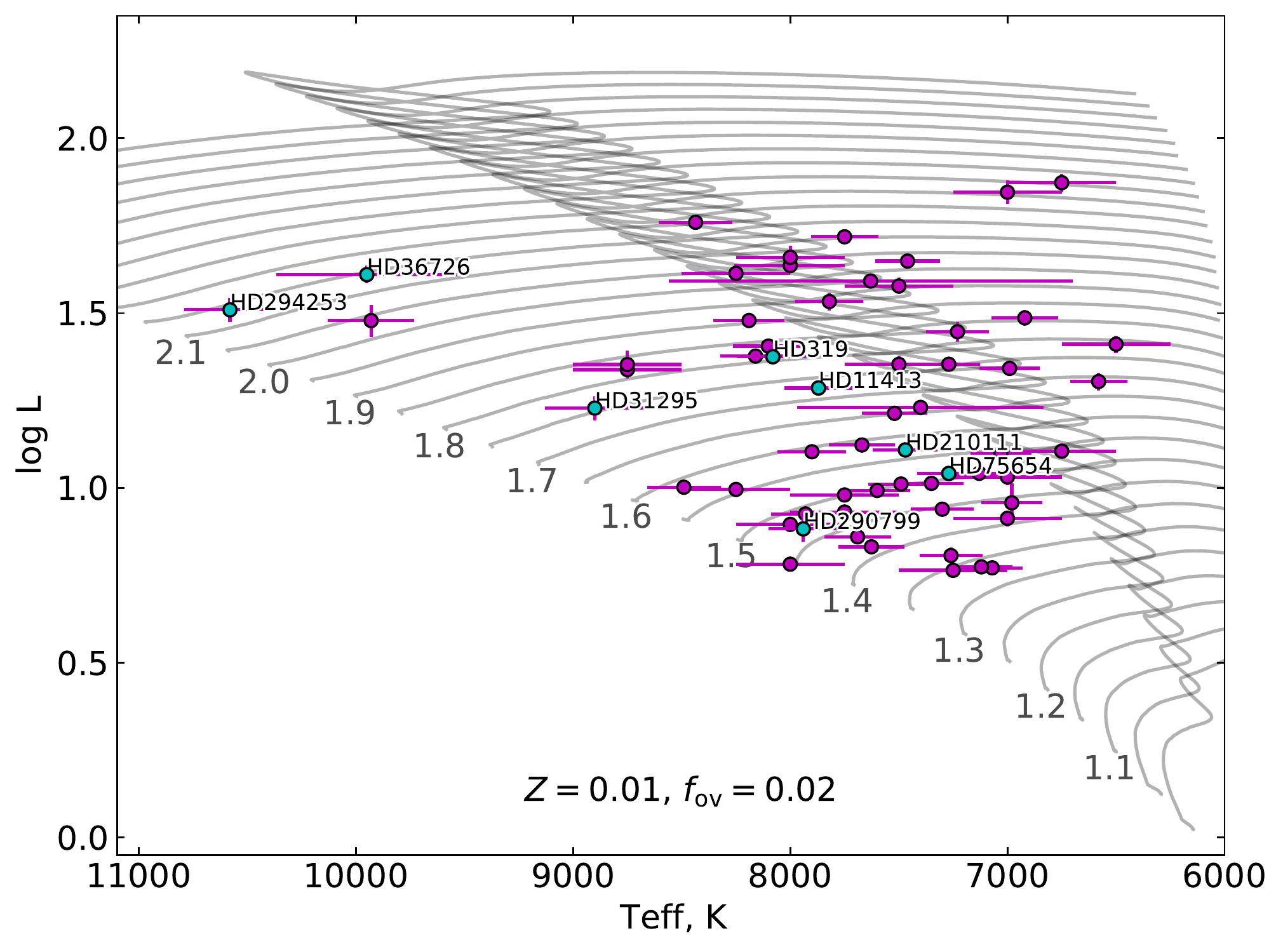}
\includegraphics[width=1\columnwidth]{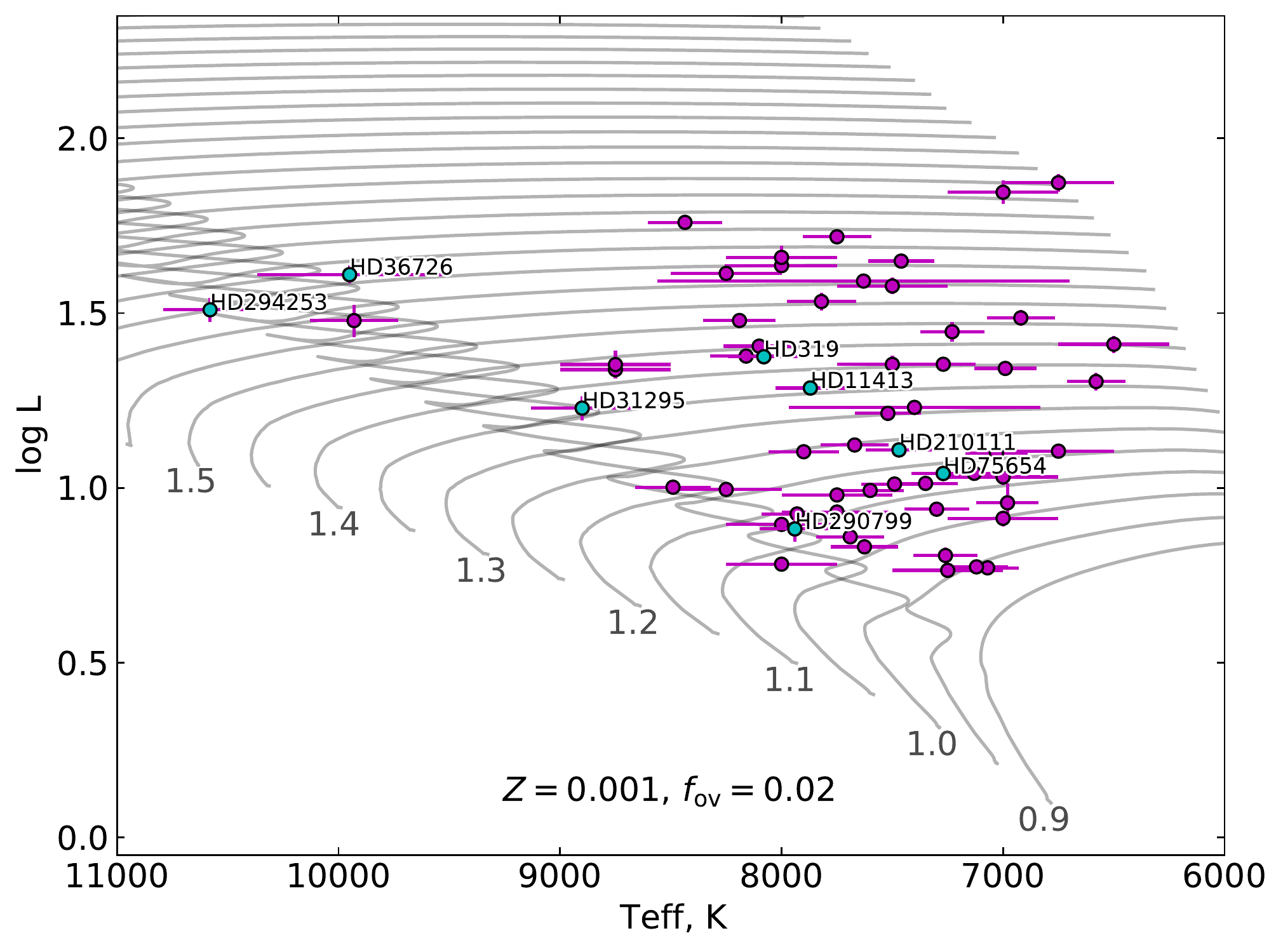}
	\caption{HR diagrams with stellar evolution tracks of different metallicity $Z$ and overshooting $f_{\rm ov}$. The positions of the 58 stars with reliable atmospheric parameters are shown with their 1$\sigma$ uncertainties in magenta. The eight stars studied in detail in this Section (\ref{sec:models}) are also plotted in cyan and labelled by their HD numbers. Evolutionary models all use initial hydrogen abundance $X_{\rm {ini}}=0.7$, rotational velocity $V_{\rm {rot}}=60$ km\,s$^{-1}$, and are shown at intervals of 0.05\,M$_{\odot}$. Mass labels in M$_{\odot}$ are given at the ZAMS for every second track.}
	\label{fig:HR1}
\end{center}
\end{figure*}

Figure\,\ref{fig:HR1} shows that the stars can be found in a range of evolutionary stages, with the results depending strongly on metallicity and overshooting efficiency. For $Z=0.010$ and $f_{\rm {ov}}=0.02$, most of the 58 stars in the wider sample and all of the 8-star subsample are located on the main sequence.  For the same metallicity but with no overshooting, only half of the wider sample and half of the eight-star subsample are definitely on the main sequence, while the others could be on the main sequence, in the post-main-sequence contraction phase, or after the loop and crossing the Hertzsprung Gap.
   
For low metallicity ($Z=0.001$) and without overshooting from the convective core, all stars but one are beyond the main sequence. The only exception, HD\,294253, is ambiguously located in one of the three different stages. If we add overshooting with $f_{\rm {ov}}=0.02$, the majority of the wider sample are crossing the Hertzsprung Gap and the rest are ambiguously located.

From the evolutionary modelling it follows that if the $\lambda$\,Boo stars have intrinsically low metallicity of order $-1$\,dex, then most of them must have evolved beyond the main sequence. Arguments concerning the timescales of the evolutionary phases show this to be unlikely. Using the evolutionary models, we calculated the time taken for a 1.7\,M$_{\odot}$ star to evolve 100\,K (between 7100 and 7000\,K) along its evolutionary track at the main-sequence, the contraction phase and the after-the-loop phase. These durations are 25.8, 1.6 and 0.7\,Myr, respectively. Since the stars are almost equally luminous in these phases, no Malmquist bias is present. It can be surmised that the stars are 16 to 37 times more likely to be on the main-sequence than in the other phases. The evolutionary models present a strong argument that the stars are not globally metal poor.

It is also known from TESS data that several $\lambda$\,Boo stars have pulsation properties consistent with young main-sequence ages (Bedding et al. 2020, in press; Appendix\,\ref{sec:notes}, here). We conclude with high confidence that $\lambda$\,Boo stars are only superficially metal-weak.

\subsection{Pulsation models}
\label{ssec:puls_models}

Another way to test whether $\lambda$\,Boo stars are globally metal weak is to construct pulsation models using two metallicities ($Z=0.01$ and $Z=0.001$) and to see which is a better fit to the modes observed with TESS. We used the eight-star subsample (Table\:\ref{tab:starlist}) for this, constructing models that fit their observed $T_{\rm eff}$, $L$, and $v\sin i$.
    
The model parameters are given in Table\,\ref{tab:models1}. In all but one case we considered models within the corresponding error box. The only exception was HD\,294253 when calculated with $Z=0.010$, because at this metallicity star lies beneath the ZAMS. In that one case we adopted a slightly lower value of the effective temperature.

Depending on the values of metallicity and overshooting, a given star can be located in one of the three aforementioned evolutionary phases. Hence, by varying metallicity and overshooting, a given HR position based on $T_{\rm eff}$ and $L$ results in quite different model parameters such as mass and age (Table\,\ref{tab:models1}). Therefore, it is impossible to independently constrain the masses and ages of the stars. The value of stellar radius, however, is insensitive to the metallicity, overshooting and mass because it is determined from the $T_{\rm eff}$ and $L$ alone. The radii thereby determined for the eight target stars are given in Table\:\ref{tab:starlist}.

In general, the models suggest slightly higher surface gravities than the spectroscopic values, although most agree within the typical uncertainty of 0.2\,dex. There are two exceptions, HD\,290799 and HD\,294253. Their models have lower $\log g$ than the observations, but the observed values of $\log{g}=4.5$ are unphysical because they place the star beneath the main sequence at $Z=0.01$.

The pulsation properties are shown in Figs.\,\ref{fig:nuetaHD319} -- \ref{fig:nuetaHD294253}, where we plot the instability parameter, $\eta$, as a function of frequency for all considered stellar models. The instability parameter is a measure of the energy gained/lost during the pulsation cycle, normalised to the range [-1,1] \citep{stellingwerf1978}. If $\eta>0$, then the corresponding mode is excited in a model.  In our calculations we used the customised non-adiabatic pulsation code of \cite{dziembowski1977a}. While models with and without overshooting were calculated, we found that overshooting affected only the possible evolutionary stages (as described in Sec.\,\ref{ssec:evo_models}), with negligible difference to the excitation (Fig.\,\ref{fig:nuetaHD319}). Hence from Fig.\,\ref{fig:nuetaHD11413} onwards, only the models with overshooting are shown. Although helium is predicted to be overabundant in $\lambda$\,Boo stars, there are no helium lines in A stars to use for helium abundance measurements. We did not enhance the helium abundance in our models (Figs.\,\ref{fig:nuetaHD319} -- \ref{fig:nuetaHD294253}), which would have increased the pulsational driving, but on a case-by-case basis we do examine the driving effect of helium enhancement.

\begin{table*}
	\begin{center}
		\caption{Models of the eight target stars, fitting their observed values of effective temperature, luminosity and rotational velocity. The model parameters given in the columns are: evolutionary stage of the star (MS = main sequence, CP = contraction phase, AL = after the loop), metallicity, initial hydrogen abundance, the extent of overshooting, initial rotational velocity on the zero-age main sequence, current rotational velocity, stellar mass, age, effective temperature, luminosity, radius, and surface gravity.}
		\begin{tabular}{cccccccccHHcc}
			\hline
			Star & ES & $Z$ & $X_{\rm {ini}}$ & $f_{\rm {ov}}$ & $V_{\rm {rot,ini}}$  &$V_{\rm {rot}}$ & $M$ & star age  & $\log{T_{\rm {eff}}} $ & $\log{L/L_{\sun}}$ & $R$  & $\log{g}$ \\
			&    &     &                 &                &   (km\,s$^{-1}$)     &(km\,s$^{-1}$) &  ($M_{\sun}$)  & (years)& & &($R_{\sun}$) &  \\\hline
			
			HD\,319&
			MS & 0.01 & 0.7 & 0.00 & 79 & 60 & 1.865 & 7.287$\times 10^{8}$ &  3.9096 & 1.373 & 2.4578 & 3.9274 \\
			&CP & 0.01 & 0.7 & 0.00 & 76 & 60 & 1.806 & 8.799$\times 10^{8}$ & 3.9096 & 1.373 & 2.4577 & 3.9137 \\
			&AL & 0.01 & 0.7 & 0.00 & 74 & 60 & 1.734 & 9.985$\times 10^{8}$ & 3.9096 & 1.373 & 2.4578 & 3.8958 \\
			&MS & 0.01 & 0.7 & 0.02 & 76 & 60 & 1.811 & 8.795$\times 10^{8}$ & 3.9096 & 1.373 & 2.4577 & 3.9148 \\
			&AL & 0.001 & 0.7 & 0.00 & 75 & 60 & 1.338 & 1.528$\times 10^{9}$ & 3.9096 & 1.373 & 2.4577 & 3.7834 \\
			&AL & 0.001 & 0.7 & 0.02 & 76 & 60 & 1.331 & 1.630$\times 10^{9}$ & 3.9096 & 1.373 & 2.4577 & 3.7810 \\

HD\,11413      & MS   & 0.010 & 0.7 & 0.00 & 147 & 125 & 1.782 & 8.081$\times10^8$ & 3.896    & 1.286 & 2.3676 & 3.9403 \\
                     & CP   & 0.010 & 0.7 & 0.00 & 138 & 125 & 1.707 & 1.034$\times10^9$ & 3.896    & 1.286 & 2.3673 & 3.9217 \\
                     & AL   & 0.010 & 0.7 & 0.00 & 136 & 125 & 1.662 & 1.122$\times10^9$ & 3.896    & 1.286 & 2.3673 & 3.9099 \\
                     & MS  & 0.010 & 0.7 & 0.02 & 142 & 125 & 1.736 & 9.605$\times10^8$ & 3.896    & 1.286 & 2.3675 & 3.9288 \\
                     & AL   & 0.001 & 0.7 & 0.00 & 128 & 125 & 1.264 & 1.858$\times10^9$ & 3.896    & 1.286 & 2.3673 & 3.7912 \\
                     & AL   & 0.001 & 0.7 & 0.02 & 131 & 125 & 1.267 & 1.896$\times10^9$ & 3.896    & 1.286 & 2.3673 & 3.7923 \\

HD\,31295 & MS & 0.01 & 0.7 & 0.00 & 123 & 115 &  1.786 & 4.327$\times10^8$ & 3.9494 & 1.2281 & 1.7318 & 4.2128\\
               &  MS & 0.01 & 0.7 & 0.02 & 121 & 115 & 1.767 & 4.953$\times10^8$ & 3.9494 & 1.2281 & 1.7318 & 4.2082 \\

               & AL & 0.001 & 0.7 & 0.00 & 109 & 115 & 1.276 & 1.717$\times10^9$ & 3.9494 & 1.2281 & 1.7318 & 4.0668 \\
               & MS & 0.001 & 0.7 & 0.02 & 121 & 115 & 1.374 & 1.393$\times10^9$ & 3.9494 & 1.2281 & 1.7319 & 4.0988 \\
               & CP & 0.001 & 0.7 & 0.02 & 117 & 115 & 1.342 & 1.558$\times10^9$ & 3.9494 & 1.2281 & 1.7318 & 4.0887 \\
               & AL & 0.001 & 0.7 & 0.02 & 113 & 115 & 1.269 & 1.835$\times10^9$ & 3.9494 & 1.2281 & 1.7319 & 4.0642 \\

HD\,36726 & MS & 0.01 & 0.7 & 0.00 & 94 & 80 & 2.189 & 3.450$\times10^8$ & 3.9978 & 1.61 & 2.1510 & 4.1128 \\
                & MS & 0.01 & 0.7 & 0.02 & 92 & 80 & 2.149 & 4.075$\times10^8$ & 3.9978 & 1.61 & 2.1511 & 4.1047 \\

                & AL & 0.001 & 0.7 & 0.00 & 95 & 80 & 1.613 & 8.367$\times10^8$ & 3.9978 & 1.61 & 2.1513 & 3.9802 \\
                & AL & 0.001 & 0.7 & 0.02 & 94 & 80 & 1.573 & 1.017$\times10^9$ & 3.9978 & 1.61 & 2.1511 & 3.9693 \\
			
			HD\,75654&
			MS & 0.01 & 0.7 & 0.00 & 57 & 45 & 1.547 & 1.207$\times 10^{9}$ & 3.8588 & 1.0445 & 2.1275 & 3.9716 \\
			&CP & 0.01 & 0.7 & 0.00 & 55 & 45 & 1.494 & 1.491$\times 10^{9}$ & 3.8588 & 1.0445 & 2.1276 & 3.9565 \\
			&AL & 0.01 & 0.7 & 0.00 & 53 & 45 & 1.434 & 1.713$\times 10^{9}$ & 3.8588 & 1.0445 & 2.1277 & 3.9387 \\
			&MS & 0.01 & 0.7 & 0.02 & 56 & 45 & 1.512 & 1.413$\times 10^{9}$ & 3.8588 & 1.0445 & 2.1276 & 3.9616 \\
			&AL & 0.001 & 0.7 & 0.00 & 50 & 45 & 1.072 & 3.168$\times 10^{9}$ & 3.8588 & 1.0445 & 2.1276 & 3.8124 \\
			&AL & 0.001 & 0.7 & 0.02 & 51 & 45 & 1.082 & 3.083$\times 10^{9}$ & 3.8588 & 1.0445 & 2.1276 & 3.8162 \\
			
			HD\,210111&
			MS & 0.01 & 0.7 & 0.00 & 40 & 30 & 1.613 & 1.081$\times 10^{9}$ & 3.8692 & 1.1188 & 2.2095 & 3.9570 \\
			&CP & 0.01 & 0.7 & 0.00 & 39 & 30 & 1.565 & 1.301$\times 10^{9}$ & 3.8692 & 1.1188 & 2.2093 & 3.9439 \\
			&AL & 0.01 & 0.7 & 0.00 & 38 & 30 & 1.494 & 1.527$\times 10^{9}$ & 3.8692 & 1.1188 & 2.2094 & 3.9237 \\
			&MS & 0.01 & 0.7 & 0.02 & 39 & 30 & 1.573 & 1.281$\times 10^{9}$ & 3.8692 & 1.1188 & 2.2093 & 3.9461 \\
			&AL & 0.001 & 0.7 & 0.00 & 37 & 30 & 1.123 & 2.710$\times 10^{9}$ & 3.8692 & 1.1188 & 2.2093 & 3.7999 \\
			&AL & 0.001 & 0.7 & 0.02 & 38 & 30 & 1.134 & 2.637$\times 10^{9}$ & 3.8692 & 1.1188 & 2.2092 & 3.8041 \\
			
			HD\,290799&
			MS & 0.01 & 0.7 & 0.00 & 71 & 70 & 1.472 & 3.401$\times 10^{8}$ & 3.9031 & 0.853 & 1.3910 & 4.3191 \\
			&MS & 0.01 & 0.7 & 0.02 & 71 & 70 & 1.468 & 3.791$\times 10^{8}$ & 3.9031 & 0.853 & 1.3909 & 4.3180 \\
			&AL & 0.001 & 0.7 & 0.00 & 64 & 70 & 1.051 & 3.063$\times 10^{9}$ & 3.9031 & 0.853 & 1.3910 & 4.1727 \\
			&MS & 0.001 & 0.7 & 0.02 & 70 & 70 & 1.116 & 2.423$\times 10^{9}$ & 3.9031 & 0.853 & 1.3910 & 4.1990 \\
			&CP & 0.001 & 0.7 & 0.02 & 66 & 70 & 1.073 & 2.939$\times 10^{9}$ & 3.9031 & 0.853 & 1.3910 & 4.1819 \\
			&AL & 0.001 & 0.7 & 0.02 & 64 & 70 & 1.038 & 3.257$\times 10^{9}$ & 3.9031 & 0.853 & 1.3911 & 4.1674 \\
			
			HD\,294253&
			MS & 0.01 & 0.7 & 0.00 & 72 & 70 & 2.120 & 9.592$\times 10^{6}$ & 4.0340 & 1.452 & 1.5180 & 4.4018 \\
			&MS & 0.01 & 0.7 & 0.02 & 72 & 70 & 2.120 & 1.188$\times 10^{7}$ & 4.0340 & 1.452 & 1.5180 & 4.4016 \\
			&MS & 0.001 & 0.7 & 0.00 & 80 & 70 & 1.682 & 5.818$\times 10^{8}$ & 4.0374 & 1.452 & 1.4945 & 4.3148 \\
			&CP & 0.001 & 0.7 & 0.00 & 75 & 70 & 1.611 & 7.521$\times 10^{8}$ & 4.0374 & 1.452 & 1.4944 & 4.2960 \\
			&AL & 0.001 & 0.7 & 0.00 & 74 & 70 & 1.570 & 8.097$\times 10^{8}$ & 4.0374 & 1.452 & 1.4944 & 4.2848 \\
			&MS & 0.001 & 0.7 & 0.02 & 77 & 70 & 1.640 & 6.912$\times 10^{8}$ & 4.0374 & 1.452 & 1.4944 & 4.3038 \\
			
			\hline
		\end{tabular}
		\label{tab:models1}
	\end{center}
\end{table*}

\subsubsection{Pulsation model analysis}

\begin{figure}
	\includegraphics[width=\columnwidth]{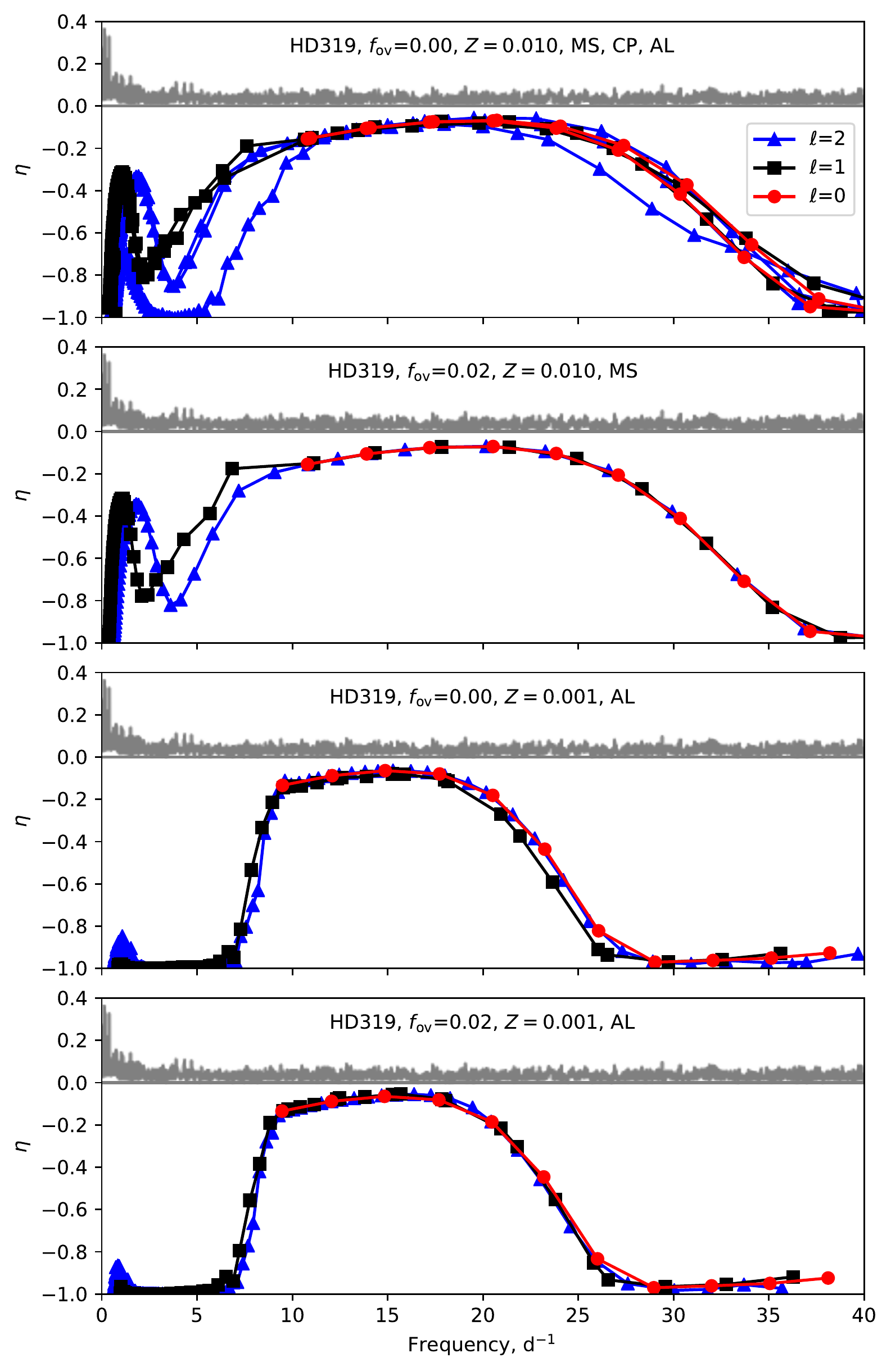}
	\caption{Instability parameter, $\eta$, as a function of the pulsation frequency for the seismic models of HD\,319. Mode driving exceeds damping and modes are excited where $\eta>0$, but no modes are observed or predicted to be excited in HD\,319. Mode degrees $\ell=0$, 1 and 2 were considered. The top panel shows models calculated with $Z=0.010$ and $f_{\rm {ov}}=0.00$. In the next panel we show the overshooting effect with $f_{\rm{ov}}=0.02$. In the bottom two panels the models are calculated with lower metallicity ($Z=0.001$) and two values of $f_{\rm{ov}}$ (0.00 and 0.02). All possible evolutionary stages were considered (MS = main sequence, CP = contraction phase, AL = after the loop).}
	\label{fig:nuetaHD319}
\end{figure}

\paragraph*{HD\,319:}
The pulsation models of HD\,319 predict no excited modes (Fig.\,\ref{fig:nuetaHD319}). The highest value of the instability parameter, $\eta$, is about $-0.1$ for modes with frequencies 15--20 d$^{-1}$. The evolutionary phase only has an important impact on the low frequencies (g\:modes) but in no model are these excited. These results are consistent with the observations (Fig.\,\ref{fig:nuetaHD319}, Table\:\ref{tab:puls}, and Appendices\,\ref{sec:notes}--\ref{sec:FTs}), where no variability was seen.

\begin{figure}
	\includegraphics[width=\columnwidth]{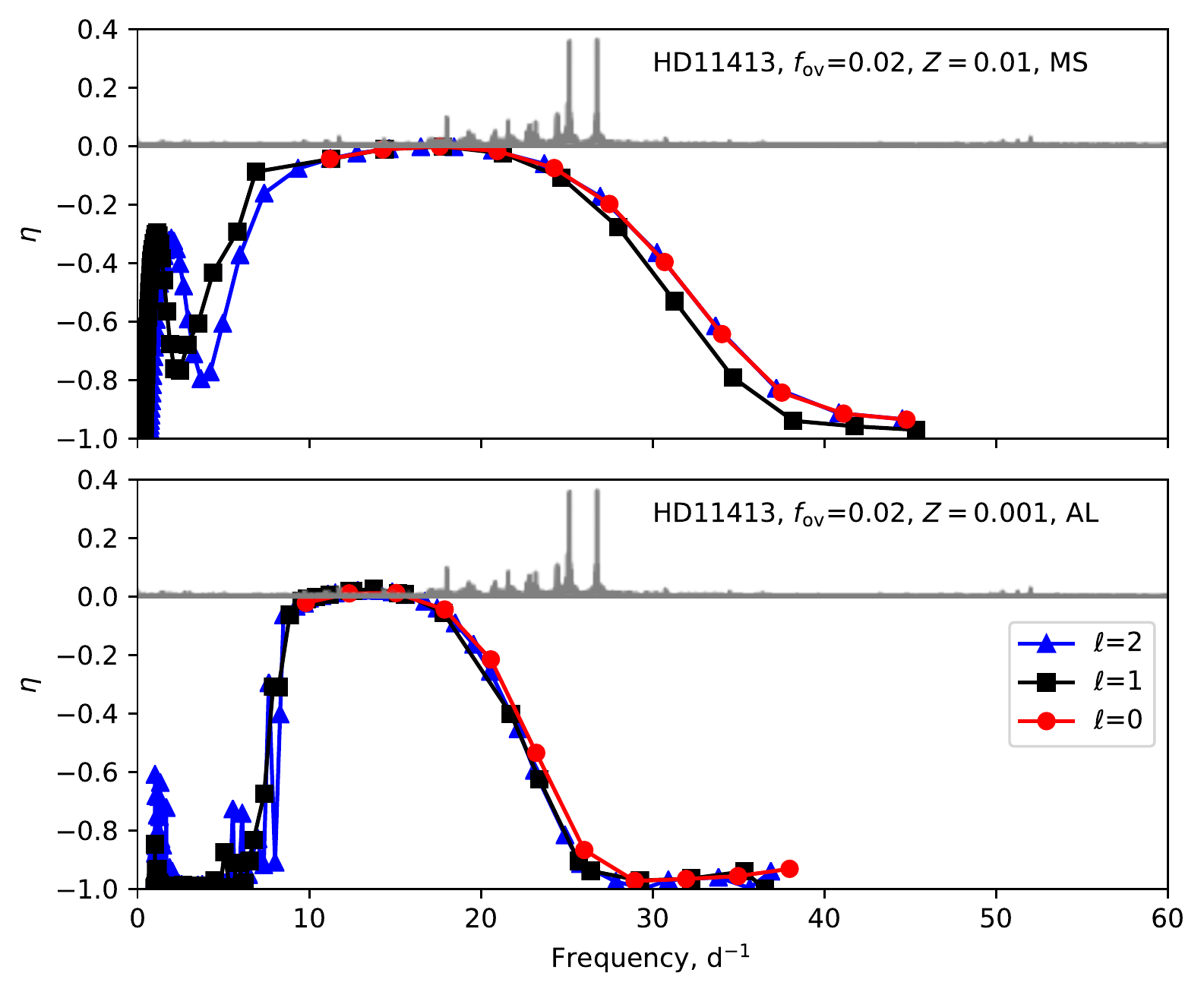}
	\caption{The same as in Fig.\,\ref{fig:nuetaHD319}, but for HD\,11413 and showing only the models with overshooting.}
	\label{fig:nuetaHD11413}
\end{figure}

\paragraph*{HD\,11413:} Models calculated with $Z=0.01$ sit just below the threshold for excitation, with $\eta\sim-0.001$ for frequencies of $\sim$20 d$^{-1}$ (Fig.\,\ref{fig:nuetaHD11413}). For lower metallicity, $Z=0.001$, the models are much more evolved, after the loop, and radial and dipole modes with frequencies of 10--15 d$^{-1}$ reach $\eta\sim0.012$. The high frequency modes are very effectively damped in these post main sequence models, so the observed frequencies, which mostly lie in the range 20--30 d$^{-1}$, are slightly better matched by the higher metallicity models, even though $\eta$ is slightly negative in the latter. We note that an increase in the helium abundance by about $0.1$ produces excited modes in the observed frequency range in the higher metallicity models.

\begin{figure}
	\includegraphics[width=\columnwidth]{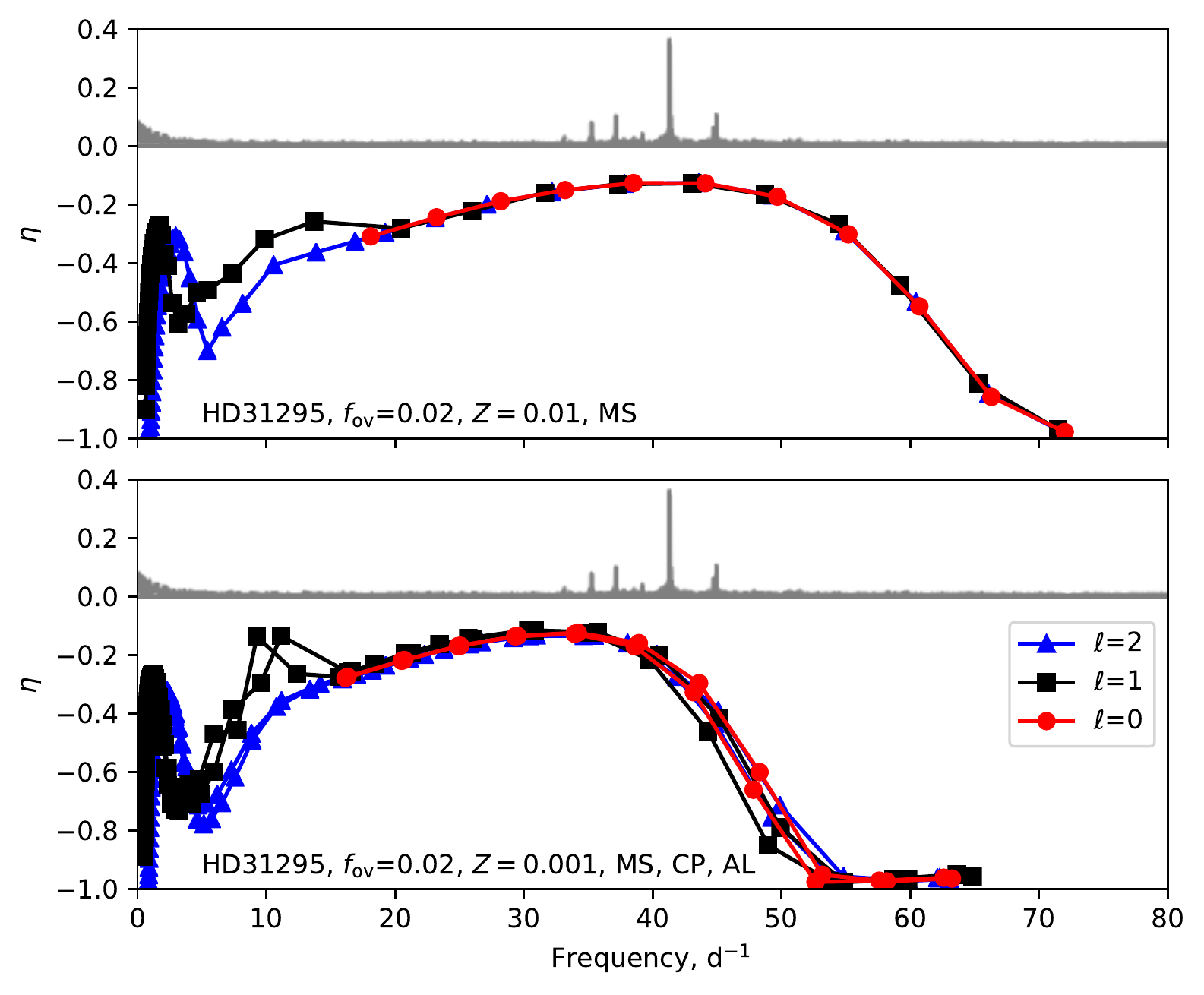}
	\caption{The same as in Fig.\,\ref{fig:nuetaHD11413}, but for HD\,31295}
	\label{fig:nuetaHD31295}
\end{figure}

\paragraph*{HD\,31295:} The models do not predict any excited modes, regardless of metallicity, overshooting and evolutionary phase (Fig.\,\ref{fig:nuetaHD31295}). This is in contradiction with the observations, which indicate pulsations with high frequencies of the order of 40\,d$^{-1}$. Interestingly, in this frequency region the instability coefficient $\eta$ does have a maximum. However, an increase of the helium abundance to $Y=0.39$ is needed in order to excite modes at $\sim$40\,d$^{-1}$. The higher metallicity model is slightly preferred because it has $\eta$ reaching a maximum closer to the observed frequency of maximum power.

\begin{figure}
	\includegraphics[width=\columnwidth]{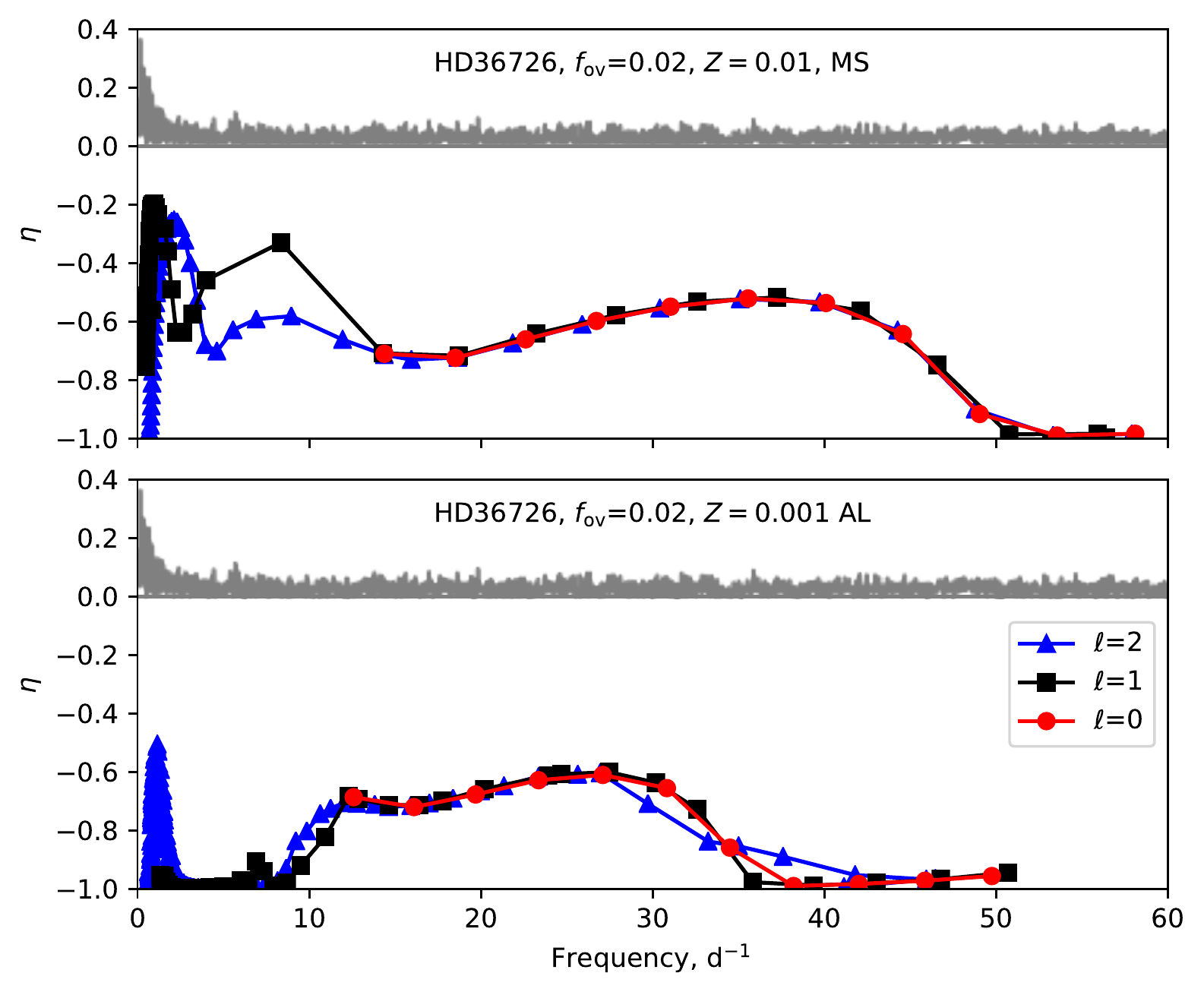}
	\caption{The same as in Fig.\,\ref{fig:nuetaHD11413}, but for HD\,36726}
	\label{fig:nuetaHD36726}
\end{figure}

\paragraph*{HD\,36726:} Neither the models nor the observations show excited modes (Fig.\,\ref{fig:nuetaHD36726}). The instability parameter is the highest for g-modes, but its value is still no greater than $-0.2$.

\begin{figure}
	\includegraphics[width=\columnwidth]{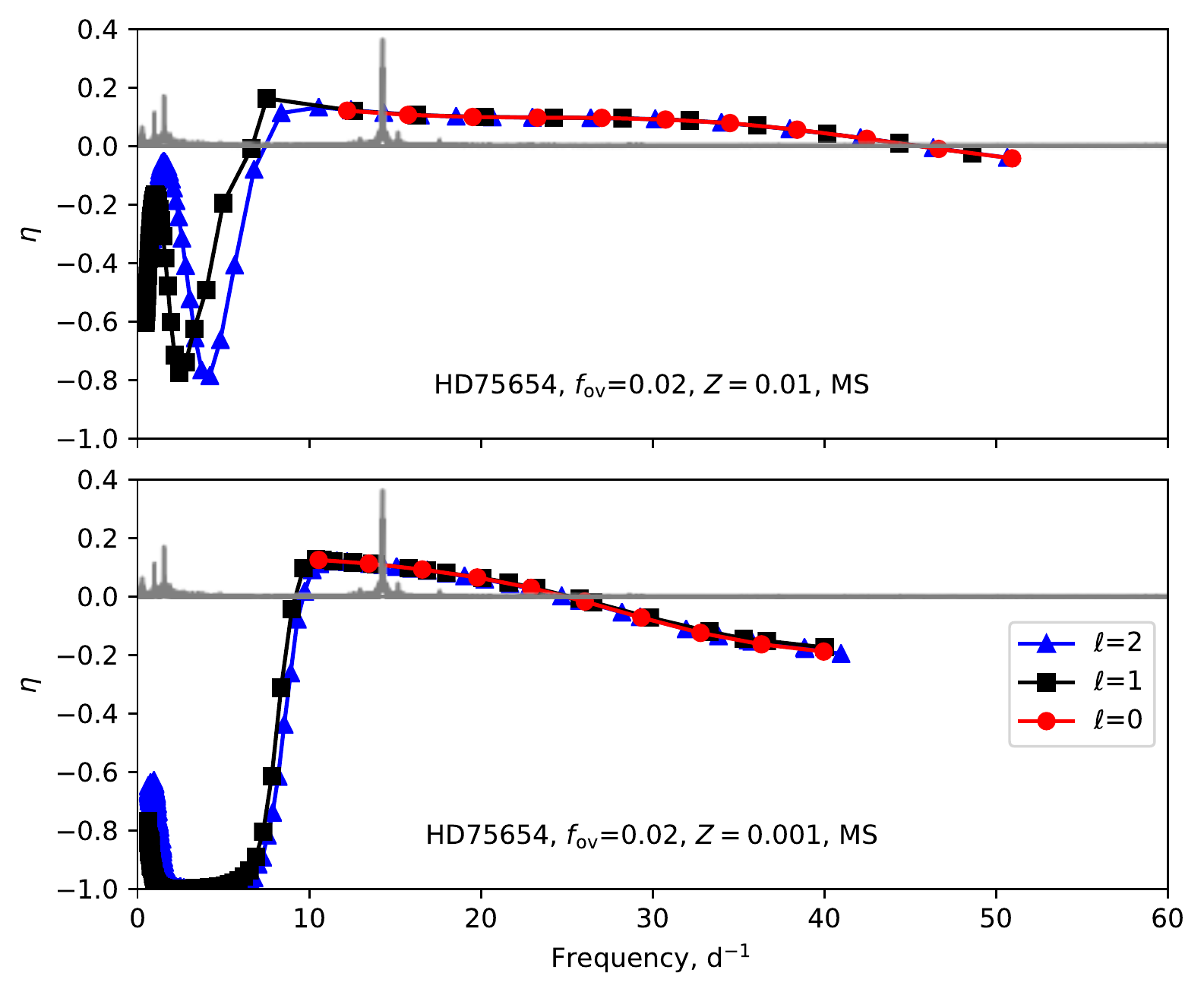}
	\caption{The same as in Fig.\,\ref{fig:nuetaHD11413}, but for HD\,75654}
	\label{fig:nuetaHD75654}
\end{figure}

\paragraph*{HD\,75654:} A very clear instability occurs for all considered models of HD\,75654 (Fig.\,\ref{fig:nuetaHD75654}). Models calculated with $Z=0.010$ have unstable modes with frequencies from about 7 to 45\,d$^{-1}$. For lower metallicity, $Z=0.001$, the unstable modes occur over a smaller frequency range (10--25\,d$^{-1}$). Our models are computed with frozen convection, which is not strictly valid for high oscillation frequencies or for stars like this where as much as 90\% of the flux is transported by convection. Nonetheless, the observed frequencies of order 14\,d$^{-1}$ are consistent with all inspected models. There is no clear preference for metallicity, since the higher metallicity over-predicts instability at higher frequencies ($\sim$30\,d$^{-1}$) but also comes closer to describing the observed lower frequencies ($\sim$1\,d$^{-1}$).

\begin{figure}
	\includegraphics[width=\columnwidth]{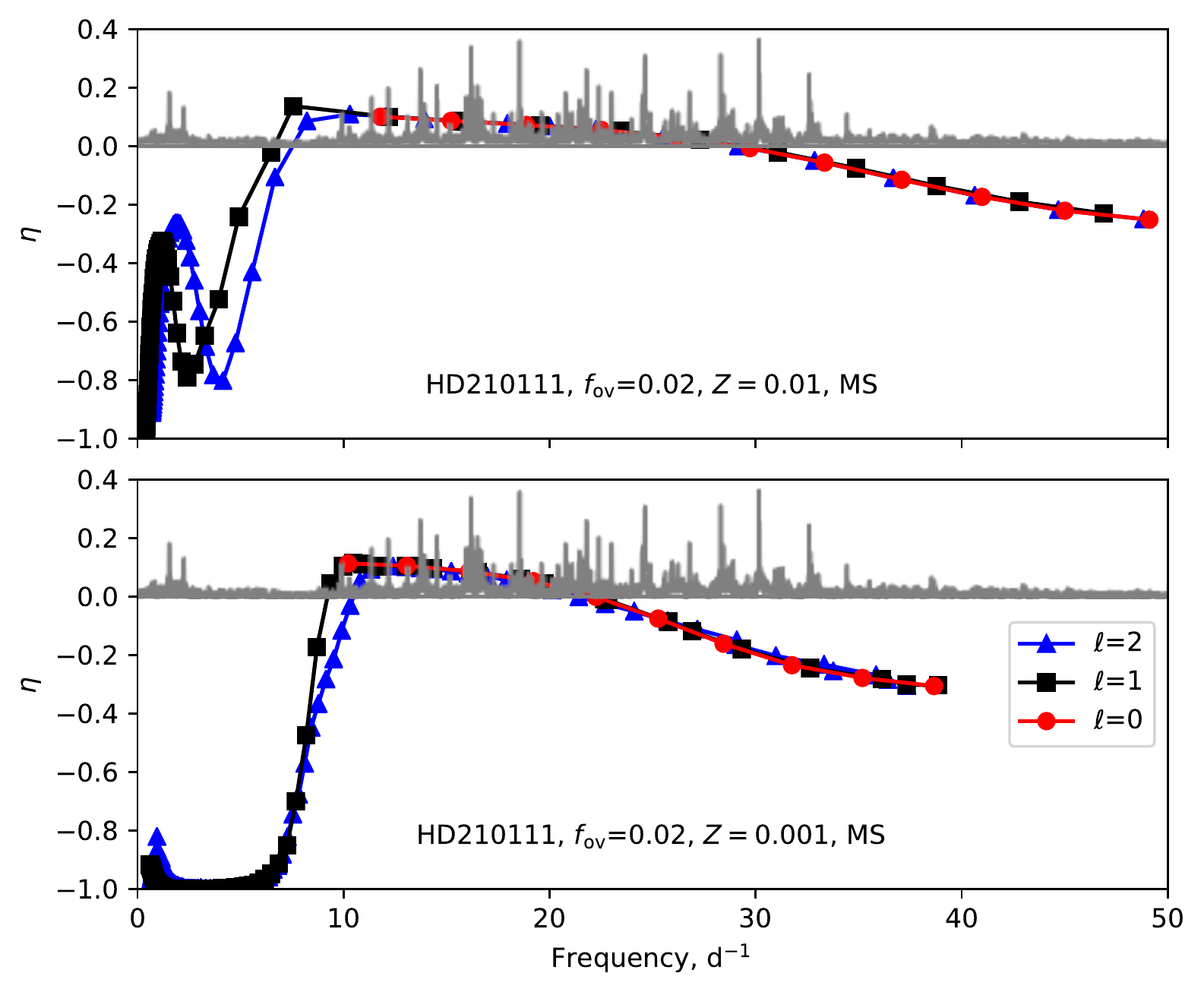}
	\caption{The same as in Fig.\,\ref{fig:nuetaHD11413}, but for HD\,210111}
	\label{fig:nuetaHD210111}
\end{figure}

\paragraph*{HD\,210111:} Models predict unstable modes between $\sim$7 and 30\,d$^{-1}$ for $Z=0.010$, and from $\sim$10 to 20\,d$^{-1}$ for $Z=0.001$ (Fig.\,\ref{fig:nuetaHD210111}). These results agree well with the high frequencies seen in the TESS data (see also Appendix\,\ref{sec:notes}). The detected low frequencies cannot be explained by post-main-sequence models since they are strongly damped ($\eta=-0.8$) in the advanced evolution phases. They are also damped on the main sequence, but with $\eta\sim-0.2$. The higher metallicity model is a better match to both the higher and lower frequencies.

\begin{figure}
	\includegraphics[width=\columnwidth]{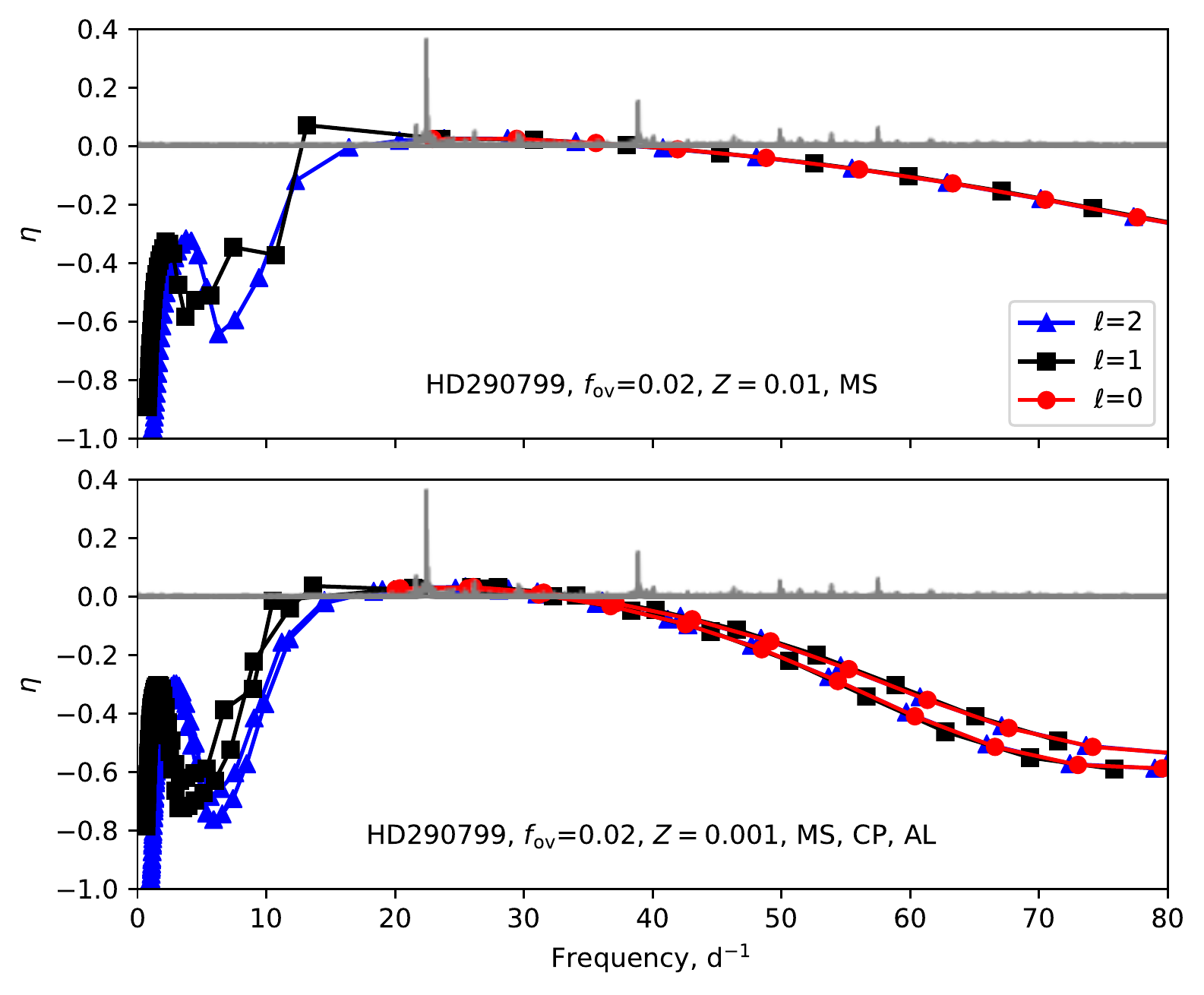}
	\caption{The same as in Fig.\,\ref{fig:nuetaHD11413}, but for HD\,290799}
	\label{fig:nuetaHD290799}
\end{figure}

\paragraph*{HD\,290799:} The instability parameter is slightly positive for this star (Fig.\,\ref{fig:nuetaHD290799}). The excited modes are at frequencies in the range 15--40\,d$^{-1}$ for higher metallicity models and 15--30\,d$^{-1}$ for lower metallicity models. TESS observations show pulsation frequencies in a wide range of 20--60\,d$^{-1}$. The high frequencies fit better to the main sequence models and a preference towards high metallicity was found. However, caution is again required by the use of the frozen convection approximation. It is noteworthy that Bedding et al. (2020, in press) found a very high $\Delta\nu$ of 7.55\,d$^{-1}$, suggesting this star is very young (see Appendix\,\ref{sec:notes}). 

\begin{figure}
	\includegraphics[width=\columnwidth]{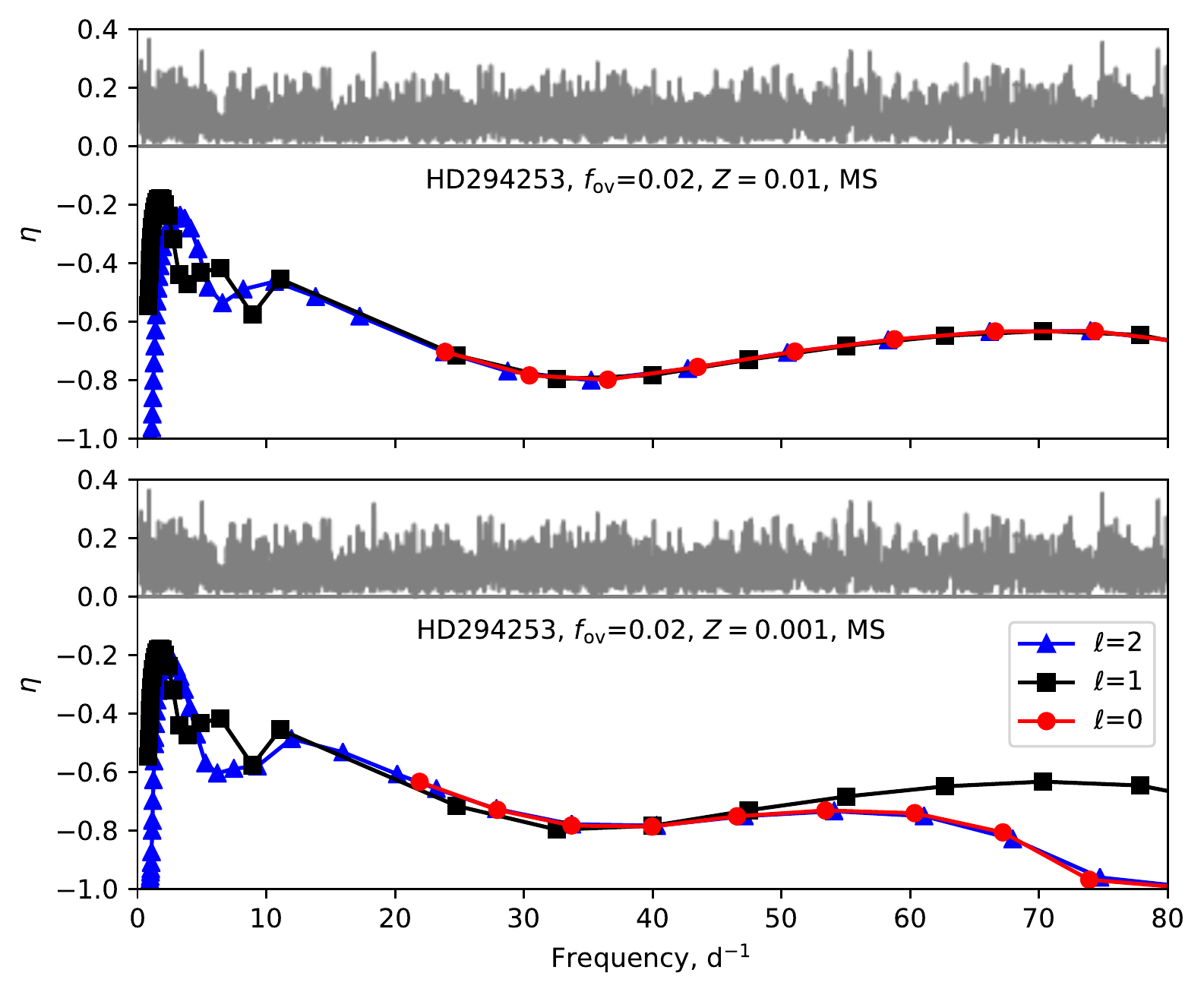}
	\caption{The same as in Fig.\,\ref{fig:nuetaHD11413}, but for HD\,294253}
	\label{fig:nuetaHD294253}
\end{figure}

\paragraph*{HD\,294253:} This is the most massive star we analysed. Its models predict no unstable modes, regardless of the chemical composition and evolution phase (Fig.\,\ref{fig:nuetaHD294253}). No variability was observed with TESS, either (Table\:\ref{tab:puls}, Appendices\,\ref{sec:notes} and \ref{sec:FTs}).

\subsubsection{Pulsation model summary}

Three of the eight stars for which we constructed pulsation models are non-pulsators. The pulsation models accurately reproduce this. In three of the remaining five targets, excited modes were reproduced by the models. Where any metallicity was preferred, it was always the higher metallicity ($Z=0.01$) model, though HD\,210111 is the only target for which this case can be made strongly. While the frozen convection approximation and the simplicity of the analysis are no substitute for dedicated modelling, we conclude that such analyses are valuable in evaluating the global metallicity of $\lambda$\,Boo stars, which in this case has hinted that these $\lambda$\,Boo stars are unlikely to be metal-poor throughout.

%%%%%%%%%%%%%%%%%%%%%%%%%%%%%
%%%%%%%%%%%%%%%%%%%%%%%%%%%%%

\section{Conclusions}
\label{sec:results}

We analysed the TESS light curves of 70 southern $\lambda$\,Boo stars to evaluate whether they pulsate as $\delta$\,Sct stars. We discovered four objects to be binaries: HD\,31508 and HD\,94390 are heartbeat stars, and HD\,168947 and HD\,74423 are eclipsing binaries. Sixteen of the studied $\lambda$\,Boo stars are not $\delta$\,Sct stars at the precision of TESS data, although many of these lie far from the $\delta$\,Sct instability strip, where pulsation is not expected. By reference to a sample of 1826 \textit{Kepler} $\delta$\,Sct stars, we computed the expected pulsator fraction, given the location of stars with respect to the instability strip. We found that the fraction of $\lambda$\,Boo stars that show $\delta$\,Sct pulsation is about twice that of normal stars. This has important implications: firstly, having more pulsating $\lambda$\,Boo stars opens up more possibility for detailed asteroseismic modelling; and secondly, it suggests that the mechanism that produces the chemical peculiarity may also make the stars more susceptible to pulsation. Indeed, this link was already hypothesized \citep{murphy2014}.

As a step towards asteroseismology of the $\lambda$\,Boo stars in this sample, we have attempted to identify the fundamental mode based on the period--luminosity relationship, \'echelle diagrams and Petersen diagrams (period ratios). We identified the fundamental mode in 20 of the 40 pulsating $\lambda$\,Boo stars that have 2-min TESS data and reliable atmospheric parameters. In an additional 8 stars, we can conclude that the stars are not pulsating in the fundamental mode, while in the remaining 12 it was not possible to identify the fundamental mode unambiguously. 
We confirmed the finding by Bedding et al (2020, in press) that some $\lambda$\,Boo stars have high-frequency pulsations with remarkably regular frequency patterns.
These patterns facilitate mode identification, allowing radial ($\ell=0$) and dipole ($\ell=1$) modes to be distinguished, and the radial order ($n$) to be inferred. This enables a precise stellar age to be calculated and supports the hypothesis that $\lambda$\,Boo stars are young (e.g. \citealt{kamaetal2015}).

An important question is whether $\lambda$\,Boo stars are globally metal-poor, or just metal-poor at the surface. It was hoped that asteroseismology would answer this question, but the first major step towards this was at odds with other analysis methods. Specifically, the asteroseismic investigation of 29\,Cyg --- the first $\lambda$\,Boo star discovered to be variable \citep{gies&percy1977} --- suggested a global metal weakness \citep{casasetal2009}, whereas evolution models for several other $\lambda$\,Boo stars suggested they were of global solar abundance \citep{paunzenetal2015}. We used \textit{Gaia} DR2 parallaxes to provide precise localisations on the HR diagram for the $\lambda$\,Boo stars. By comparing with stellar evolution models with a range of metallicities, we found it implausible that $\lambda$\,Boo stars are globally metal poor. The most likely conclusion is that they are all main-sequence stars of approximately solar metallicity. Otherwise, if the $\lambda$\,Boo stars of our sample are metal-poor throughout, the majority of them would have to be in the extremely short-lived phase of evolution after the main-sequence, where stars cross the Hertzsprung gap.

We also calculated pulsation models for a subset of eight stars. We found that the observed pulsations tend to be better matched to the higher metallicity models ($Z=0.01$) than the low metallicity models ($Z=0.001$), albeit with small-number statistics. However, the simplicity of the models leaves much room for refinement. Future modelling could use time-dependent convection, incorporate turbulent convection into the driving \citep{houdek2000,antocietal2014}, use a higher helium abundance, and be based on mode identification as far as possible. We provided a start for this mode identification, and physical parameters on which to base future pulsation models. This opportunity is an exciting prospect for $\lambda$\,Boo star research.

%%%%%%%%%%%%%%%%%%%%%%%%%%%%%
%%%%%%%%%%%%%%%%%%%%%%%%%%%%%

\section*{Acknowledgements}
This work was supported by the Australian Research Council through Discovery Project DP170104160 and DECRA DE180101104, by the National Aeronautics and Space Administration (80NSSC18K1585, 80NSSC19K0379) awarded through the TESS Guest Investigator Program and by the National Science Foundation (AST-1717000), by the Polish National Science Centre grants 2018/29/B/ST9/02803 and 2018/29/B/ST9/01940, and by the Alfred P. Sloan Foundation. Calculations have been partly carried out using resources provided by Wroclaw Centre for Networking and Supercomputing \url{http://www.wcss.pl}, grant No. 265. This research made use of {\sc Lightkurve}, a Python package for Kepler and TESS data analysis \citep{lightkurvecollaboration2018}. TESS full-frame images were extracted with {\sc eleanor} \citep{feinsteinetal2019}, which uses the \href{https://mast.stsci.edu/tesscut/}{TESScut} service from MAST at STScI. This work has made use of data from the European Space Agency (ESA) mission {\it Gaia} (\url{https://www.cosmos.esa.int/gaia}), processed by the {\it Gaia} Data Processing and Analysis Consortium (DPAC, \url{https://www.cosmos.esa.int/web/gaia/dpac/consortium}). Funding for the DPAC has been provided by national institutions, in particular the institutions participating in the {\it Gaia} Multilateral Agreement. This work also made use of ARI's Gaia Services at \url{http://gaia.ari.uni-heidelberg.de/} for RUWE values.

%%%%%%%%%%%%%%%%%%%%%%%%%%%%%
%%%%%%%%%%%%%%%%%%%%%%%%%%%%%

\appendix

\section{Notes on the Fourier transforms of individual stellar light curves}
\label{sec:notes}

TESS has made possible a much broader investigation of pulsating $\lambda$\,Boo stars than the six already known (Table\:\ref{tab:known_puls}). In this appendix, we describe the features of TESS light curves (and their Fourier transforms) of southern $\lambda$\,Boo stars.

\begin{table}
\centering
\caption{Known pulsation properties of $\lambda$\,Boo stars from the literature. Only 
objects with long time series are included.}
\begin{tabular}{rrrrr}
\hline  
HD & Frequency & Ampl. & Filter & \multicolumn{1}{c}{Reference} \\
   & d$^{-1}$ & mmag & & \\       
\hline
11413 & 25.206 & 9.1 & $V$ & \citet{koenetal2003} \\
& 24.099 & 5.3 & & \\
& 24.493 & 5.0 & & \\
83041 & 14.5293 & 1.6 & SWASP & \citet{paunzenetal2015} \\
111786 & 31.01 & 8.0 & $V$ & \citet{paunzenetal1998b} \\
& 20.31 & 6.0 & & \\
& 33.36 & 5.8 & & \\
184779 & 12.5687 & 13.9 & SWASP & \citet{paunzenetal2015} \\
& 13.8351 & 6.8 & & \\
& 10.4756 & 3.0 & & \\ 
210111 & 28.47 & 9.5 & $V$ & \citet{bregeretal2006} \\
& 20.81 & 2.6 & & \\
& 18.59 & 2.6 & & \\
290799 & 22.5300 & 7.2 & SWASP & \citet{paunzenetal2015} \\
\hline  
\end{tabular}
\label{tab:known_puls}
\end{table}

\subsection*{HD\,319, sector 2} 
This non-$\delta$\,Sct star is also discussed in Sec.\,\ref{sec:models}. No p\:modes are evident at the 0.01\,mmag level, even though the star is located within the instability strip. The Fourier transform features peaks at low frequency ($<$0.5\,d$^{-1}$), with amplitudes $<$0.30\,mmag, which could be rotational or instrumental in origin.

 \subsection*{HD\,3922, sector 1--2}
 The low frequencies ($<$1\,d$^{-1}$) are caused by trends in the data near the sector boundaries, and are not pulsational in origin. The p\:modes are unaffected by this. The star does not appear to be a $\gamma$\,Dor variable, unless pulsation is masked by the low-frequency trends at the $\sim$0.3\,mmag level. This star lies near the TAMS, which explains the low frequencies of the strongest p\:modes (5--10\,d$^{-1}$). Identifying these modes is difficult. $f_1$ is a good candidate for the first overtone mode because the period ratio with the peak at 11.15\,d$^{-1}$ coincides with the expected one for the second radial overtone mode. No peak clearly lines up with the predicted frequency of the fundamental mode from the P--L relation (7.91\,d$^{-1}$), or from frequency ratios using the three strongest modes.

\subsection*{HD\,4158, sector 3}
Period ratios suggest that $f_2=7.106$\,d$^{-1}$ is the fundamental mode and $f_3=9.169$ is the first radial overtone (the ratio $f_2/f_3$ = 0.775).  In addition to the p\:modes, there are also peaks at 0.57 and 1.70 \,d$^{-1}$, with amplitudes of 0.2 and 0.3\,mmag. These could be r and g modes, respectively. The similarity of the p\:mode spectrum to that of HD\,3922 is noteworthy -- the two stars seem to be equally evolved but HD\,4158 is $\sim$500\,K hotter.

\subsection*{HD\,6870, sectors 1--3}
The pulsations are dominated by three p\:modes with amplitudes between 1 and 2\,mmag, but there are also groups of peaks between 1.1 and 4.6\,d$^{-1}$ that are likely g\:modes and perhaps prograde sectoral modes. No pair from the three dominant peaks has a frequency ratio near 0.77 that would correspond to the fundamental mode and first radial overtone, but the strongest (and highest frequency) of the three, $f_1=17.87$\,d$^{-1}$, lies very close to the predicted value from the P-L relation of 17.80\,d$^{-1}$.

\subsection*{HD\,7908, sector 3}
Not a $\delta$\,Sct star at the 0.02\,mmag level. Broad power excesses at 1.0--1.5\,d$^{-1}$ and at 2.4--2.9\,d$^{-1}$ suggest this is a $\gamma$\,Dor star. It lies just inside the red edge of the theoretical $\delta$\,Sct instability strip \citep{dupretetal2005b} but outside the observed one \citep{murphyetal2019}.

\subsection*{HD\,11413, sectors 2--3}
Pulsation models of this star are presented in Sec.\,\ref{sec:models}. We measure a characteristic spacing for the p\:modes of 3.6--3.8\,d$^{-1}$ that could be the large spacing, $\Delta\nu$ (see Bedding et al. 2020, in press). For a spacing of 3.8\,d$^{-1}$, the fundamental mode is probably the peak at 11.8\,d$^{-1}$, but there is not independent support from the P--L relation for this. Period ratios suggest the peaks at 14.424 and 18.06\,d$^{-1}$ could be the first and second radial overtone modes (measured ratio = 0.799), but these are not compatible with the fundamental mode measured from the \'echelle. The peaks near 50\,d$^{-1}$ are harmonics and combinations of the strong peaks. 

\subsection*{HD\,13755, sector 3}
The Fourier spectrum is quite dense with p\:modes at relatively low frequencies (5--15\,d$^{-1}$). The predicted fundamental mode frequency is 8.00\,d$^{-1}$, which is close to $f_1$ at 7.93\,d$^{-1}$. It then follows from period ratios that $f_2=12.66$\,d$^{-1}$ is probably the second radial overtone, which we predict to lie at $\sim$12.75\,d$^{-1}$. No g\:modes are obvious.

\subsection*{HD\,17341, sectors 3--4}
\begin{figure*}
\begin{center}
\includegraphics[width=0.33\textwidth]{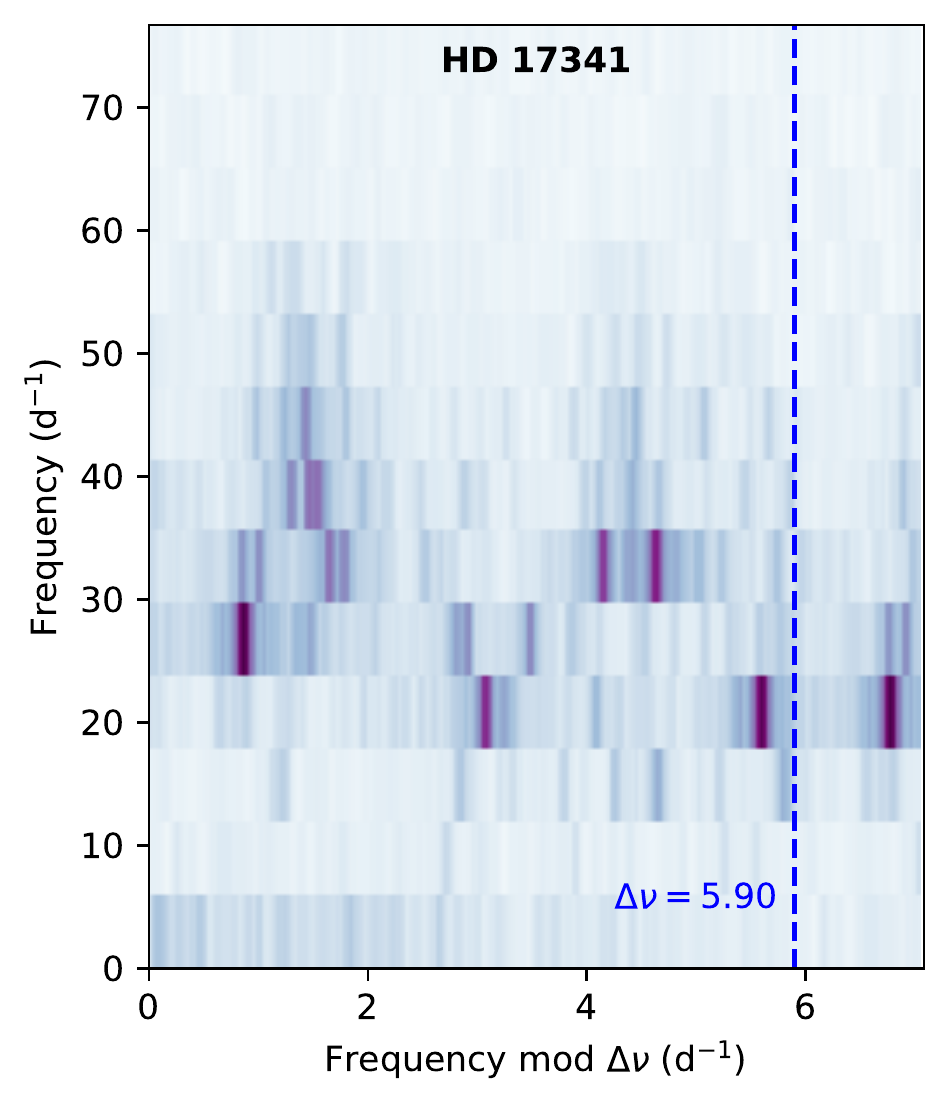}
\includegraphics[width=0.33\textwidth]{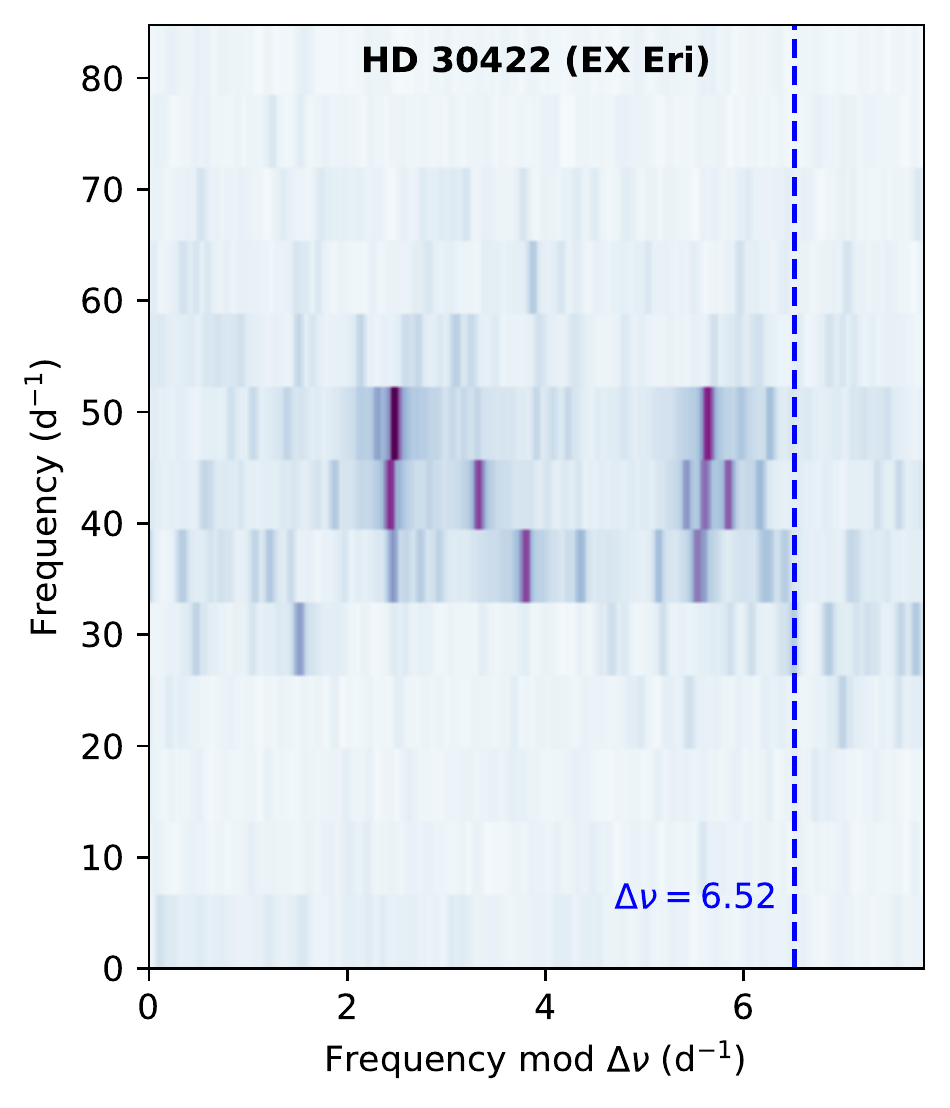}
\includegraphics[width=0.33\textwidth]{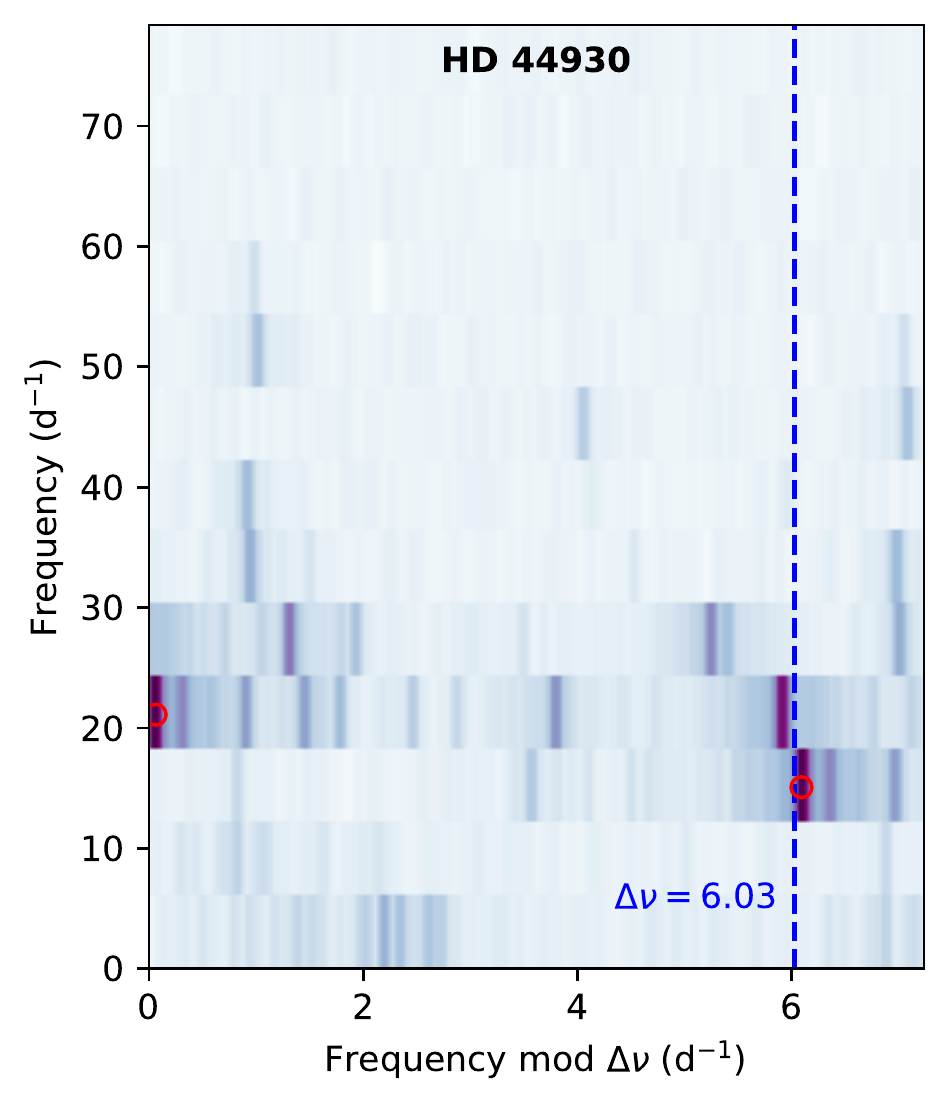}\\
\caption{\'Echelle diagrams of HD\,17341, HD\,30422 and HD\,44930. While HD\,17341 shows two nearly vertical ridges at $x\approx1.5$ and $x\approx4.3$\,d$^{-1}$, there is no clear candidate for the fundamental mode. By contrast, the fundamental mode is clear in HD\,44930 (red circle).}
\label{fig:echelles_1}
\end{center}
\end{figure*}

The p\:modes span a broad frequency range. This star lies on the ZAMS and has a measured $\Delta\nu = 5.90$ (Bedding et al., 2020, in press). The P--L relation suggests a fundamental mode frequency of 16.3\,d$^{-1}$, which does fall close to an observed but weak peak at 16.4\,d$^{-1}$. However, there are other weak peaks at similar frequencies and the \'echelle diagram shows no clear candidate for the fundamental mode (Fig\,\ref{fig:echelles_1}). Some low-frequency peaks are evident and could be g-mode oscillations.

\subsection*{HD\,23392, sector 4}
Not a $\delta$\,Sct star at the 0.02\,mmag level, which is unsurprising given its location well beyond the blue edge of the $\delta$\,Sct instability strip. The star does not appear to pulsate in g\:modes, either. The light curve drifts on time-scales of one sector, dipping towards sector boundaries, so the peaks at low frequency are largely instrumental.

\subsection*{HD\,28490, sector 5}
The oscillation frequencies are at the low end of the range expected for $\delta$\,Sct p\:modes, consistent with the star's location near the TAMS. None of the dominant frequencies shows the expected period ratios for radial modes. Despite the high luminosity, an \'echelle diagram does show a ridge (using $\Delta\nu=2.32$\,d$^{-1}$), which implies the peak at $f_2 = 6.95$\,d$^{-1}$ is the fundamental mode. This identification is consistent with the P--L relation. The stronger peak at lower frequency ($f_1 = 5.2$\,d$^{-1}$) is presumably a mixed mode. There are no clear low frequencies (i.e. no g\:modes).

\subsection*{HD\,28548, sector 5}
The Fourier transform shows many almost-equally-spaced pulsation frequencies at high frequency ($\sim$35--65\,d$^{-1}$). Bedding et al. (2020, in press) found a very high $\Delta\nu=7.56$\,d$^{-1}$, consistent with the star's location near the ZAMS. They also matched it to a model, which had an age of 270\,Myr. The fundamental mode is weak or missing. Two clear peaks exist at low frequency, but are not harmonics of each other. Instead, they are themselves the first two peaks of a harmonic series. The harmonic nature argues against a $\gamma$\,Dor classification, but the cause is currently undetermined.
 
\subsection*{HD\,30422, sector 5}
Also known as EX\,Eri. The high-frequency p\:modes exhibit a regular spacing, with a large spacing of 6.52\,d$^{-1}$. This is one of the exemplars from Bedding et al. (2020, in press). The \'echelle suggests this star does not pulsate in the fundamental mode (Fig\,\ref{fig:echelles_1}).

\subsection*{HD\,31295, sector 5}
Pulsation models of this star are presented in Sec.\,\ref{sec:models}. Its p\:modes are at relatively high frequency ($\sim$40\,d$^{-1}$). There are few of them, resulting in a simple pulsation spectrum, and no excited modes lie near the predicted fundamental mode. The oscillation spectrum of this hot star seems to be incompatible with the period inferred from the P--L relation, which is probably because of the absence of a colour term. There are no clear g\:modes.

\subsection*{HD\,31508, sectors 1--13}
\begin{figure}
\begin{center}
\includegraphics[width=0.48\textwidth]{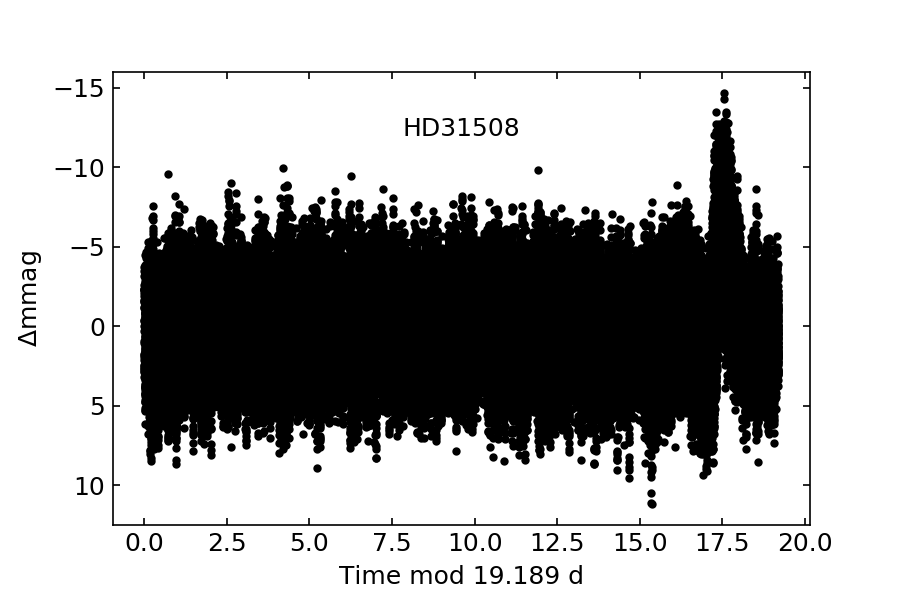}\\
\includegraphics[width=0.48\textwidth]{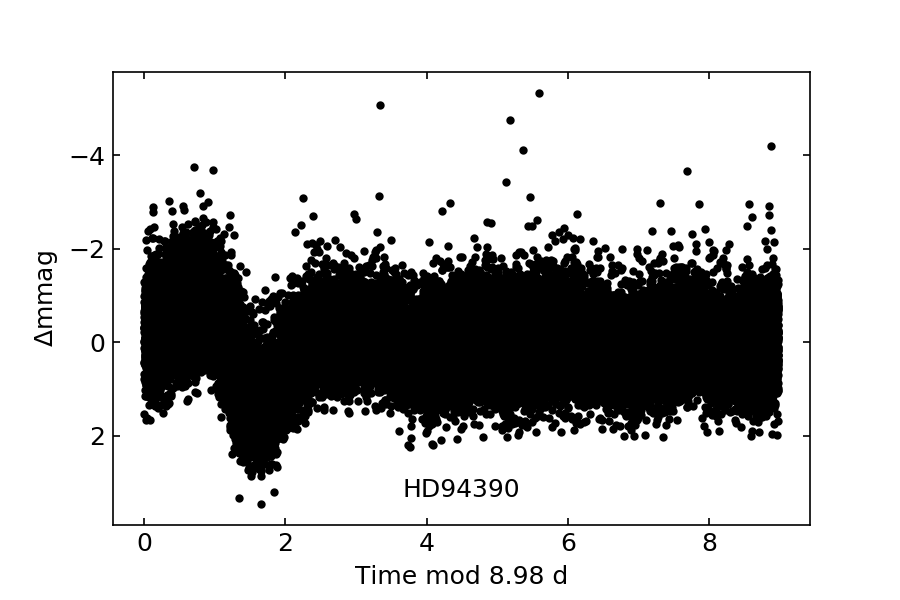}\\
\caption{TESS light curves of the $\lambda$\,Boo heartbeat stars HD\,31508 and HD\,94390, folded on their orbital periods of 19.189\,d and 8.98\,d, respectively.}
\label{fig:heartbeat}
\end{center}
\end{figure}

This target falls on silicon for sectors 1--13, only a subset of which has 2-min data available (2, 3, 6, 12, 13). We analysed the 2-min data only. The target is a heartbeat star (an eccentric ellipsoidal variable) with a period of 19.2\,d. The periastron brightening is stronger than the dip immediately preceding it (Fig.\,\ref{fig:heartbeat}).

\subsection*{HD\,36726, sector 6}
This non-pulsator is also discussed in Sec.\,\ref{sec:models}. No pulsation is evident down to a grass level of 0.025--0.030\,mmag. The star is much hotter than the blue edge of the $\delta$\,Sct instability strip.

\subsection*{HD\,37411, sector 6}
This FFI-only target is not a $\delta$\,Sct star at the 0.2\,mmag level. It does, however, show non-periodic excursions of $\sim$25\,mmag (peak-to-peak) on timescales of days. This variability is likely associated with its pre-MS nature -- \citet{mcnamara&huels1983} found HD\,37411 to be a member of the Orion Nebula Trapezium cluster with 96\% probability and \citet{gray&corbally1998} studied the spectrum of this star in detail, finding emission lines and a variable spectrum. It has near- and far-infrared excesses and is accreting matter \citep{malfaitetal1998,gradyetal1996,gray&corbally1998}. According to the star's spectral type (hA2\,Vae\,kA8mA8 $\lambda$\,Boo; \citealt{murphyetal2015b}) our $T_{\rm eff}$ of $\sim$11400\,K is probably too hot. Either way, HD\,37411 lies beyond the blue edge of the $\delta$\,Sct instability strip, so its lack of pulsation is unsurprising. We also find a large extinction of $A_V = 1.56$\,mag for this target.

\subsection*{HD\,38043, sectors 1--3, 5--6, 8--13}
The TESS light curve spans 357\,d and includes all sectors except 4 and 7. The data are at 2-min cadence in all available sectors. The p\:modes appear to be dominated by two nearly-equally spaced multiplets: the first contains the strongest three modes -- the same ones for which frequencies and amplitudes are provided in Table\,\ref{tab:puls} -- with a spacing of $\sim$0.51\,d$^{-1}$. The second is slightly lower in frequency and amplitude and has a spacing of $\sim$0.55\,d$^{-1}$. These could be rotationally-split multiplets of modes with similar but non-identical $n$ and $\ell$. This star is highly multiperiodic at lower frequencies, consistent with g-mode pulsation, though we could not identify any clear period spacings.

\subsection*{HD\,41958, sectors 1, 4--5. 7--8, 11}
The three strongest p\:modes have similar amplitudes (7.5--8.2\,mmag). They do not have any obvious harmonic relationship, nor do they have the frequency ratio (0.77--0.78) expected of the fundamental mode and first radial overtone modes.

\subsection*{HD\,42503, sectors 5--7}
The mutliperiodic nature of the light curve at frequencies near 5\,d$^{-1}$ is rather unusual. The frequency content does not strongly resemble that of $\gamma$\,Dor stars, but is also quite different from typical $\delta$\,Sct stars. No radial modes could be identified from frequency ratios of the stronger peaks. The peaks at 0.99 and 2.00\,d$^{-1}$ are not perfect harmonics, and another peak at 3.94\,d$^{-1}$ also falls on this near-harmonic pattern. 

\subsection*{HD\,43533, sectors 6--7}
The star is a $\gamma$\,Dor variable and the light curve is dominated by low-frequency variability, with dozens of significant peaks below 6\,d$^{-1}$, and this variability spills into the $\delta$\,Sct frequency domain. Excluding this variability, which presumably does not include independent p\:modes, there are only a few peaks in the Fourier transform that make this a $\delta$\,Sct star. All three extracted peaks are significant, with SNR$>$4, despite their low amplitudes. The low amplitudes perhaps suggest these peaks belong to a contaminating source, but the Gaia DR2 data show no likely contaminant. It is noteworthy that this star is located below the ZAMS compared to tracks of solar metallicity.

 \subsection*{HD\,44930, sector 5}
The Fourier transform is dominated by several p\:modes of mmag amplitude. A group of weaker peaks with frequencies between 2 and 3\,d$^{-1}$ might be g\:modes. The absence of peaks at even lower frequencies would then suggest these are $\ell=1$, $m=1$ prograde modes, which suggests the near-core region rotates rapidly. This star is in the Octans association and has a measured $\Delta\nu$ of 6.03\,d$^{-1}$ (Bedding et al. 2020, in press). The \'echelle diagram ((Fig\,\ref{fig:echelles_1}) suggests $f_1=18.15$\,d$^{-1}$ is the fundamental mode, even though this is not in agreement with the P--L prediction.

\subsection*{HD\,47425, sectors 6--7}
The two dominant p\:modes are too close in frequency to be the fundamental mode and first radial overtone. Indeed, it is not possible to clearly identify radial modes by period ratios in this star. There are many p\:modes with mmag amplitudes, though no g\:modes are evident at amplitudes similar to the p\:modes.

\subsection*{HD\,46722, sectors 6--7}
This $\delta$\,Sct star has a semi-regular series of frequencies from $\sim$30--50\,d$^{-1}$ with large spacing $\Delta\nu=6.45$\,d$^{-1}$ (Bedding et al. 2020, in press). Three humps of peaks at lower frequency suggest this is also a $\gamma$\,Dor star. The \'echelle diagram and the P--L relation suggest that $f_1=20.76$\,d$^{-1}$ is the fundamental mode.

\subsection*{HD\,68695, sectors 7--8}
In these FFI data, this $\delta$\,Sct star has only three significant p-mode peaks. The identified oscillation frequencies may be Nyquist aliases -- one of the aliases of the identified $f_1$ has a higher amplitude by 1.9\,$\upmu$mag, which may suggest the `alias' is in fact the real peak (see \citealt{murphy2015}). However, the noise level is an order of magnitude higher than this and the star is not located near the ZAMS (where oscillation frequencies tend to be higher), so with the current observations, we suggest the given frequencies are more likely to be the real ones.

\subsection*{HD\,73211, sector 8}
The peak between the two strongest peaks is $f_4 = 19.25$\,d$^{-1}$; the third strongest peak falls outside of the range of the Fourier plot in Appendix\,\ref{sec:FTs}. The triplet formed from $f_1$, $f_2$ and $f_4$ is not equally spaced, and none of these peaks falls near the predicted fundamental mode from the P--L relation.

 \subsection*{HD\,74423, sectors 9--11}
The analysis of the TESS light curve of this fascinating $\lambda$\,Boo star is described in a dedicated paper (Handler et al 2020, submitted). The entire variability can be represented with two independent frequencies and as many as 21 combination frequencies of amplitude $>$0.01\,mmag. The two independent frequencies are a binary orbital frequency, $f_{\rm orb} = 0.63262$\,d$^{-1}$, and a single p-mode oscillation at $f_1 = 8.7569$\,d$^{-1}$. Handler et al. (2020) interpret the light curve as pulsations trapped in a single hemisphere of the star by tidal forces. The reader is referred there for more information.

\subsection*{HD\,75654, sector 8--9}
Pulsation models of this star are presented in Sec.\,\ref{sec:models}. The strongest peak lies almost on top of the predicted fundamental mode frequency from the P--L relation. In addition to the p\:modes, there are also peaks at 0.36, 1.07 and 1.62\,d$^{-1}$. These are not obvious harmonics and combinations, and may be r and g\:modes.

\subsection*{HD\,76097, sector 8}
The third peak falls close to the strongest one, and their spectral windows are not well separated. The frequencies for this star were therefore extracted by simultaneous non-linear least-squares fitting, rather than by the automated algorithm. The fundamental mode cannot be unambiguously identified because of the proximity of these strong peaks, and because there are no clear candidates for the first and second radial overtone modes for reference.

\subsection*{HD\,80426, sector 8}
The star is dominated by a 10-mmag peak at $f_1=11.01$\,d$^{-1}$. Despite first appearances, the next strongest peaks are not harmonics of this frequency. The ragged peak shape is simply the result of the spectral window. Since this star is located above the TAMS and at the red edge of the instability strip, this frequency appears to be too high to be the fundamental mode.

\subsection*{HD\,83041, sectors 8--9}
In addition to the p\:modes, there are many peaks below 5\,d$^{-1}$, which suggests this star is also a $\gamma$\,Dor variable. The strongest peak agrees with the one identified in SuperWASP data by \citet{paunzenetal2015}.

\subsection*{HD\,84159, sector 8}
The broad peaks in the Fourier transform are simply due to a broad spectral window, caused by the large data gap between orbits. The p-mode amplitudes are very high. A pair of peaks near 1.8\,d$^{-1}$ could be prograde g\:modes. The group of peaks near 18\,d$^{-1}$ appear to be combination frequencies involving the strongest peak, $f_1$, which is to be expected given the strong pulsation amplitudes. No frequency pair gives a period ratio consistent with both being radial modes.

\subsection*{HD\,88554, sectors 9--10}
An unremarkable low amplitude $\delta$\,Sct star.

\subsection*{HD\,94326, sectors 9--10}
The light curve morphology gradually changes from the start of sector 9 to the end of sector 10, with the amplitude of the low-frequency variability decreasing. Independent of this, the star appears to be a $\gamma$\,Dor variable, with mmag peaks between 0.8 and 1.4\,d$^{-1}$. The strongest $\delta$\,Sct peak is at 10.1\,d$^{-1}$, and there is significant variability at many frequencies between this and the g\:modes at the $\sim$0.3 mmag level. No peak stands out as the fundamental mode.

\subsection*{HD\,94390, sectors 9--10}
This is a heartbeat star (an eccentric ellipsoidal variable) with an orbital period of 8.98\,d, perhaps exhibiting tidally excited oscillations. However, it is not a $\delta$\,Sct star at the 0.02\,mmag level.

\subsection*{HD\,98069, sector 9}
This object has both K2 and TESS (FFI) data. There has been remarkable change in pulsation amplitudes between the two data sets, beyond that which is explained by the change in passband. While the strongest mode is still at the same frequency, the relative amplitudes of other modes compared to this peak are very different. The second and third strongest modes in the K2 data are only marginal in the TESS data set. The extracted peaks given here are the TESS ones, for consistency with the other stars. We assume that the real peaks fall between 0 and the Nyquist frequency of 24\,d$^{-1}$. The temperature and luminosity of the star place it near the TAMS, so this assumption appears to be justified.

\subsection*{HD\,100546, sector 11}
The TESS light-curve is unremarkable in most regards, and show this is not a $\delta$\,Sct star. It is also too hot to lie in the $\delta$\,Sct instability strip. Some low-frequency variability is evident, even against the rising background at low frequency. There could be g\:modes, or this could be attributed to rotation, or since this star is known to be actively accreting material \citep{vieiraetal1999}, the low frequencies could be accretion-induced variability.

\subsection*{HD\,101412, sectors 10--11}
Only FFI data are available in each sector. Only one peak is (questionably) evident in the Fourier transform, at 10.679\,d$^{-1}$, but it has a SNR of 3.3, and we therefore do not consider it significant. Hence, HD\,101412 is not a $\delta$\,Sct star, despite being inside the instability strip.

\subsection*{HD\,102541, sector 10}
A known $\delta$\,Sct star (V1023\,Cen). The Fourier transform shows a series of oscillation frequencies that continue to high frequency. Closer inspection shows they form a regular pattern and have a $\Delta \nu$ of 6.52\,d$^{-1}$. The strongest peak is within 1\,d$^{-1}$ of the predicted fundamental mode frequency. This ID is confirmed by the \'echelle diagram (Fig.\,\ref{fig:echelle_102541}, Sec.\,\ref{sec:PLC}). 

\subsection*{HD\,103701, sector 10}
In addition to the $\delta$\,Sct pulsation, there are peaks at 1.35 and 1.53\,d$^{-1}$ that are likely g\:modes. The low p-mode frequencies are consistent with the star's position on the HR diagram, apparently beyond the TAMS (assuming solar metallicity and $f_{\rm ov} = 0.02$).  However, this star  has a high RUWE (2.04) in Gaia DR2, so might be an unresolved binary and a little over-luminous.

 \subsection*{HD\,109738, sector 11}
A highly multiperiodic $\delta$\,Sct star. The strongest peak is at about twice the frequency predicted by the P--L relation, making this a `second ridge' star \citep{ziaalietal2019}. We tentatively infer a large separation of 4.8--4.9\,d$^{-1}$ for this star. If the fundamental mode is excited, it is not driven strongly. A power excess around 2\,d$^{-1}$ is likely to be a group of $\ell=1$ g\:modes.
 
\subsection*{HD\,111786, sector 10}
\begin{figure*}
\begin{center}
\includegraphics[width=0.33\textwidth]{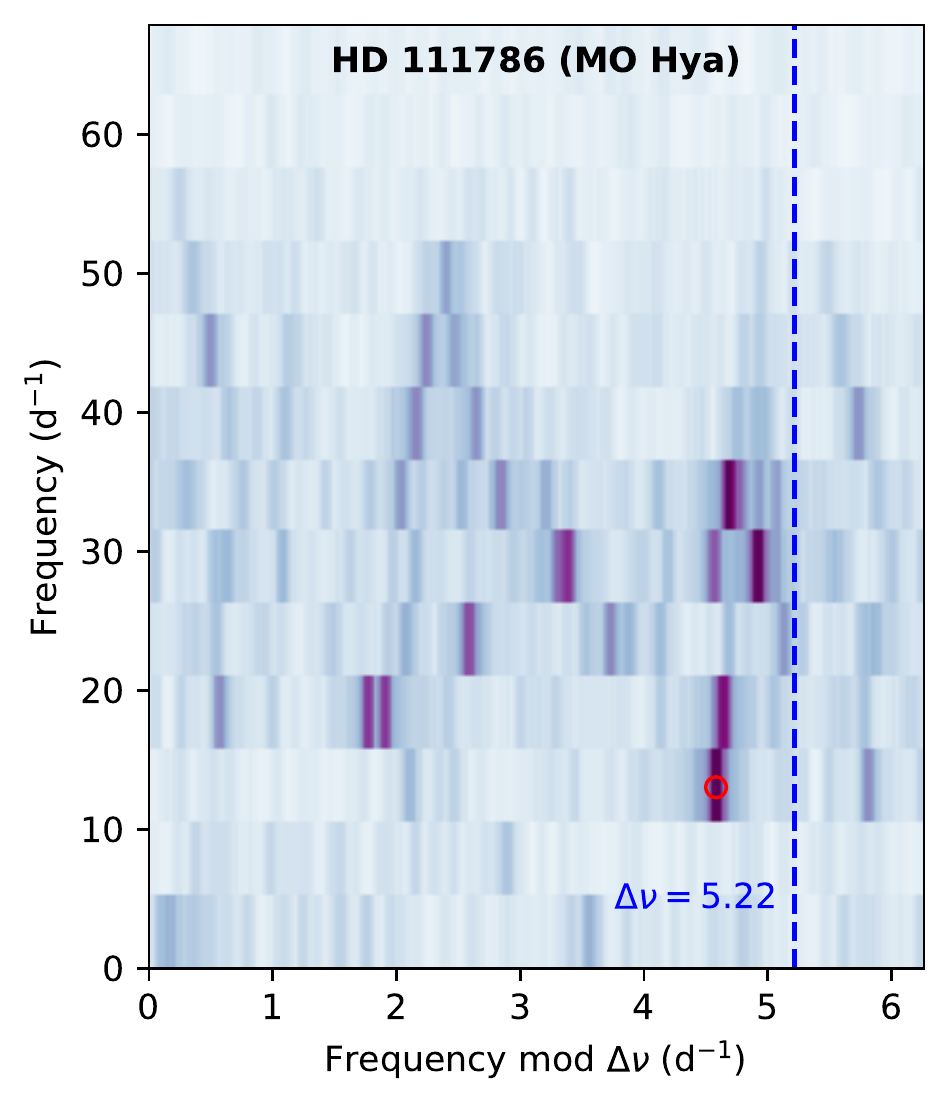}
\includegraphics[width=0.33\textwidth]{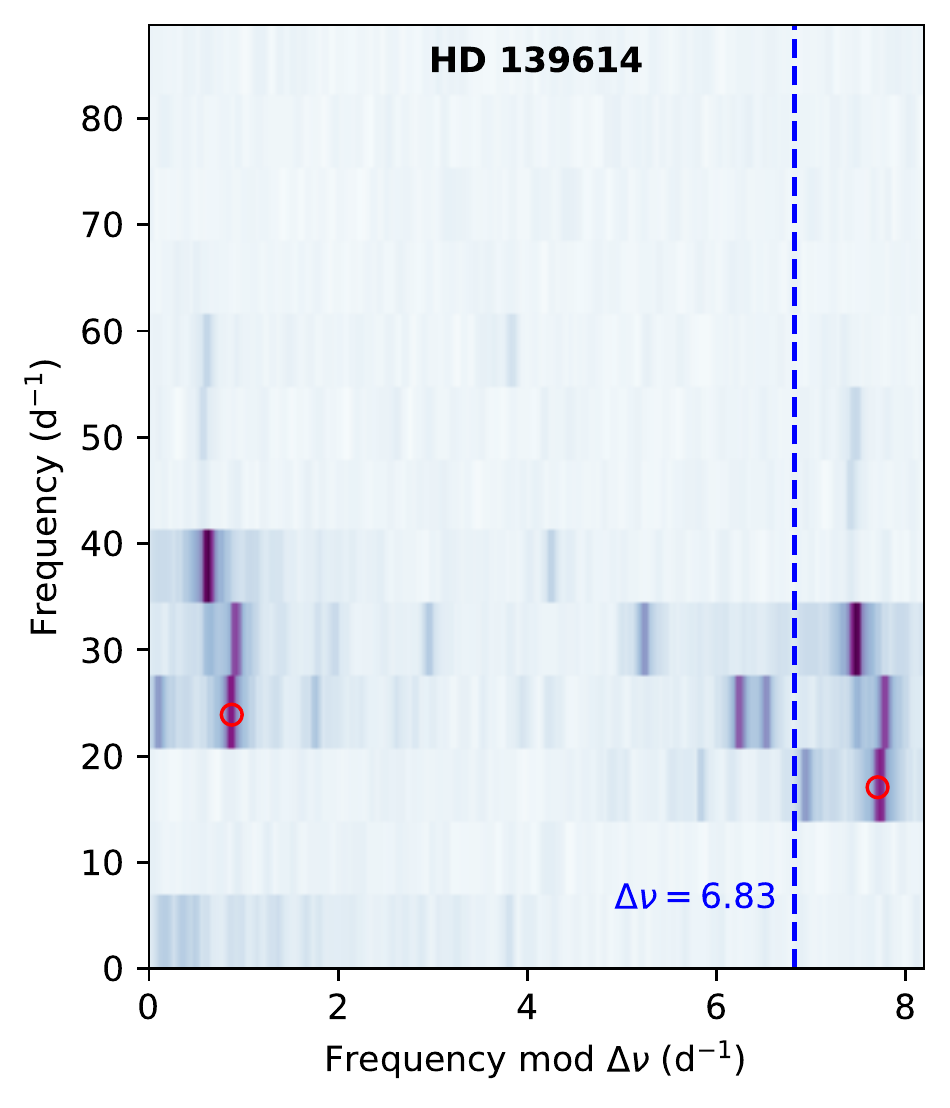}
\caption{\'Echelle diagrams of HD\,111786 and HD\,139614. Red circles show the fundamental mode. Identification of this mode in HD\,111786 is ambiguous based on the \'echelle only, but the P--L relation argues for the stronger of the peaks near the end of order 3.}
\label{fig:echelles_2}
\end{center}
\end{figure*}

A known $\delta$\,Sct star (MO\,Hya), this star exhibits pulsation over a broad range of frequencies. The strongest peak ($f_1 = 15.02$\,d$^{-1}$) lies close to the P--L prediction (15.37\,d$^{-1}$). It is noteworthy that $f_2$ is close to (but not exactly) double the frequency of $f_1$, and the two are almost equal in amplitude. That is, it would be a typical `second ridge' star on the P--L diagram \citep{ziaalietal2019} if the amplitudes were only slightly different. Ridges emerge on an \'echelle diagram using a large frequency separation of $\Delta\nu \approx 5.2$\,d$^{-1}$ (Fig.\,\ref{fig:echelles_2}).

\subsection*{HD\,112682, sector 10}
This is a non-pulsator at the 0.04\,mmag level. It lies close to the theoretical blue edge of the $\delta$\,Sct instability strip \citep{dupretetal2005b}, but comfortably inside the observational blue edge \citep{murphyetal2019}.

\subsection*{HD\,112948, sector 10}
A clear $\delta$\,Sct star. There is no obvious candidate for the fundamental mode. A solitary peak near 11\,d$^{-1}$ appears to be unresolved, and lies far from the prediction from the P--L relation (see Fig.\,\ref{fig:FT_montage}). Two very close peaks at 1.78\,d$^{-1}$ could be g-mode pulsation, but the lack of other peaks in the typical $\gamma$\,Dor frequency range is puzzling if so.

\subsection*{HD\,119896, sector 11}
Only FFI data are available. This is a clear rotational variable situated close to the red edge of the $\delta$\,Sct instability strip, but not a $\delta$\,Sct star (at the 0.05\,mmag level).

 \subsection*{HD\,120122, sector 11}
A series of peaks beginning at $f = 1.49$\,d$^{-1}$ are near harmonics of each other, but the series is imperfect. It is unclear whether they describe rotational variability (suggesting starspots, potentially on a cooler companion) or perhaps g\:modes near combination frequencies \citep{kurtzetal2015,saioetal2018b}. We consider the unmarked peak near 4.6\,d$^{-1}$ to be part of this series. There is a Fourier peak at 16.07\,d$^{-1}$, suggesting this is also a $\delta$\,Sct star, or that there is otherwise a $\delta$\,Sct star in the aperture. This peak is far from the predicted fundamental mode frequency (18.8\,d$^{-1}$), and is low in amplitude (0.09\,mmag), so we suspect it is not the fundamental mode and that this star does not pulsate in the fundamental mode. Near 8.60\,d$^{-1}$ there is a pair of poorly-resolved Fourier peaks that are possibly additional pulsation modes, one of which is the strongest peak in the $\delta$\,Sct frequency region. The star is located between the theoretical \citep{dupretetal2005b} and observational \citep{murphyetal2019} red edges of the $\delta$\,Sct instability strip.

\subsection*{HD\,125508, sector 11}
A clear $\delta$\,Sct star, also with many lower frequency peaks that could be g\:modes, though combination frequencies of the p\:modes would have to be ruled out. A peak at 0.500\,d$^{-1}$ and its harmonic are present, which are not likely to be g\:modes. With an extremely high RUWE of 16.630 and no parallax data available in DR2, HD\,125508 is almost certainly a binary or multiple system.
 
 \subsection*{HD\,126627, sectors 11--12}  
 Although this is not a $\delta$\,Sct star at around the 0.04\,mmag level, it is a high-amplitude g-mode pulsator, with the strongest g\:modes achieving amplitudes of over 10\,mmag. Because the time series is only two sectors long, the g\:modes are not resolved from one another, but the light curve shows clear beating, and the non-sinusoidal nature of the light curve suggests the star pulsates in prograde sectoral modes and exhibits combination frequencies \citep{kurtzetal2015,saioetal2018b}. The pulsational properties of the star are consistent with its location on the HR diagram, at low luminosity and between the theoretical \citep{dupretetal2005b} and observational \citep{murphyetal2019} red edges of the $\delta$\,Sct instability strip, therefore in the middle of the $\gamma$\,Dor instability strip \citep{dupretetal2005b}.
 
 \subsection*{HD\,127659, sectors 11--12} 
 The star is multiperiodic, with several peaks having amplitudes greater than 1\,mmag. It was excluded from the analyses in Sections\,\ref{sec:puls_frac}--\ref{sec:models} on the basis of its hot effective temperature (11\,900\,K), which might suggest that it is a $\beta$\,Cep variable rather than a $\delta$\,Sct variable. However, it has a hydrogen line (spectral) type of F2\,V in \citet{grayetal2017}, where it was first identified as a $\lambda$\,Boo candidate, and has a spectral type of F2\,V\,kA3mA4~$\lambda$\,Boo in Murphy et al. (submitted), who confirmed it as a $\lambda$\,Boo star. These suggest a temperature closer to 7000\,K, in which case it is indeed a $\delta$\,Sct star. It also shows low-frequency peaks in the Fourier transform, so it may be a $\gamma$\,Dor--$\delta$\,Sct hybrid. The uncertainty over its parameters justifies its exclusion from analyses in Sections\,\ref{sec:puls_frac}--\ref{sec:models}.

 \subsection*{HD\,139614, sector 12} 
  The pulsation spectrum contains relatively few modes spread over a wide frequency range. This low mode density is amenable to analysis with only one sector of TESS data. The modes have a characteristic large spacing, which we measure to be 6.83\,d$^{-1}$, consistent with its position near the ZAMS. The predicted fundamental mode at 21.6\,d$^{-1}$ falls very close to the observed peak at 21.36\,d$^{-1}$. The \'echelle diagram (Fig.\,\ref{fig:echelles_2}) confirms this is the correct identification.

 \subsection*{HD\,153747, sector 12} 
 The TESS light curve spans only one sector, and does not have sufficient frequency resolution for the multiple pulsation modes of this star, but the three extracted frequencies (Table\:\ref{tab:puls}) are resolved from each other. Other peaks of similar amplitudes and frequencies are present in the Fourier transform of the light curve. We find a characteristic spacing of $\sim$4.8\,d$^{-1}$, but no candidate for the fundamental mode.
 
\subsection*{HD\,154153, sector 12}
Not a $\delta$\,Sct pulsator, down to the grass level of around 0.015\,mmag. The lack of pulsation is surprising, given the star's location in the middle of the $\delta$\,Sct instability strip. Two weak (0.05\,mmag) low frequencies could be g\:modes.
 
\subsection*{HD\,154951, sector 12}
This $\delta$\,Sct star oscillates in a single p\:mode (at the precision of these TESS FFI data) and apparently many g modes. The light curve appears noisy or otherwise contains considerable additional variability at the start of the sector.

\subsection*{HD\,159021, sectors 12--13}
This star seems to oscillate richly in g and p\:modes. The broad power excesses around the dominant p\:modes is not a spectral window effect. This leaves the high mode density difficult to explain. The two stronger p\:modes appear to be the fundamental mode ($f_2=10.707$\,d$^{-1}$) and the first radial overtone mode ($f_1=13.825$\,d$^{-1}$) based on their observed period ratio of 0.774. This suggests that the luminosity based on the Gaia DR2 parallax is incorrect.

\subsection*{HD\,162193, sector 13}
The light curve is of good quality. The dominant mode has a 1-hr period and also shows a clear harmonic. Using the P--L relation, we predict a fundamental mode of 8.65\,d$^{-1}$, but there is no observed peak near this frequency. The closest candidate is at 10.13\,d$^{-1}$ and there are no other contenders. The star lies between the theoretical \citep{dupretetal2005b} and observational \citep{murphyetal2019} blue edges of the $\delta$\,Sct instability strip.

\subsection*{HD\,168740, sector 13}
\begin{figure}
\begin{center}
\includegraphics[width=0.48\textwidth]{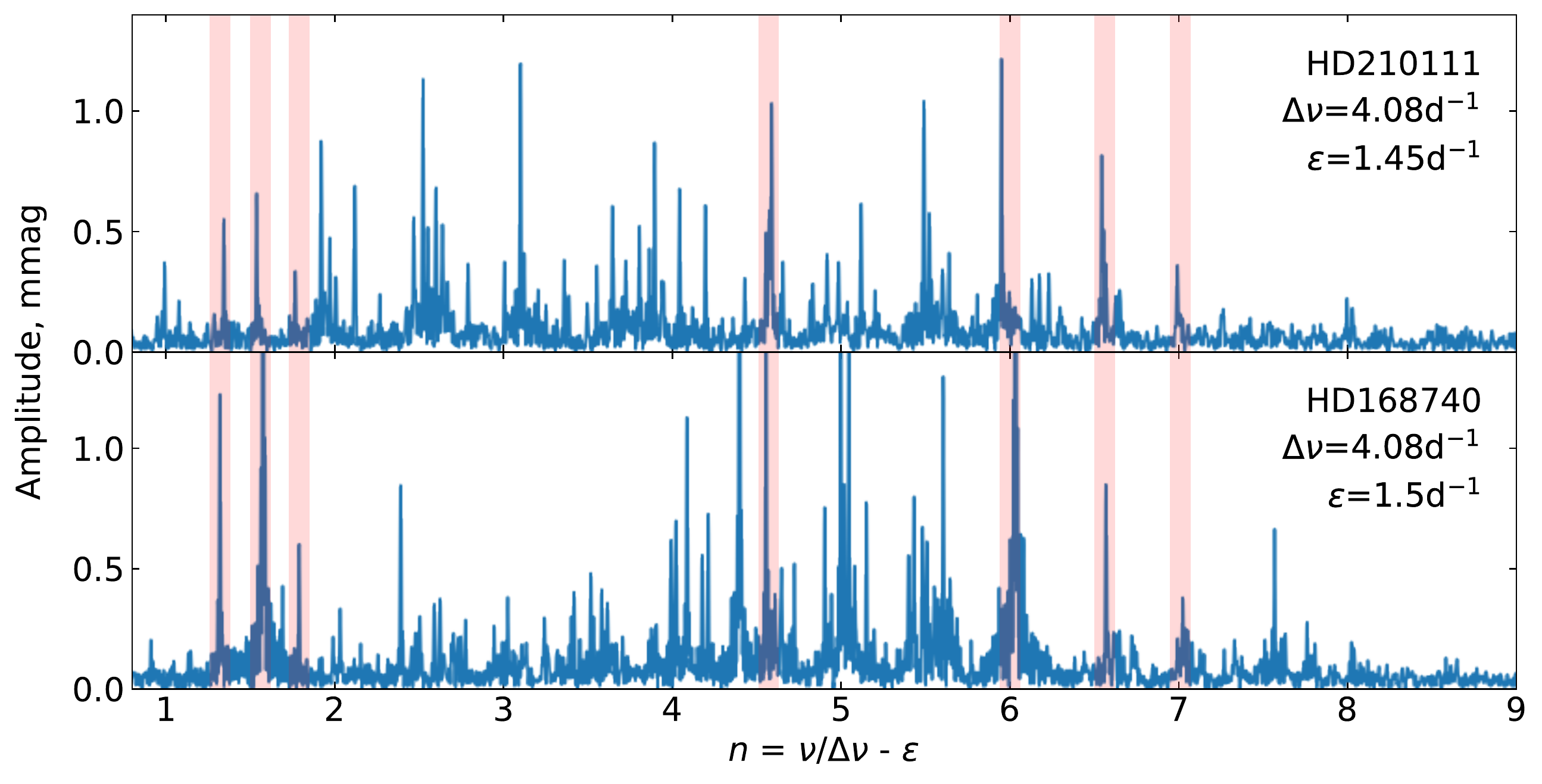}\\
\caption{Amplitude spectra of the near-twins HD\,210111 and HD\,168740. The stars have $\Delta T_{\rm eff} = 200$\,K and $\Delta \log L = 0.01$\,dex. Vertical red rectangles highlight similar features within approx 0.1\,d$^{-1}$ in the two amplitude spectra. At frequencies in the middle, the pulsation spectra are too messy to make meaningful matches.}
\label{fig:twin}
\end{center}
\end{figure}

A highly multiperiodic $\delta$\,Sct star. Using the P--L relation, we calculate the frequency of the fundamental mode to be $12.3$\,d$^{-1}$, which falls close to the rather prominent peak $f_2=12.57$\,d$^{-1}$. This is confirmed using an \'echelle diagram with a large spacing of $\Delta\nu=4.08$\,d$^{-1}$ (Fig.\,\ref{fig:echelles_3}). This rather small value $\Delta\nu$ is consistent with a star that has evolved considerably away from the ZAMS. Indeed, our observed $T_{\rm eff}$ and $\log L$ place the star about a third of the way through its main-sequence lifetime. Interestingly, a near-twin on the HR\,diagram (HD\,210111) has a very similar amplitude spectrum (Fig.\,\ref{fig:twin}). A 1-mmag peak at 0.80\,d$^{-1}$ could be rotational in origin, and would be consistent with the known $v\sin i$ of 130\,km\,s$^{-1}$ \citep{heiter2002}.

\subsection*{HD\,168947, sector 13}
\begin{figure}
\begin{center}
\includegraphics[width=0.46\textwidth]{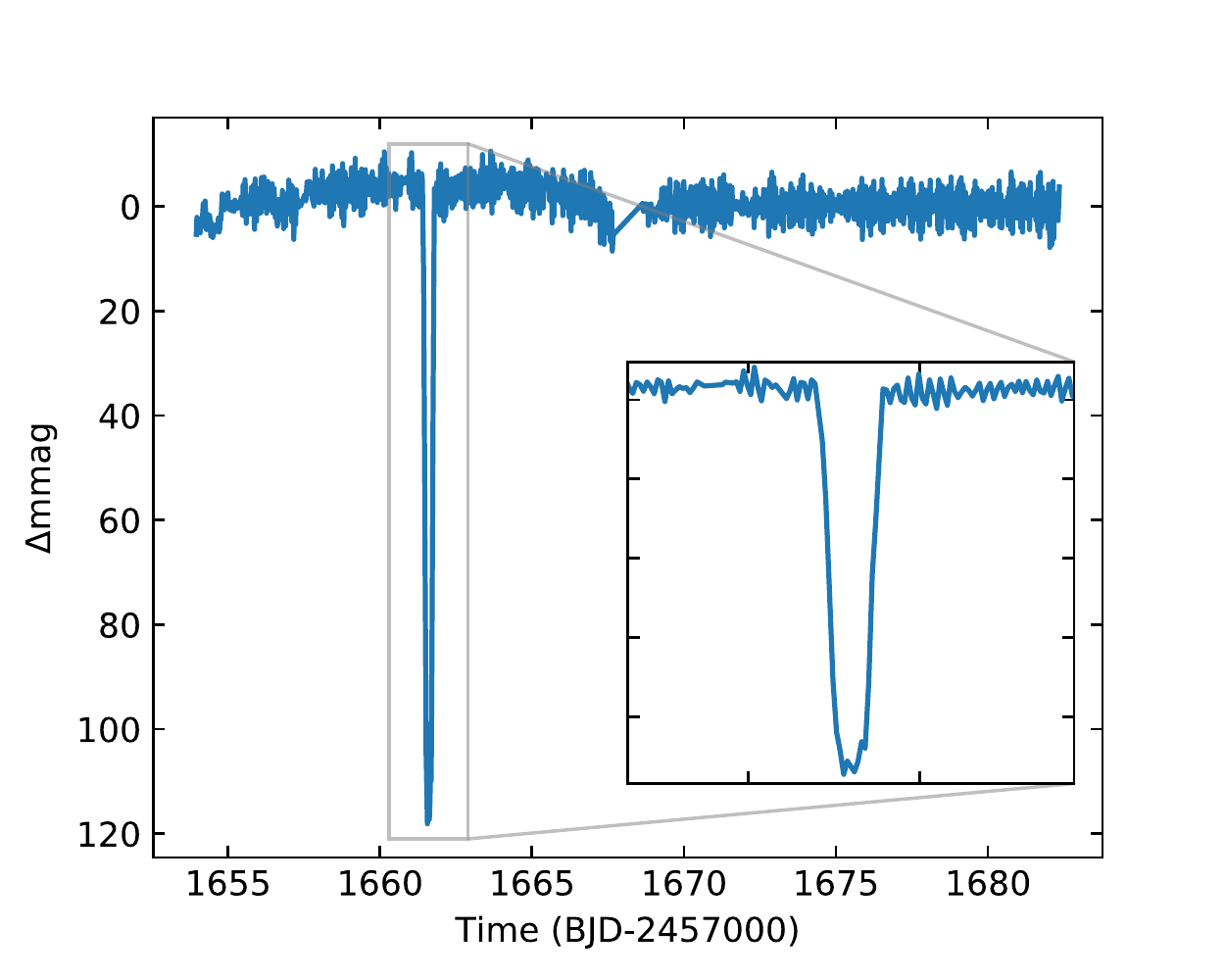}\\
\caption{TESS sector~13 FFI light curve of the $\lambda$\,Boo star HD\,168947, showing the single eclipse.}
\label{fig:eclipse}
\end{center}
\end{figure}

Only FFI data are available. The target shows a clear 0.12-mag eclipse of 0.39-d duration in the 27-d data (Fig.\,\ref{fig:eclipse}). It is difficult to determine whether the eclipse is truly flat-bottomed, but it is certainly not triangular. We infer a period of at least 21\,d since only one eclipse is observed. This, and the long eclipse duration, suggest that the companion is a late-type dwarf in a wide orbit. At least one of the stars is a $\delta$\,Sct variable. There are also some peaks near 0.7\,d$^{-1}$ that could potentially be r\:modes. Additionally, some power near 1.5\,d$^{-1}$ and 3\,d$^{-1}$ could be g\:modes, but the short light curve and low signal-to-noise make this uncertain.

\subsection*{HD\,169142, sector 13}
This target has only FFI data, the first TESS orbit of which are very poor in quality. Using only the second orbit, we infer that the star is not a $\delta$\,Sct star at the 0.15\,mmag level, but it does have some low-frequency variability, probably from g\:modes. Although there is one significant peak above 4\,d$^{-1}$, we did not extract it, as it is clearly associated with the lower frequency peaks rather than a p\:mode.

\subsection*{HD\,169346, sector 13}
This target has only FFI data. The first few days of the light curve have a significant dip, which could be the transit of a large body, or just an instrumental or data processing artefact. For the pulsation analysis we excluded these data points. The pulsation frequencies go right up to the Nyquist frequency of the FFI data, so the extracted peaks probably include aliases of super-Nyquist pulsation frequencies. The peak at $10.87$\,d$^{-1}$ is close to the predicted fundamental mode frequency (10.48\,d$^{-1}$), but it does not dominate the spectrum and the peak could be a Nyquist alias.

\subsection*{HD\,184779, sector 13}
The grouping of some p\:modes suggests they could be rotationally split multiplets. The star appears to have a large separation of $\Delta \nu \sim 4.0$\,d$^{-1}$. The dominant peaks agree with those found in SuperWASP data \citep{paunzenetal2015}. We identify the strongest peak at 12.5686\,d$^{-1}$ as the fundamental mode, which for roughly solar metallicity suggests a first radial overtone mode would lie at around 16.3\,d$^{-1}$ \citep{petersen&christensen-dalsgaard1996}. However, no strong peak occurs at this frequency, leaving the mode identification as tentative. The low frequency peaks at around 0.65 and 1.90\,d$^{-1}$ could be r and g\:modes, respectively. 

\subsection*{HD\,188164, sector 13}
Only FFI data are available, and the light curve is not particularly clean. Nonetheless, it shows the star is not a $\delta$\,Sct star, at the 0.015\,mmag level, despite the star sitting comfortably within the $\delta$\,Sct instability strip.

\subsection*{HD\,198160 and HD\,198161, sector 13}
Form a close binary with a separation of a few arcseconds (3.2\,arcsec in the WDS; 4.38\,arcsec in Gaia DR2), and a magnitude difference of 0.27\,mag in Gaia G \citep{gaiacollaboration2018a}. Both are known $\lambda$\,Boo stars \citep{murphyetal2015b}. These stars are unresolved by TESS so the light curve contains the pulsation properties of both stars (without guaranteeing that both are pulsators). According to Gaia DR2 \citep{gaiacollaboration2018a}, both stars have the same parallax (15.3\,mas) within the uncertainties. If this parallax is correct and not affected by the possible binarity of the system, then using the DR2 separation the stars are $\sim$300\,au apart and hence bound. Their proper motions, while similar, are not identical within the uncertainties, which likely arises from binary orbital motion. Throughout this paper, we consider only HD\,198160 in our statistics, and we have 70 targets rather than 71.

\subsection*{HD\,203709, sector 1}
This is a definite $\delta$\,Sct variable, and an unresolved group of peaks at 1.5--2.2\,d$^{-1}$ suggests it is also a $\gamma$\,Dor variable. The Fourier transform reveals an unusual pulsation property, which is an almost comb-like series of peaks between 4.1 and 8.1\,d$^{-1}$. The signal to noise of these peaks, and spacing between them at $\sim$0.7\,d$^{-1}$$\approx$81\,$\upmu$Hz, are both far too large for these to belong to a red giant companion. The photometry is also inconsistent with one component being a red giant. There is no obvious pattern in an \'echelle diagram.

\subsection*{HD\,210111, sector 1}
\begin{figure*}
\begin{center}
\includegraphics[width=0.33\textwidth]{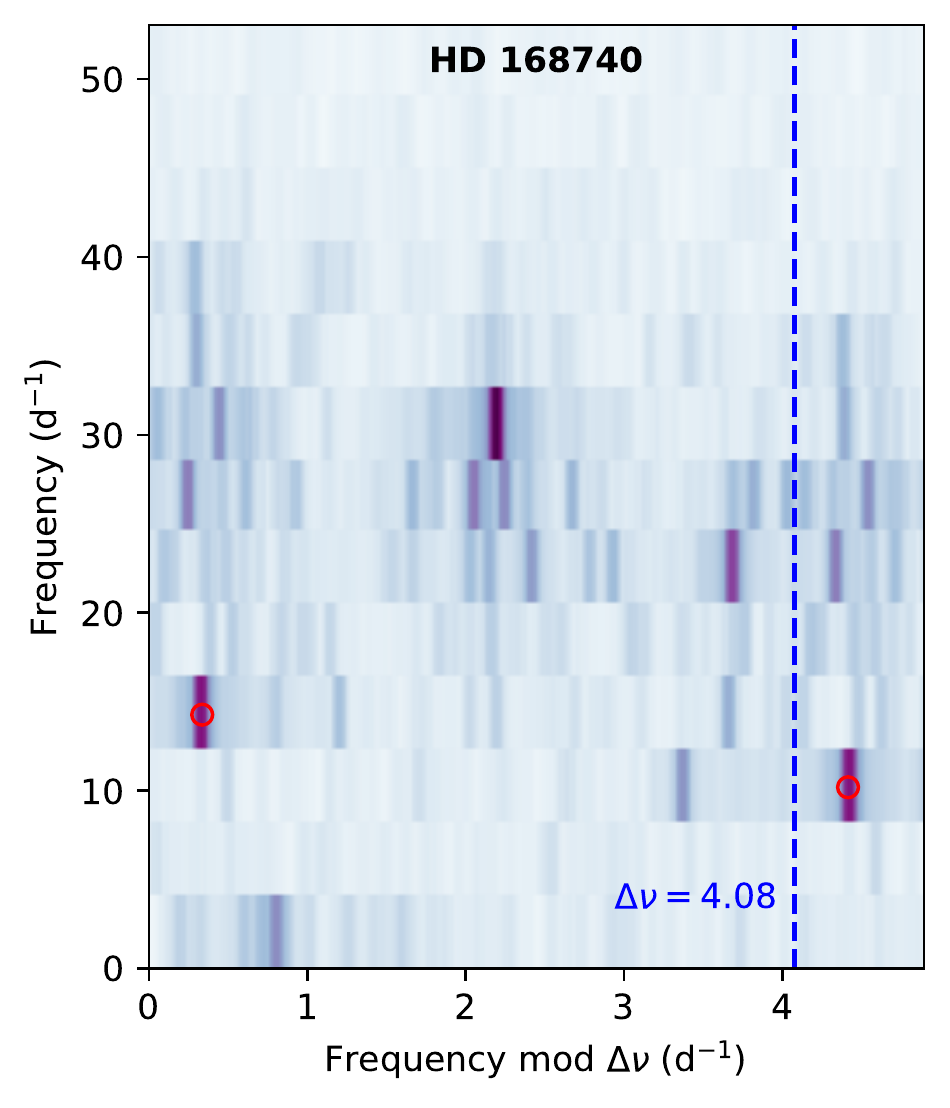}
\includegraphics[width=0.33\textwidth]{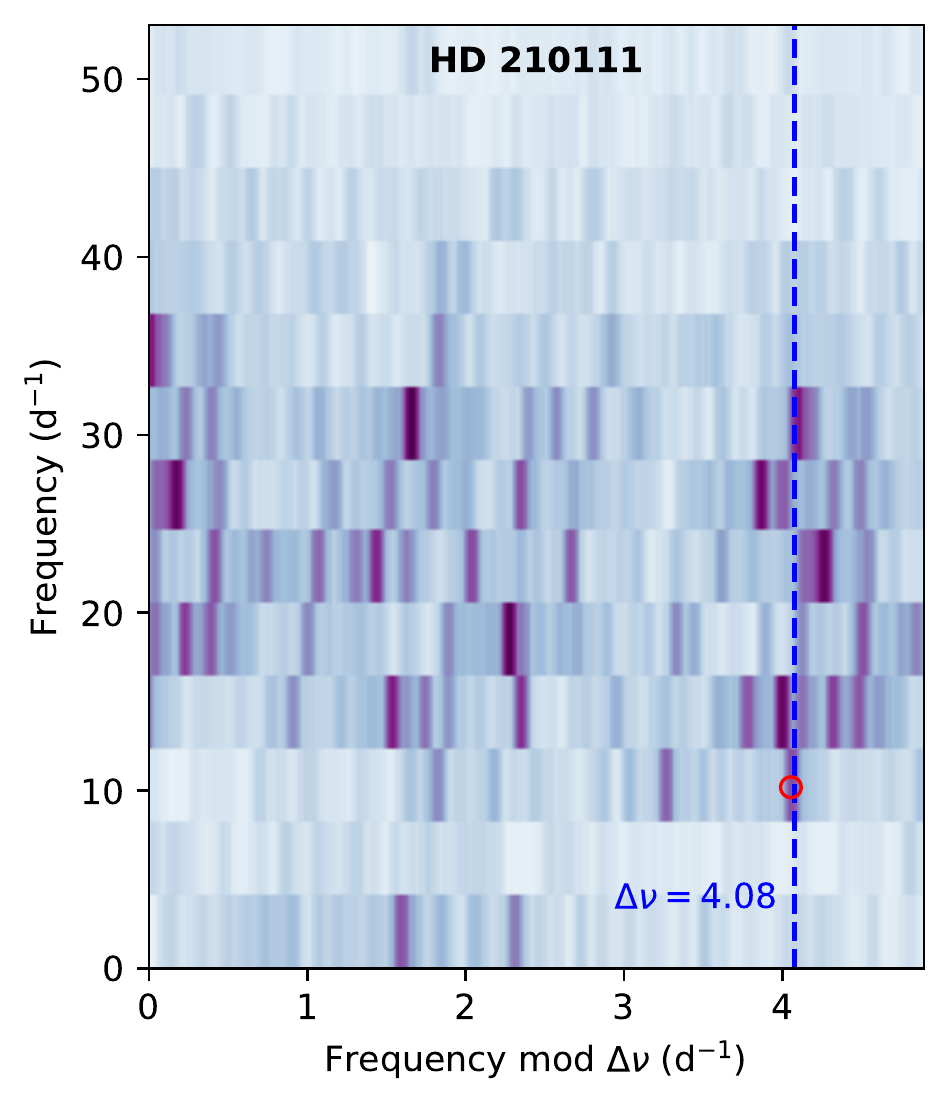}
\includegraphics[width=0.33\textwidth]{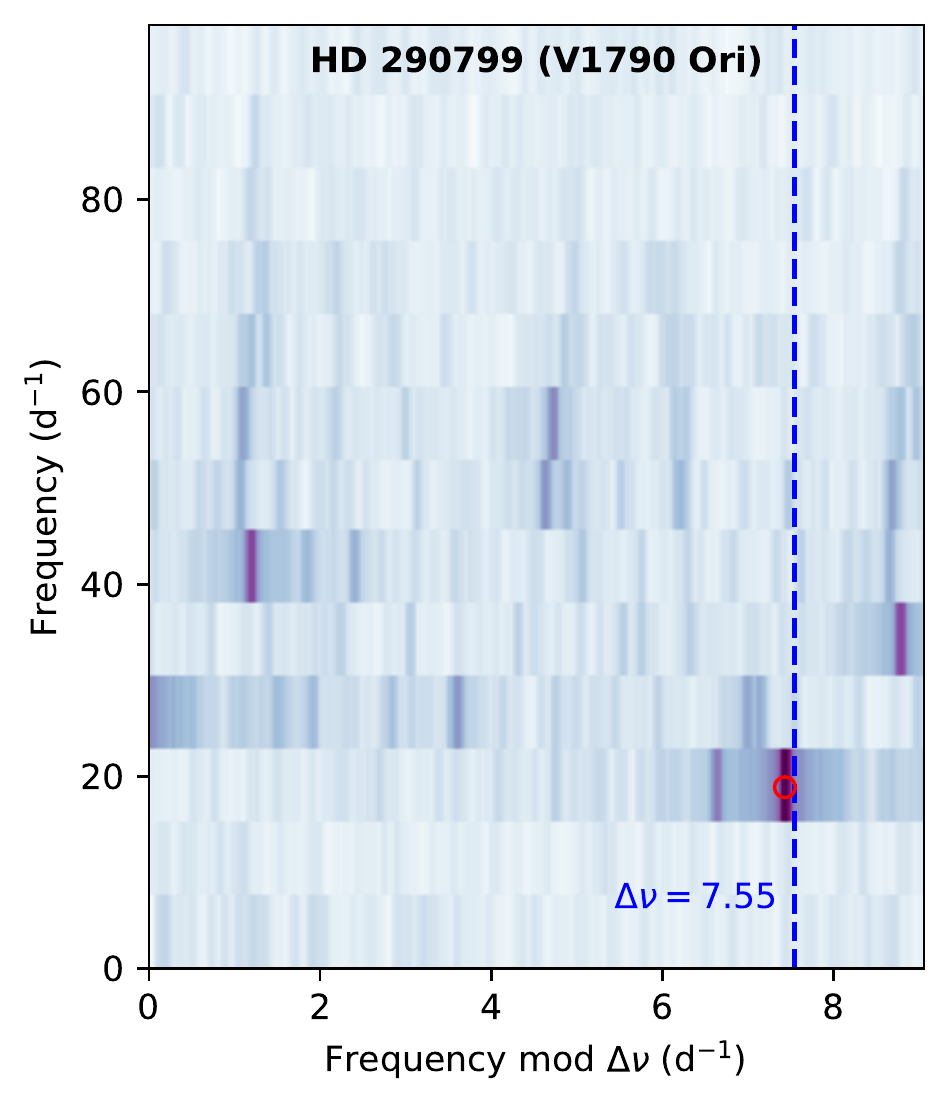}\\
\caption{\'Echelle diagrams of HD\,168740, HD\,210111 and HD\,290799, all of which have a clear fundamental mode (red circles).}
\label{fig:echelles_3}
\end{center}
\end{figure*}

This is a very highly multiperiodic $\delta$\,Sct star, which is also discussed in Sec.\,\ref{sec:models}. The TESS light curve has been described by \citep{antocietal2019}. There seems to be a spacing of about 4.1\,d$^{-1}$ in the \'echelle diagram of this TESS light curve (Fig.\,\ref{fig:echelles_3}), which also suggests the peak at 12.2\,d$^{-1}$ is the fundamental mode. The P--L relation agrees with this. The star is a near twin of HD\,168740 (Fig.\,\ref{fig:twin}). The two low-frequency peaks (at 1.60 and 2.31\,d$^{-1}$) are not harmonics of each other, which suggests one of the stars is likely a $\gamma$\,Dor star, too.

\subsection*{HD\,213669, sector 1}
The two strong peaks, $f_2$ and $f_3$, near 15\,d$^{-1}$ do not have a frequency ratio consistent with them both being low-order radial modes. The predicted fundamental mode frequency lies between the two, making it difficult to determine which is the correct identification. We tentatively identify that $f_2$ as the stronger and lower frequency of the two peaks is the fundamental mode. We note that $f_1$ is not a harmonic of $f_2$, but they are close. Given the frequency of $f_1$, this would be a `second ridge' star in the P--L diagram \citep{ziaalietal2019}.  It is noteworthy that the frequency content in this region of the FT is remarkably similar to HD\,38043, except that the multiplet spacing is smaller in HD\,38043. The low frequency peaks between 1.5 and 2.1\,d$^{-1}$ are almost certainly unresolved $\ell=1$ g\:modes.

\subsection*{HD\,214582, sector 1}
A 8.7-mmag peak at 1.26\,d$^{-1}$ dominates the Fourier transform of the light curve, and is accompanied by other low frequencies of mmag amplitude. Aside from this strong low-frequency content, two p\:modes are evident. Both are significant (SNR=10.3, 5.4, respectively), and only appear low in amplitude because of the contrast against the lower frequencies. This star has the highest predicted frequency from the P--L relation of all stars in the sample, but the prediction doesn't match the observed peaks well. We tentatively identify $f_1=19.76$\,d$^{-1}$ as the fundamental mode.
 
\subsection*{HD\,223352, sector 2}
The light curve contains a lot of noise, particularly of instrumental origin, but it still is possible to exclude $\delta$\,Sct pulsation at the 0.02\,mmag level. This is not surprising, given the $T_{\rm eff}$ of 9900\,K, far from the $\delta$\,Sct instability strip blue edge.

\subsection*{HD\,290799, sector 6}
A known $\delta$\,Sct star (V1790\,Ori). A clear $\delta$\,Sct star with oscillations covering a wide frequency range, also discussed in Sec.\,\ref{sec:models} and by Bedding et al. (2020, in press). The strongest peak ($f_1=22.53$\,d$^{-1}$) agrees with the one in SuperWASP data \citep{paunzenetal2015}; both differ by 1\,d$^{-1}$ from the one found by \citet[][22.53\,d$^{-1}$]{paunzenetal2002a}, suggesting the latter peak was a diurnal alias. Both an \'echelle diagram (Fig.\,\ref{fig:echelles_3}) and consideration of the P--L relation suggest that this strongest peak is the fundamental mode.

\subsection*{HD\,294253, sector 6}
A non-variable star at the 0.05\,mmag level, across all frequencies. A truly constant star, as is to be expected at this spectral type (A0). This star is also discussed in Sec.\,\ref{sec:models}.
 
\subsection*{BD-11\,1239, sector 6}
This FFI-only target is highly multi-periodic with many p\:modes and some lower frequencies that are most likely g\:modes. Some of the extracted peaks could by Nyquist aliases of the real ones.

\subsection*{T\,Ori, sector 6}
T\,Ori appears to be a $\delta$\,Sct star pulsating in a single mode, at the precision of these TESS FFI data. It lies between the theoretical \citep{dupretetal2005b} and observational \citep{murphyetal2019} red edges of the $\delta$\,Sct instability strip. T\,Ori also exhibits long-period variability on timescales of 5\,d, not dissimilar in appearance to RR\,Lyr pulsations.

%%%%%%%%%%%%%%%%%%%%%%%%%%%%%
%%%%%%%%%%%%%%%%%%%%%%%%%%%%%

\section{Fourier transforms of $\lambda$\,Boo star light curves}
\label{sec:FTs}

In this online-only\footnote{\url{https://doi.org/10.5281/zenodo.3727290}} appendix we show Fourier transforms of the TESS light curves. For all targets, a panel from 0 to 80\,d$^{-1}$ is shown. For stars with only FFI data, integer multiples of the Nyquist frequency are shown with vertical dashed red lines. For $\delta$\,Sct stars, a zoomed plot focussing around the area of the strongest p\:mode is also shown adjacent to the first panel, using that same amplitude range (y-axis limit) for consistency. The extracted frequencies given in Table\:\ref{tab:puls} of the main text are indicated with red diamonds, where possible. Note that peaks corresponding to the harmonics of g\:modes are not identified as p\:modes in any case.

\bibliographystyle{mnras}
\interlinepenalty=10000 % stops citations being broken over a page boundary, which can otherwise cause compilation to fail.
\bibliography{sjm_bibliography} % name of .bib file, without extension

\bsp	% typesetting comment
%\label{lastpage}
\end{document}

%% file: main_table.tex
\centering
\begin{tabular}{rrrrrrrccc}
\toprule
Name	&	$f_1$	&	$a_1$	&	$f_2$	&	$a_2$	&	$f_3$	&	$a_3$	&	FFI	&	limit	&	TESS Sectors	\\
& d$^{-1}$ & mmag & d$^{-1}$ & mmag & d$^{-1}$ & mmag & only & mmag & \\ 
\midrule
HD\,319	&		&		&		&		&		&		&		&	0.01	&	2	\\
HD\,3922	&	8.542	&	6.487	&	5.020	&	5.158	&	8.720	&	4.754	&		&		&	1--2	\\
HD\,4158	&	8.715	&	3.328	&	7.106	&	2.578	&	9.169	&	1.233	&		&		&	3	\\
HD\,6870	&	17.873	&	1.873	&	15.429	&	1.490	&	15.860	&	1.252	&		&		&	1--2	\\
HD\,7908	&		&		&		&		&		&		&		&	0.02	&	3	\\
HD\,11413	&	26.804	&	8.450	&	25.215	&	8.342	&	25.107	&	3.303	&		&		&	2--3	\\
HD\,13755	&	7.925	&	7.172	&	12.662	&	4.989	&	15.080	&	4.850	&		&		&	2--3	\\
HD\,17341	&	24.451	&	2.419	&	23.276	&	2.247	&	20.768	&	1.654	&		&		&	3--4	\\
HD\,23392	&		&		&		&		&		&		&		&	0.02	&	4	\\
HD\,28490	&	5.232	&	7.406	&	6.948	&	4.068	&	7.487	&	2.836	&		&		&	5	\\
HD\,28548	&	61.517	&	3.573	&	65.246	&	3.475	&	57.559	&	2.341	&		&		&	5	\\
HD\,30422	&	48.091	&	1.243	&	51.242	&	0.922	&	41.531	&	0.839	&		&		&	5	\\
HD\,31295	&	41.390	&	0.634	&	44.999	&	0.184	&	37.196	&	0.177	&		&		&	5	\\
HD\,31508	&	19.040	&	1.447	&	20.278	&	0.973	&	14.321	&	0.861	&		&		&	1--13	\\
HD\,36726	&		&		&		&		&		&		&		&	0.04	&	6	\\
HD\,37411	&		&		&		&		&		&		&	FFI	&	0.2	&	6	\\
HD\,38043	&	14.004	&	2.800	&	14.510	&	1.348	&	15.034	&	0.854	&		&		&	1--3, 5--6, 8--13	\\
HD\,41958	&	12.429	&	8.161	&	6.390	&	7.815	&	11.754	&	7.478	&		&		&	1, 4--5, 7--8, 11	\\
HD\,42503	&	7.178	&	4.315	&	5.250	&	4.231	&	6.185	&	3.219	&		&		&	5--7	\\
HD\,43533	&	23.558	&	0.267	&	30.937	&	0.076	&	26.193	&	0.043	&		&		&	6--7	\\
HD\,44930	&	18.152	&	2.924	&	23.994	&	2.436	&	25.418	&	1.067	&		&		&	6	\\
HD\,46722	&	20.764	&	8.981	&	43.360	&	5.084	&	20.233	&	4.377	&		&		&	6--7	\\
HD\,47425	&	14.201	&	8.945	&	15.841	&	8.648	&	5.931	&	3.292	&		&		&	6--7	\\
HD\,68695	&	17.172	&	2.078	&	13.827	&	0.438	&	17.172	&	0.167	&	FFI	&		&	7--8	\\
HD\,73211	&	17.588	&	0.089	&	21.346	&	0.059	&	5.324	&	0.056	&		&		&	8	\\
HD\,74423	&	8.757	&	1.901	&		&		&		&		&		&		&	8--11	\\
HD\,75654	&	14.340	&	3.289	&	15.240	&	0.401	&	13.017	&	0.216	&		&		&	8--9	\\
HD\,76097	&	8.933	&	11.666	&	9.872	&	4.747	&	8.800	&	3.237	&		&		&	8	\\
HD\,80426	&	11.016	&	10.076	&	39.305	&	0.622	&	31.271	&	0.431	&		&		&	8	\\
HD\,83041	&	14.528	&	1.060	&	14.085	&	0.314	&	15.862	&	0.295	&		&		&	8--9	\\
HD\,84159	&	9.139	&	43.124	&	10.900	&	8.558	&	8.003	&	7.580	&		&		&	8	\\
HD\,88554	&	5.648	&	0.081	&	7.939	&	0.070	&		&		&	FFI	&		&	9--10	\\
HD\,94326	&	10.130	&	0.778	&	17.607	&	0.428	&	10.383	&	0.404	&		&		&	9--10	\\
HD\,94390	&		&		&		&		&		&		&		&	0.02	&	9--10	\\
HD\,98069	&	13.094	&	3.529	&	15.426	&	2.615	&	16.225	&	2.067	&	FFI	&		&	9	\\
HD\,100546	&		&		&		&		&		&		&	FFI	&	0.2	&	11	\\
HD\,101412	&		&		&		&		&		&		&	FFI	&	0.08	&	10--11	\\
HD\,102541	&	19.898	&	7.326	&	19.490	&	1.126	&	37.793	&	1.060	&		&		&	10	\\
HD\,103701	&	6.188	&	3.504	&	4.653	&	3.399	&	12.376	&	0.695	&		&		&	10	\\
HD\,109738	&	33.769	&	3.623	&	12.083	&	1.688	&	33.328	&	1.443	&		&		&	11	\\
HD\,111786	&	15.023	&	1.728	&	31.020	&	1.690	&	36.006	&	1.622	&		&		&	10	\\
HD\,112682	&		&		&		&		&		&		&		&	0.04	&	10	\\
HD\,112948	&	29.755	&	3.449	&	31.825	&	2.332	&	27.040	&	2.113	&		&		&	10	\\
HD\,119896	&		&		&		&		&		&		&	FFI	&	0.05	&	11	\\
HD\,120122	&	8.555	&	0.113	&	16.066	&	0.093	&		&		&		&		&	11	\\
HD\,125508	&	16.066	&	1.028	&	9.257	&	0.949	&	8.931	&	0.736	&		&		&	11	\\
HD\,126627	&		&		&		&		&		&		&		&	0.04	&	11--12	\\
HD\,127659	&	6.537	&	4.144	&	6.884	&	3.351	&	11.904	&	2.442	&		&		&	11--12	\\
HD\,139614	&	34.777	&	1.557	&	21.365	&	1.142	&	28.241	&	0.881	&		&		&	12	\\
HD\,153747	&	21.471	&	6.266	&	19.879	&	2.862	&	22.809	&	2.283	&		&		&	12	\\
HD\,154153	&		&		&		&		&		&		&		&	0.015	&	12	\\
HD\,154951	&	9.335	&	0.228	&		&		&		&		&	FFI	&		&	12	\\
HD\,159021	&	13.825	&	0.931	&	10.707	&	0.819	&	14.156	&	0.521	&		&		&	12--13	\\
HD\,162193	&	24.068	&	7.926	&	19.342	&	3.286	&	20.566	&	1.311	&		&		&	13	\\
HD\,168740	&	30.757	&	4.565	&	12.576	&	3.528	&	24.082	&	2.639	&		&		&	13	\\
HD\,168947	&	14.594	&	2.339	&	17.644	&	1.518	&	15.242	&	0.929	&	FFI	&		&	13	\\
HD\,169142	&		&		&		&		&		&		&	FFI	&	0.15	&	13	\\
HD\,169346	&	23.441	&	1.341	&	20.120	&	0.994	&	21.225	&	0.905	&	FFI	&		&	13	\\
HD\,184779	&	12.566	&	9.401	&	13.837	&	4.812	&	22.875	&	3.315	&		&		&	13	\\
HD\,188164	&		&		&		&		&		&		&	FFI	&	0.015	&	13	\\
\bottomrule
\end{tabular}

%% file: main_table_part2.tex
\centering
\begin{tabular}{rrrrrrrccc}
\toprule
Name	&	$f_1$	&	$a_1$	&	$f_2$	&	$a_2$	&	$f_3$	&	$a_3$	&	FFI	&	limit	&	TESS Sectors	\\
& d$^{-1}$ & mmag & d$^{-1}$ & mmag & d$^{-1}$ & mmag & only & mmag & \\ 
\midrule
HD\,198160	&	35.961	&	0.967	&	33.302	&	0.945	&	41.137	&	0.767	&		&		&	13	\\
%HD\,198161	&	35.961	&	0.967	&	33.302	&	0.945	&	41.137	&	0.767	&		&		&	13	\\
HD\,203709	&	14.110	&	2.362	&	16.083	&	1.803	&	6.277	&	1.787	&		&		&	1	\\
HD\,210111	&	30.207	&	1.211	&	18.593	&	1.192	&	16.233	&	1.128	&		&		&	1	\\
HD\,213669	&	28.242	&	12.647	&	14.017	&	10.228	&	15.430	&	7.520	&		&		&	1	\\
HD\,214582	&	19.763	&	0.542	&	18.185	&	0.214	&	nan	&	nan	&		&		&	1	\\
HD\,223352	&		&		&		&		&		&		&		&	0.02	&	2	\\
HD\,290799	&	22.534	&	4.955	&	38.940	&	2.059	&	21.734	&	0.947	&		&		&	6	\\
HD\,294253	&		&		&		&		&		&		&		&	0.05	&	6	\\
BD-11\,1239	&	13.150	&	1.630	&	15.835	&	0.844	&	10.189	&	0.765	&	FFI	&		&	6	\\
T\,Ori	&	10.961	&	0.496	&		&		&		&		&	FFI	&		&	6	\\
\bottomrule
\end{tabular}

%% file: southern_tess_LB_params_v5_part1.tex
\centering
\begin{tabular}{rrr@{}lr@{}lr@{}lr@{}lr@{}lr@{}lrrcrr}
\toprule
HD  & $V$ &  \multicolumn{2}{c}{$T_{\rm eff}$}  &  \multicolumn{2}{c}{$L$}  &  \multicolumn{2}{c}{$M_V$}  &  \multicolumn{2}{c}{$\log g$}  &  \multicolumn{2}{c}{[Fe/H]}  &  \multicolumn{2}{c}{$M$}  &  $A_V$  &  RUWE  &  good  &  $\Delta m$  &  Sep.\\
  & mag & \multicolumn{2}{c}{K}  &  \multicolumn{2}{c}{L$_{\odot}$}  &  \multicolumn{2}{c}{mag}  &  \multicolumn{2}{c}{}  &  \multicolumn{2}{c}{}  &  \multicolumn{2}{c}{M$_{\odot}$}  &  mag  & --- &   &  mag  &  arcsec\\
\midrule
\vspace{1.5mm}
HD\,319  & 5.92 & 8080 &  $\pm$162  & 23.72 &  $^{+1.13}_{-1.06}$  & 1.28 &  $^{+0.05}_{-0.05}$  & 3.95 &  $^{+0.04}_{-0.04}$  & 0.00 &  $^{+0.12}_{-0.10}$  & 2.02 &  $^{+0.06}_{-0.05}$  & 0.01 & 1.33 & 1 & 5.10 &  2.20\\
\vspace{1.5mm}
HD\,3922  & 8.34 & 7270 &  $\pm$145  & 22.57 &  $^{+1.02}_{-0.96}$  & 1.30 &  $^{+0.05}_{-0.05}$  & 3.77 &  $^{+0.04}_{-0.03}$  & -0.01 &  $^{+0.08}_{-0.09}$  & 1.96 &  $^{+0.04}_{-0.04}$  & 0 & 1.01 & 1 &  ---  &  --- \\
\vspace{1.5mm}
HD\,4158  & 9.54 & 7820 &  $\pm$156  & 34.12 &  $^{+2.06}_{-1.93}$  & 0.84 &  $^{+0.06}_{-0.06}$  & 3.77 &  $^{+0.04}_{-0.04}$  & 0.02 &  $^{+0.07}_{-0.07}$  & 2.17 &  $^{+0.04}_{-0.04}$  & 0.58 & 0.98 & 1 &  ---  &  --- \\
\vspace{1.5mm}
HD\,6870  & 7.45 & 7300 &  $\pm$146  & 8.69 &  $^{+0.36}_{-0.35}$  & 2.34 &  $^{+0.04}_{-0.04}$  & 4.11 &  $^{+0.04}_{-0.05}$  & -0.01 &  $^{+0.15}_{-0.16}$  & 1.61 &  $^{+0.08}_{-0.09}$  & 0 & 1.08 & 1 &  ---  &  --- \\
\vspace{1.5mm}
HD\,7908  & 7.30 & 7120 &  $\pm$142  & 5.94 &  $^{+0.25}_{-0.23}$  & 2.73 &  $^{+0.04}_{-0.04}$  & 4.20 &  $^{+0.04}_{-0.05}$  & -0.02 &  $^{+0.12}_{-0.16}$  & 1.48 &  $^{+0.07}_{-0.08}$  & 0 & 1.02 & 1 &  ---  &  --- \\
\vspace{1.5mm}
HD\,11413  & 5.93 & 7870 &  $\pm$157  & 19.31 &  $^{+0.80}_{-0.76}$  & 1.45 &  $^{+0.04}_{-0.04}$  & 3.98 &  $^{+0.04}_{-0.05}$  & 0.02 &  $^{+0.12}_{-0.22}$  & 1.95 &  $^{+0.06}_{-0.13}$  & 0.01 & 0.93 & 1 &  ---  &  --- \\
\vspace{1.5mm}
HD\,13755  & 7.83 & 6990 &  $\pm$140  & 21.96 &  $^{+0.94}_{-0.89}$  & 1.31 &  $^{+0.04}_{-0.05}$  & 3.73 &  $^{+0.04}_{-0.04}$  & 0.09 &  $^{+0.09}_{-0.21}$  & 1.99 &  $^{+0.04}_{-0.16}$  & 0 & 0.99 & 1 &  ---  &  --- \\
\vspace{1.5mm}
HD\,17341  & 9.32 & 8250 &  $\pm$250  & 9.90 &  $^{+0.48}_{-0.45}$  & 2.22 &  $^{+0.05}_{-0.05}$  & 4.25 &  $^{+0.04}_{-0.05}$  & -0.14 &  $^{+0.13}_{-0.16}$  & 1.66 &  $^{+0.07}_{-0.08}$  & 0.10 & 1.01 & 1 &  ---  &  --- \\
\vspace{1.5mm}
HD\,23392  & 8.24 & 10140 &  $\pm$203  & 28.69 &  $^{+1.61}_{-1.52}$  & 1.37 &  $^{+0.06}_{-0.06}$  & 4.26 &  $^{+0.03}_{-0.04}$  & -0.24 &  $^{+0.10}_{-0.15}$  & 2.14 &  $^{+0.05}_{-0.09}$  & 0.16 & 1.24 & 0 &  ---  &  --- \\
\vspace{1.5mm}
HD\,28490  & 9.53 & 7230 &  $\pm$145  & 27.96 &  $^{+1.91}_{-1.77}$  & 1.06 &  $^{+0.07}_{-0.07}$  & 3.68 &  $^{+0.05}_{-0.04}$  & -0.09 &  $^{+0.10}_{-0.08}$  & 1.98 &  $^{+0.04}_{-0.04}$  & 0.05 & 0.89 & 1 &  ---  &  --- \\
\vspace{1.5mm}
HD\,28548  & 9.22 & 8490 &  $\pm$170  & 10.04 &  $^{+0.48}_{-0.46}$  & 2.22 &  $^{+0.05}_{-0.05}$  & 4.29 &  $^{+0.03}_{-0.03}$  & -0.25 &  $^{+0.13}_{-0.15}$  & 1.64 &  $^{+0.07}_{-0.07}$  & 0.12 & 1.06 & 1 &  ---  &  --- \\
\vspace{1.5mm}
HD\,30422  & 6.17 & 7930 &  $\pm$159  & 8.42 &  $^{+0.35}_{-0.33}$  & 2.38 &  $^{+0.04}_{-0.04}$  & 4.25 &  $^{+0.04}_{-0.04}$  & -0.13 &  $^{+0.12}_{-0.14}$  & 1.60 &  $^{+0.06}_{-0.07}$  & 0 & 0.90 & 1 &  ---  &  --- \\
\vspace{1.5mm}
HD\,31295  & 4.65 & 8900 &  $\pm$231  & 16.91 &  $^{+1.42}_{-1.29}$  & 1.71 &  $^{+0.09}_{-0.09}$  & 4.22 &  $^{+0.04}_{-0.05}$  & -0.12 &  $^{+0.14}_{-0.16}$  & 1.90 &  $^{+0.08}_{-0.09}$  & 0.09 & 0 & 1 &  ---  &  --- \\
\vspace{1.5mm}
HD\,31508  & 9.57 & 6750 &  $\pm$250  & 28.88 &  $^{+1.54}_{-1.45}$  & 1.04 &  $^{+0.06}_{-0.06}$  & 3.60 &  $^{+0.04}_{-0.05}$  & 0.03 &  $^{+0.11}_{-0.12}$  & 2.05 &  $^{+0.06}_{-0.14}$  & 0.05 & 1.07 & 0 &  ---  &  --- \\
\vspace{1.5mm}
HD\,36726  & 8.85 & 9950 &  $\pm$418  & 40.73 &  $^{+2.47}_{-2.31}$  & 0.95 &  $^{+0.06}_{-0.06}$  & 4.13 &  $^{+0.06}_{-0.10}$  & -0.07 &  $^{+0.15}_{-0.18}$  & 2.36 &  $^{+0.09}_{-0.16}$  & 0.29 & 0.92 & 1 &  ---  &  --- \\
\vspace{1.5mm}
HD\,37411  & 9.79 & 11462 &  $\pm$229  &  ---  &    &  ---  &    &  ---  &    &  ---  &    &  ---  &    & 1.56 & 2.92 & 0 &  ---  &  --- \\
\vspace{1.5mm}
HD\,38043  & 9.45 & 7750 &  $\pm$250  & 8.53 &  $^{+0.41}_{-0.39}$  & 2.36 &  $^{+0.05}_{-0.05}$  & 4.21 &  $^{+0.05}_{-0.06}$  & -0.07 &  $^{+0.13}_{-0.15}$  & 1.61 &  $^{+0.07}_{-0.08}$  & 0.06 & 1.08 & 1 &  ---  &  --- \\
\vspace{1.5mm}
HD\,41958  & 8.78 & 8000 &  $\pm$250  & 43.23 &  $^{+2.11}_{-1.98}$  & 0.61 &  $^{+0.05}_{-0.05}$  & 3.69 &  $^{+0.06}_{-0.05}$  & -0.14 &  $^{+0.13}_{-0.10}$  & 2.19 &  $^{+0.06}_{-0.05}$  & 0.21 & 1.11 & 1 &  ---  &  --- \\
\vspace{1.5mm}
HD\,42503  & 7.43 & 7460 &  $\pm$149  & 44.53 &  $^{+1.98}_{-1.87}$  & 0.57 &  $^{+0.05}_{-0.05}$  & 3.59 &  $^{+0.03}_{-0.03}$  & 0.00 &  $^{+0.07}_{-0.07}$  & 2.27 &  $^{+0.04}_{-0.03}$  & 0 & 1.04 & 1 &  ---  &  --- \\
\vspace{1.5mm}
HD\,43533  & 9.61 & 8000 &  $\pm$250  & 6.05 &  $^{+0.28}_{-0.27}$  & 2.74 &  $^{+0.05}_{-0.05}$  & 4.31 &  $^{+0.04}_{-0.04}$  & -0.28 &  $^{+0.14}_{-0.19}$  & 1.44 &  $^{+0.06}_{-0.06}$  & 0.14 & 1.13 & 1 &  ---  &  --- \\
\vspace{1.5mm}
HD\,44930  & 9.42 & 7260 &  $\pm$145  & 6.41 &  $^{+0.35}_{-0.33}$  & 2.67 &  $^{+0.06}_{-0.06}$  & 4.21 &  $^{+0.04}_{-0.05}$  & -0.04 &  $^{+0.12}_{-0.15}$  & 1.50 &  $^{+0.07}_{-0.08}$  & 0.06 & 0.72 & 1 &  ---  &  --- \\
\vspace{1.5mm}
HD\,46722  & 9.29 & 8000 &  $\pm$250  & 7.86 &  $^{+0.35}_{-0.33}$  & 2.46 &  $^{+0.05}_{-0.05}$  & 4.27 &  $^{+0.04}_{-0.05}$  & -0.16 &  $^{+0.12}_{-0.14}$  & 1.57 &  $^{+0.05}_{-0.07}$  & 0.07 & 0.96 & 1 &  ---  &  --- \\
\vspace{1.5mm}
HD\,47425  & 9.53 & 8250 &  $\pm$250  & 41.06 &  $^{+2.57}_{-2.40}$  & 0.67 &  $^{+0.07}_{-0.07}$  & 3.78 &  $^{+0.11}_{-0.07}$  & -0.14 &  $^{+0.35}_{-0.14}$  & 2.16 &  $^{+0.25}_{-0.06}$  & 0.19 & 0.85 & 1 &  ---  &  --- \\
\vspace{1.5mm}
HD\,68695  & 9.87 & 8436 &  $\pm$169  & 57.47 &  $^{+2.21}_{-2.12}$  &  ---  &    & 3.76 &  $^{+0.03}_{-0.04}$  & 0.16 &  $^{+0.07}_{-0.08}$  & 2.55 &  $^{+0.04}_{-0.04}$  & 0.10 & 1.08 & 1 &  ---  &  --- \\
\vspace{1.5mm}
HD\,73211  & 9.02 & 6750 &  $\pm$250  & 12.73 &  $^{+0.63}_{-0.59}$  & 1.94 &  $^{+0.05}_{-0.05}$  & 3.84 &  $^{+0.07}_{-0.06}$  & 0.01 &  $^{+0.15}_{-0.15}$  & 1.71 &  $^{+0.09}_{-0.08}$  & 0.64 & 1.16 & 1 &  ---  &  --- \\
\vspace{1.5mm}
HD\,74423  & 8.61 & 8250 &  $\pm$250  & 69.30 &  $^{+3.74}_{-3.54}$  & 0.10 &  $^{+0.06}_{-0.06}$  & 3.65 &  $^{+0.04}_{-0.05}$  & 0.11 &  $^{+0.09}_{-0.09}$  & 2.62 &  $^{+0.05}_{-0.05}$  & 0.10 & 1.12 & 0 &  ---  &  --- \\
\vspace{1.5mm}
HD\,75654  & 6.37 & 7270 &  $\pm$145  & 11.00 &  $^{+0.45}_{-0.43}$  & 2.10 &  $^{+0.04}_{-0.04}$  & 4.03 &  $^{+0.05}_{-0.05}$  & 0.00 &  $^{+0.15}_{-0.17}$  & 1.68 &  $^{+0.09}_{-0.09}$  & 0 & 0.92 & 1 &  ---  &  --- \\
\vspace{1.5mm}
HD\,76097  & 9.56 & 7500 &  $\pm$250  & 22.56 &  $^{+1.38}_{-1.28}$  & 1.30 &  $^{+0.06}_{-0.06}$  & 3.83 &  $^{+0.08}_{-0.06}$  & -0.03 &  $^{+0.19}_{-0.14}$  & 1.94 &  $^{+0.12}_{-0.06}$  & 0.35 & 1.06 & 1 &  ---  &  --- \\
\vspace{1.5mm}
HD\,80426  & 9.16 & 6500 &  $\pm$250  & 25.75 &  $^{+1.50}_{-1.41}$  & 1.18 &  $^{+0.06}_{-0.06}$  & 3.60 &  $^{+0.05}_{-0.07}$  & 0.13 &  $^{+0.11}_{-0.13}$  & 2.07 &  $^{+0.06}_{-0.11}$  & 0.89 & 1.39 & 1 &  ---  &  --- \\
\vspace{1.5mm}
HD\,83041  & 8.79 & 7630 &  $\pm$930  & 39.03 &  $^{+1.91}_{-1.79}$  & 0.70 &  $^{+0.05}_{-0.05}$  & 3.71 &  $^{+0.23}_{-0.11}$  & 0.00 &  $^{+0.15}_{-0.16}$  & 2.23 &  $^{+0.10}_{-0.11}$  & 0.66 & 1.07 & 1 &  ---  &  --- \\
\vspace{1.5mm}
HD\,84159  & 9.51 & 7000 &  $\pm$250  & 70.08 &  $^{+5.68}_{-5.20}$  & 0.06 &  $^{+0.08}_{-0.09}$  & 3.36 &  $^{+0.06}_{-0.05}$  & -0.04 &  $^{+0.17}_{-0.17}$  & 2.29 &  $^{+0.08}_{-0.09}$  & 0.48 & 1.05 & 1 &  ---  &  --- \\
\vspace{1.5mm}
HD\,88554  & 9.32 & 6920 &  $\pm$153  & 30.63 &  $^{+1.63}_{-1.53}$  & 0.96 &  $^{+0.06}_{-0.06}$  & 3.60 &  $^{+0.03}_{-0.03}$  & 0.02 &  $^{+0.10}_{-0.09}$  & 2.06 &  $^{+0.05}_{-0.05}$  & 0.14 & 1.24 & 1 &  ---  &  --- \\
\vspace{1.5mm}
HD\,94326  & 7.76 & 7750 &  $\pm$155  & 52.25 &  $^{+2.41}_{-2.27}$  & 0.39 &  $^{+0.05}_{-0.05}$  & 3.60 &  $^{+0.03}_{-0.02}$  & -0.03 &  $^{+0.11}_{-0.09}$  & 2.28 &  $^{+0.05}_{-0.05}$  & 0.05 & 0.93 & 1 &  ---  &  --- \\
\vspace{1.5mm}
HD\,94390  & 8.96 & 7000 &  $\pm$250  & 21.11 &  $^{+1.00}_{-0.94}$  & 1.37 &  $^{+0.05}_{-0.05}$  & 3.75 &  $^{+0.06}_{-0.07}$  & 0.04 &  $^{+0.14}_{-0.18}$  & 1.94 &  $^{+0.07}_{-0.14}$  & 0.42 & 1.03 & 0 &  ---  &  --- \\
\vspace{1.5mm}
HD\,98069  & 8.16 & 7500 &  $\pm$250  & 37.8 &  $^{+2.09}_{-1.98}$  & 0.73 &  $^{+0.06}_{-0.06}$  & 3.68 &  $^{+0.05}_{-0.05}$  & 0.06 &  $^{+0.10}_{-0.10}$  & 2.24 &  $^{+0.05}_{-0.06}$  & 0.06 & 1.03 & 1 &  ---  &  --- \\
\vspace{1.5mm}
HD\,100546  & 6.30 & 10720 &  $\pm$214  & 33.71 &  $^{+1.44}_{-1.36}$  & 1.31 &  $^{+0.04}_{-0.04}$  & 4.29 &  $^{+0.03}_{-0.03}$  & -0.37 &  $^{+0.10}_{-0.12}$  & 2.17 &  $^{+0.04}_{-0.05}$  & 0.20 & 1.19 & 0 & 10.10 &  4.50\\
\vspace{1.5mm}
HD\,101412  & 9.29 & 8101 &  $\pm$162  & 25.43 &  $^{+1.27}_{-1.19}$  & 1.17 &  $^{+0.05}_{-0.05}$  & 3.93 &  $^{+0.04}_{-0.04}$  & 0.00 &  $^{+0.11}_{-0.09}$  & 2.04 &  $^{+0.06}_{-0.05}$  & 0.09 & 0.95 & 1 &  ---  &  --- \\
\vspace{1.5mm}
HD\,102541  & 7.94 & 7690 &  $\pm$154  & 7.24 &  $^{+0.32}_{-0.30}$  & 2.54 &  $^{+0.05}_{-0.05}$  & 4.25 &  $^{+0.03}_{-0.04}$  & -0.12 &  $^{+0.11}_{-0.14}$  & 1.54 &  $^{+0.06}_{-0.07}$  & 0.09 & 1.11 & 1 &  ---  &  --- \\
\vspace{1.5mm}
HD\,103701  & 8.73 & 7000 &  $\pm$250  & 41.41 &  $^{+3.39}_{-3.09}$  & 0.66 &  $^{+0.08}_{-0.09}$  & 3.56 &  $^{+0.05}_{-0.06}$  & 0.13 &  $^{+0.11}_{-0.13}$  & 2.31 &  $^{+0.06}_{-0.15}$  & 0.29 & 2.04 & 0 &  ---  &  --- \\
\vspace{1.5mm}
HD\,109738  & 8.30 & 7600 &  $\pm$152  & 9.83 &  $^{+0.43}_{-0.41}$  & 2.19 &  $^{+0.05}_{-0.05}$  & 4.14 &  $^{+0.04}_{-0.04}$  & -0.02 &  $^{+0.14}_{-0.16}$  & 1.67 &  $^{+0.08}_{-0.09}$  & 0.09 & 0.90 & 1 &  ---  &  --- \\
\vspace{1.5mm}
HD\,111786  & 6.14 & 7490 &  $\pm$150  & 10.26 &  $^{+0.45}_{-0.42}$  & 2.15 &  $^{+0.05}_{-0.05}$  & 4.11 &  $^{+0.05}_{-0.04}$  & -0.01 &  $^{+0.15}_{-0.16}$  & 1.68 &  $^{+0.09}_{-0.09}$  & 0.02 & 1.98 & 1 &  ---  &  --- \\
\bottomrule
\end{tabular}

%% file: southern_tess_LB_params_v5_part2.tex
\centering
\begin{tabular}{rrr@{}lr@{}lr@{}lr@{}lr@{}lr@{}lrrcrr}
\toprule
HD  &  $V$ & \multicolumn{2}{c}{$T_{\rm eff}$}  &  \multicolumn{2}{c}{$L$}  &  \multicolumn{2}{c}{$M_V$}  &  \multicolumn{2}{c}{$\log g$}  &  \multicolumn{2}{c}{[Fe/H]}  &  \multicolumn{2}{c}{$M$}  &  $A_V$  &  RUWE  &  good  &  $\Delta m$  &  Sep.\\
  & mag &  \multicolumn{2}{c}{K}  &  \multicolumn{2}{c}{L$_{\odot}$}  &  \multicolumn{2}{c}{mag}  &  \multicolumn{2}{c}{}  &  \multicolumn{2}{c}{}  &  \multicolumn{2}{c}{M$_{\odot}$}  &  mag  & --- &   &  mag  &  arcsec\\
\midrule
\vspace{1.5mm}
HD\,112682 & 9.67 & 8000 &   $\pm$250   & 45.62 &   $^{+3.63}_{-3.32}$   & 0.52 &   $^{+0.08}_{-0.08}$   & 3.68 &   $^{+0.08}_{-0.06}$   & -0.14 &   $^{+0.16}_{-0.11}$   & 2.19 &   $^{+0.08}_{-0.05}$   & 0.26 & 1.04 & 1 &   ---   &   --- \\
\vspace{1.5mm}
HD\,112948 & 9.35 & 8750 &   $\pm$250   & 22.52 &   $^{+2.19}_{-1.95}$   & 1.38 &   $^{+0.10}_{-0.10}$   & 4.11 &   $^{+0.06}_{-0.06}$   & -0.04 &   $^{+0.14}_{-0.16}$   & 2.02 &   $^{+0.09}_{-0.10}$   & 0.07 & 1.13 & 1 &   ---   &   --- \\
\vspace{1.5mm}
HD\,119896 & 8.22 & 6580 &   $\pm$132   & 20.15 &   $^{+1.23}_{-1.15}$   & 1.45 &   $^{+0.06}_{-0.06}$   & 3.65 &   $^{+0.04}_{-0.04}$   & 0.03 &   $^{+0.10}_{-0.08}$   & 1.89 &   $^{+0.05}_{-0.05}$   & 0.37 & 1.00 & 1 &   ---   &   --- \\
\vspace{1.5mm}
HD\,120122 & 9.11 & 7000 &   $\pm$250   & 8.18 &   $^{+0.44}_{-0.42}$   & 2.41 &   $^{+0.06}_{-0.06}$   & 4.06 &   $^{+0.07}_{-0.08}$   & 0.00 &   $^{+0.15}_{-0.17}$   & 1.57 &   $^{+0.09}_{-0.09}$   & 0.07 & 1.15 & 1 &   ---   &   --- \\
\vspace{1.5mm}
HD\,125508 & 9.14 & 7250 &   $\pm$250   &   ---   &      &   ---   &      &   ---   &      &   ---   &      &   ---   &      &   ---   & 16.63 & 0 &   ---   &   --- \\
\vspace{1.5mm}
HD\,126627 & 9.00 & 7070 &   $\pm$141   & 5.91 &   $^{+0.26}_{-0.24}$   & 2.75 &   $^{+0.05}_{-0.05}$   & 4.19 &   $^{+0.04}_{-0.05}$   & -0.01 &   $^{+0.13}_{-0.16}$   & 1.48 &   $^{+0.07}_{-0.08}$   & 0.09 & 0.90 & 1 &   ---   &   --- \\
\vspace{1.5mm}
HD\,127659 & 9.31 & 11900 &   $\pm$435   & 109.42 &   $^{+5.14}_{-4.84}$   & 0.28 &   $^{+0.05}_{-0.05}$   & 4.05 &   $^{+0.05}_{-0.05}$   & -0.23 &   $^{+0.12}_{-0.14}$   & 2.85 &   $^{+0.07}_{-0.06}$   & 1.34 & 0.93 & 0 &   ---   &   --- \\
\vspace{1.5mm}
HD\,139614 & 8.24 & 7626 &   $\pm$153   & 6.78 &   $^{+0.30}_{-0.28}$   & 2.59 &   $^{+0.05}_{-0.05}$   & 4.26 &   $^{+0.03}_{-0.04}$   & -0.12 &   $^{+0.11}_{-0.14}$   & 1.52 &   $^{+0.06}_{-0.08}$   & 0.02 & 1.09 & 1 & 3.53 &   1.40\\
\vspace{1.5mm}
HD\,153747 & 7.42 & 8190 &   $\pm$164   & 30.11 &   $^{+1.43}_{-1.34}$   & 1.01 &   $^{+0.05}_{-0.05}$   & 3.88 &   $^{+0.05}_{-0.04}$   & -0.06 &   $^{+0.12}_{-0.08}$   & 2.09 &   $^{+0.06}_{-0.04}$   & 0.12 & 0.90 & 1 &   ---   &   --- \\
\vspace{1.5mm}
HD\,154153 & 6.18 & 7400 &   $\pm$567   & 16.99 &   $^{+0.73}_{-0.69}$   & 1.60 &   $^{+0.04}_{-0.05}$   & 3.93 &   $^{+0.13}_{-0.15}$   & -0.01 &   $^{+0.15}_{-0.17}$   & 1.86 &   $^{+0.09}_{-0.10}$   & 0.34 & 1.30 & 1 &   ---   &   --- \\
\vspace{1.5mm}
HD\,154951 & 8.78 & 7030 &   $\pm$141   & 12.57 &   $^{+0.60}_{-0.57}$   & 1.94 &   $^{+0.05}_{-0.05}$   & 3.92 &   $^{+0.04}_{-0.05}$   & 0.00 &   $^{+0.15}_{-0.19}$   & 1.72 &   $^{+0.08}_{-0.10}$   & 0.23 & 0.98 & 1 &   ---   &   --- \\
\vspace{1.5mm}
HD\,159021 & 8.67 & 6750 &   $\pm$250   & 74.62 &   $^{+4.56}_{-4.27}$   & -0.01 &   $^{+0.06}_{-0.06}$   & 3.33 &   $^{+0.06}_{-0.05}$   & -0.02 &   $^{+0.20}_{-0.19}$   & 2.30 &   $^{+0.09}_{-0.10}$   & 0.05 & 1.09 & 1 &   ---   &   --- \\
\vspace{1.5mm}
HD\,162193 & 8.66 & 8750 &   $\pm$250   & 21.79 &   $^{+1.10}_{-1.03}$   & 1.42 &   $^{+0.05}_{-0.05}$   & 4.12 &   $^{+0.05}_{-0.05}$   & -0.04 &   $^{+0.14}_{-0.16}$   & 2.01 &   $^{+0.09}_{-0.10}$   & 0.09 & 1.14 & 1 &   ---   &   --- \\
\vspace{1.5mm}
HD\,168740 & 6.12 & 7670 &   $\pm$153   & 13.26 &   $^{+0.56}_{-0.53}$   & 1.86 &   $^{+0.04}_{-0.04}$   & 4.06 &   $^{+0.04}_{-0.05}$   & 0.00 &   $^{+0.15}_{-0.17}$   & 1.77 &   $^{+0.09}_{-0.09}$   & 0.02 & 0.98 & 1 &   ---   &   --- \\
\vspace{1.5mm}
HD\,168947 & 8.11 & 7510 &   $\pm$150   & 29.91 &   $^{+1.44}_{-1.35}$   & 1.00 &   $^{+0.05}_{-0.05}$   & 3.76 &   $^{+0.04}_{-0.06}$   & 0.14 &   $^{+0.10}_{-0.37}$   & 2.17 &   $^{+0.05}_{-0.24}$   & 0.12 & 0.95 & 0 &   ---   &   --- \\
\vspace{1.5mm}
HD\,169142 & 8.16 & 6700 &   $\pm$322   & 5.93 &   $^{+0.33}_{-0.31}$   & 2.87 &   $^{+0.14}_{-0.14}$   & 4.65 &   $^{+0.02}_{-0.03}$   & -0.79 &   $^{+0.13}_{-0.14}$   & 1.43 &   $^{+0.03}_{-0.03}$   & 0 & 0.86 & 0 & 6.50 &   0.10\\
\vspace{1.5mm}
HD\,169346 & 9.27 & 7520 &   $\pm$150   & 16.34 &   $^{+0.85}_{-0.80}$   & 1.66 &   $^{+0.05}_{-0.06}$   & 3.95 &   $^{+0.04}_{-0.04}$   & -0.01 &   $^{+0.17}_{-0.17}$   & 1.83 &   $^{+0.10}_{-0.10}$   & 0.13 & 0.82 & 1 &   ---   &   --- \\
\vspace{1.5mm}
HD\,184779 & 8.90 & 7130 &   $\pm$143   & 11.01 &   $^{+0.57}_{-0.53}$   & 2.07 &   $^{+0.05}_{-0.05}$   & 3.99 &   $^{+0.05}_{-0.04}$   & 0.00 &   $^{+0.15}_{-0.17}$   & 1.68 &   $^{+0.09}_{-0.09}$   & 0 & 1.02 & 1 &   ---   &   --- \\
\vspace{1.5mm}
HD\,188164 & 6.37 & 8160 &   $\pm$163   & 23.82 &   $^{+1.02}_{-0.97}$   & 1.26 &   $^{+0.04}_{-0.05}$   & 3.97 &   $^{+0.05}_{-0.04}$   & -0.02 &   $^{+0.17}_{-0.10}$   & 2.02 &   $^{+0.11}_{-0.05}$   & 0.08 & 1.18 & 1 &   ---   &   --- \\
\vspace{1.5mm}
HD\,198160 & 6.21 & 7930 &   $\pm$159   & 13.44 &   $^{+0.56}_{-0.53}$   & 1.85 &   $^{+0.04}_{-0.04}$   & 4.12 &   $^{+0.04}_{-0.05}$   & -0.02 &   $^{+0.14}_{-0.17}$   & 1.79 &   $^{+0.09}_{-0.09}$   & 0.02 & 1.11 & 0 & 0.35 &   3.20\\
\vspace{1.5mm}
HD\,198161 & 6.55 & 7900 &   $\pm$158   & 13.72 &   $^{+0.58}_{-0.55}$   & 1.85 &   $^{+0.04}_{-0.04}$   & 4.10 &   $^{+0.04}_{-0.05}$   & -0.01 &   $^{+0.15}_{-0.17}$   & 1.80 &   $^{+0.08}_{-0.10}$   & 0.02 & 1.24 & 0 &   ---   &   --- \\
\vspace{1.5mm}
HD\,203709 & 9.58 & 7750 &   $\pm$250   & 9.54 &   $^{+0.47}_{-0.45}$   & 2.23 &   $^{+0.05}_{-0.05}$   & 4.18 &   $^{+0.05}_{-0.06}$   & -0.05 &   $^{+0.13}_{-0.15}$   & 1.66 &   $^{+0.07}_{-0.09}$   & 0.07 & 0.98 & 1 &   ---   &   --- \\
\vspace{1.5mm}
HD\,210111 & 6.39 & 7470 &   $\pm$149   & 12.84 &   $^{+0.54}_{-0.51}$   & 1.92 &   $^{+0.04}_{-0.04}$   & 4.02 &   $^{+0.04}_{-0.04}$   & -0.01 &   $^{+0.16}_{-0.17}$   & 1.75 &   $^{+0.09}_{-0.09}$   & 0 & 0.98 & 1 &   ---   &   --- \\
\vspace{1.5mm}
HD\,213669 & 7.42 & 7350 &   $\pm$147   & 10.30 &   $^{+0.44}_{-0.42}$   & 2.15 &   $^{+0.04}_{-0.05}$   & 4.07 &   $^{+0.05}_{-0.05}$   & 0.00 &   $^{+0.15}_{-0.17}$   & 1.66 &   $^{+0.09}_{-0.08}$   & 0 & 1.10 & 1 &   ---   &   --- \\
\vspace{1.5mm}
HD\,214582 & 9.48 & 7250 &   $\pm$250   & 5.81 &   $^{+0.29}_{-0.27}$   & 2.77 &   $^{+0.05}_{-0.05}$   & 4.22 &   $^{+0.05}_{-0.06}$   & -0.07 &   $^{+0.14}_{-0.16}$   & 1.47 &   $^{+0.06}_{-0.08}$   & 0.02 & 1.21 & 1 &   ---   &   --- \\
\vspace{1.5mm}
HD\,223352 & 4.57 & 9930 &   $\pm$199   & 30.09 &   $^{+3.46}_{-3.00}$   & 1.27 &   $^{+0.11}_{-0.12}$   & 4.22 &   $^{+0.04}_{-0.04}$   & -0.14 &   $^{+0.11}_{-0.18}$   & 2.19 &   $^{+0.09}_{-0.13}$   & 0.06 & 0.00 & 1 & 7.03 &   3.40\\
\vspace{1.5mm}
HD\,290799 & 10.80 & 7940 &   $\pm$159   & 7.63 &   $^{+0.68}_{-0.62}$   & 2.47 &   $^{+0.09}_{-0.09}$   & 4.28 &   $^{+0.03}_{-0.04}$   & -0.18 &   $^{+0.12}_{-0.14}$   & 1.55 &   $^{+0.07}_{-0.08}$   & 0.02 & 1.29 & 1 &   ---   &   --- \\
\vspace{1.5mm}
HD\,294253 & 9.67 & 10580 &   $\pm$212   & 32.31 &   $^{+2.75}_{-2.50}$   & 1.33 &   $^{+0.09}_{-0.09}$   & 4.28 &   $^{+0.04}_{-0.03}$   & -0.33 &   $^{+0.11}_{-0.14}$   & 2.17 &   $^{+0.07}_{-0.08}$   & 0.23 & 1.18 & 0 &   ---   &   --- \\
\vspace{1.5mm}
BD-11\,1239 & 9.70 & 7000 &   $\pm$250   & 10.75 &   $^{+0.59}_{-0.56}$   & 2.11 &   $^{+0.06}_{-0.06}$   & 3.96 &   $^{+0.07}_{-0.08}$   & 0.00 &   $^{+0.15}_{-0.17}$   & 1.66 &   $^{+0.09}_{-0.09}$   & 0.15 & 1.12 & 1 &   ---   &   --- \\
\vspace{1.5mm}
T\,Ori & 11.25 &  6980  &   $\pm$140   &  9.06  &   $^{+1.22}_{-1.08}$   &  2.30  &   $^{+0.14}_{-0.14}$   &  4.03  &   $^{+0.07}_{-0.06}$   &  0.00  &   $^{+0.15}_{-0.17}$   &  1.58  &   $^{+0.10}_{-0.10}$   &  0.66  &  1.20  &  1  &   ---   &   --- \\
\bottomrule
\end{tabular}